\documentclass[aip,jcp,amsmath,reprint,citeautoscript]{revtex4-1}

\usepackage{mathptmx}
\usepackage{afthd}
\usepackage{textcomp,graphicx,dcolumn}
\usepackage{hyperref}
\hypersetup{bookmarksnumbered=true,bookmarksopen,bookmarksopenlevel=1,
            colorlinks,linkcolor=black,citecolor=black,urlcolor=black,
            pdfauthor={V. Holten, C. E. Bertrand, M. A. Anisimov, and J. V. Sengers},
			bookmarksdepth=2,pdfpagemode=UseOutlines,pdftitle={Thermodynamics of supercooled water},
            pdfprintscaling=None}
\usepackage[T1]{fontenc}
\usepackage[ansinew]{inputenc}

\newcommand{\ea}{\textit{et al.}}
\newcommand{\figref}[1]{Fig.~\ref{#1}}
\newcommand{\ffigref}[1]{Figure~\ref{#1}}
\newcommand{\figsref}[2]{Figs.~\ref{#1} and~\ref{#2}}
\renewcommand*{\eqref}[1]{Eq.~(\ref{#1})}
\newcommand*{\eeqref}[1]{Equation~(\ref{#1})}
\newcommand*{\eqsref}[2]{Eqs.~(\ref{#1}) and (\ref{#2})}
\newcommand*{\eeqsref}[2]{Equations~(\ref{#1}) and (\ref{#2})}
\newcommand{\tabref}[1]{Table~\ref{#1}}
\newcommand{\pypxl}[2]{{\partial{#1}/\partial{#2}}}
\newcommand{\pypxd}[3]{\left(\frac{\partial{#1}}{\partial{#2}}\right)_{#3}}
\newcommand{\di}{\mathrm{d}}
\newcommand*{\chem}[1]{\ensuremath{\mathrm{#1}}}

\newcommand*{\Dt}{\Delta \hat{T}}
\newcommand*{\Dm}{\Delta \hat{\mu}}
\newcommand*{\Dp}{\Delta \hat{P}}
\newcommand*{\Th}{\hat{T}}
\newcommand*{\Vh}{\hat{V}}
\newcommand*{\Ph}{\hat{P}}
\newcommand*{\Pc}{P_\text{c}}
\newcommand*{\Tc}{T_\text{c}}
\newcommand*{\Vc}{V_\text{c}}
\newcommand*{\mh}{\hat{\mu}}
\newcommand*{\kap}{\hat{\kappa}_T}
\newcommand*{\alp}{\hat{\alpha}_V}
\newcommand*{\Sh}{\hat{S}}
\newcommand*{\Cph}{\hat{C}_P}
\newcommand*{\Cvh}{\hat{C}_V}
\newcommand*{\mr}{\Delta\hat{\mu}^\text{r}}

\begin{document}
\title{Thermodynamics of supercooled water}

\affiliation{Institute for Physical Science and Technology and
  Department of Chemical and Biomolecular Engineering,\\
  University of Maryland, College Park, Maryland 20742, USA}
\author{V. Holten}
\author{C. E. Bertrand}
\altaffiliation[Current address: ]{Department of Nuclear Science and Engineering,
Massachusetts Institute of Technology, Cambridge, Massachusetts 02139, USA}
\author{M. A. Anisimov}
\email[Author to whom correspondence should be addressed. Electronic mail: ]
{anisimov@umd.edu}
\author{J. V. Sengers}
\date{\today}

\begin{abstract}
We review the available experimental information on the thermodynamic properties of
supercooled ordinary and heavy water and demonstrate the possibility of modeling these
thermodynamic properties on a theoretical basis. We show that by assuming the existence
of a liquid--liquid critical point in supercooled water, the theory of critical phenomena
can give an accurate account of the experimental thermodynamic-property data up to a
pressure of 150~MPa. In addition, we show that a phenomenological extension of the
theoretical model can account for all currently available experimental data in the
supercooled region, up to 400~MPa. The stability limit of the liquid state and possible
coupling between crystallization and liquid--liquid separation are discussed. It is
concluded that critical-point thermodynamics describes the available thermodynamic data
for supercooled water within experimental accuracy, thus establishing a benchmark for
further developments in this area.
\end{abstract}

\maketitle

\section{Introduction}
The peculiar thermodynamic behavior of supercooled water continues to receive
considerable attention, despite several decades of work in this field. Upon supercooling,
water exhibits an anomalous increase of its isobaric heat capacity and its isothermal
compressibility, and an anomalous decrease of its expansivity coefficient.\cite{deben03}
One explanation of these anomalies, originally proposed by Poole \ea,\cite{poole1992} is
based on the presumed existence of a liquid--liquid critical point (LLCP) in water in the
deeply supercooled region. The hypothesis of the existence of a critical point in
metastable water has been considered by many authors, as recently reviewed by Bertrand
and Anisimov.\cite{bertrand2011} In particular, several authors have made attempts to
develop a thermodynamic model for the thermodynamic properties of supercooled water based
on the LLCP scenario.
\cite{bertrand2011,jeffery1999,kiselev2001,kiselev2002,fuentevilla2006,kalova2010} The
existence of a liquid--liquid critical point in supercooled water is still being debated,
especially in view of some recent simulations.\cite{limmer2011,wikfeldt2011} The purpose
of this article is to demonstrate that a theoretical model based on the presumed
existence of a second critical point in water is capable of representing, accurately and
consistently, all available experimental thermodynamic property data for supercooled
ordinary and heavy water. While being conceptually close to the previous works by
Fuentevilla and Anisimov\cite{fuentevilla2006} and Bertrand and
Anisimov,\cite{bertrand2011} this work is the first comprehensive correlation of
thermodynamic properties of supercooled water, incorporating non-critical backgrounds in
a thermodynamically consistent way.

This article is organized as follows. In Sec.~\ref{sec:expdata} we review the currently
available experimental information for the thermodynamic properties of supercooled water.
We also provide an assessment of the IAPWS-95 formulation \cite{iapws95,wag02nonote} for
the thermodynamic properties of H$_2$O (designed for water at temperatures above the
melting temperature) when extrapolated into the supercooled region, confirming the
practical need for an improved equation of state for supercooled water. In
Sec.~\ref{sec:theory} we formulate a thermodynamic model for supercooled water by
adopting suitable physical scaling fields with respect to the location of a
liquid--liquid critical point in supercooled water. In Sec.~\ref{sec:physmodel} we show
that this theoretical model yields an accurate representation of the thermodynamic
property data of both supercooled H$_2$O and D$_2$O up to pressures of 150~MPa. In
Sec.~\ref{sec:extended} we present a less-restricted phenomenological extension of the
theoretical model and show that this extension allows a representation of all currently
available experimental data for supercooled H$_2$O up to the pressure of 400~MPa. The
article concludes with a discussion of the results and of some unresolved theoretical
issues in Sec.~\ref{sec:interpretation} and \ref{sec:discussion}.

\section{Review of experimental data}\label{sec:expdata}
Experimental data on the properties of supercooled water have been reviewed by Angell in
1982,\cite{angell1982book,angell83} by Sato \ea\ in 1991,\cite{sato1991} and by
Debenedetti in 2003.\cite{deben03} Therefore, in this review, we focus on data published
after 2003, and restrict ourselves to bulk thermodynamic properties. For a complete
overview of the older data the reader is referred to the earlier reviews. Besides
reviewing experiments, we also assess the performance in the supercooled region of the
current reference correlation for the properties of water and steam, the ``IAPWS
Formulation 1995 for the Thermodynamic Properties of Ordinary Water Substance for General
and Scientific Use,'' or IAPWS-95 for short.\cite{iapws95,wag02nonote} Such an assessment
has been carried out before,\cite{wag02nonote,feistel2008} but not with all property data
that are now available. Most of the data discussed here are provided in numerical form in
the supplemental material.\cite{SCWsupplement}

\subsection{Density}\label{sec:densitydata}
\begin{figure*}
\includegraphics{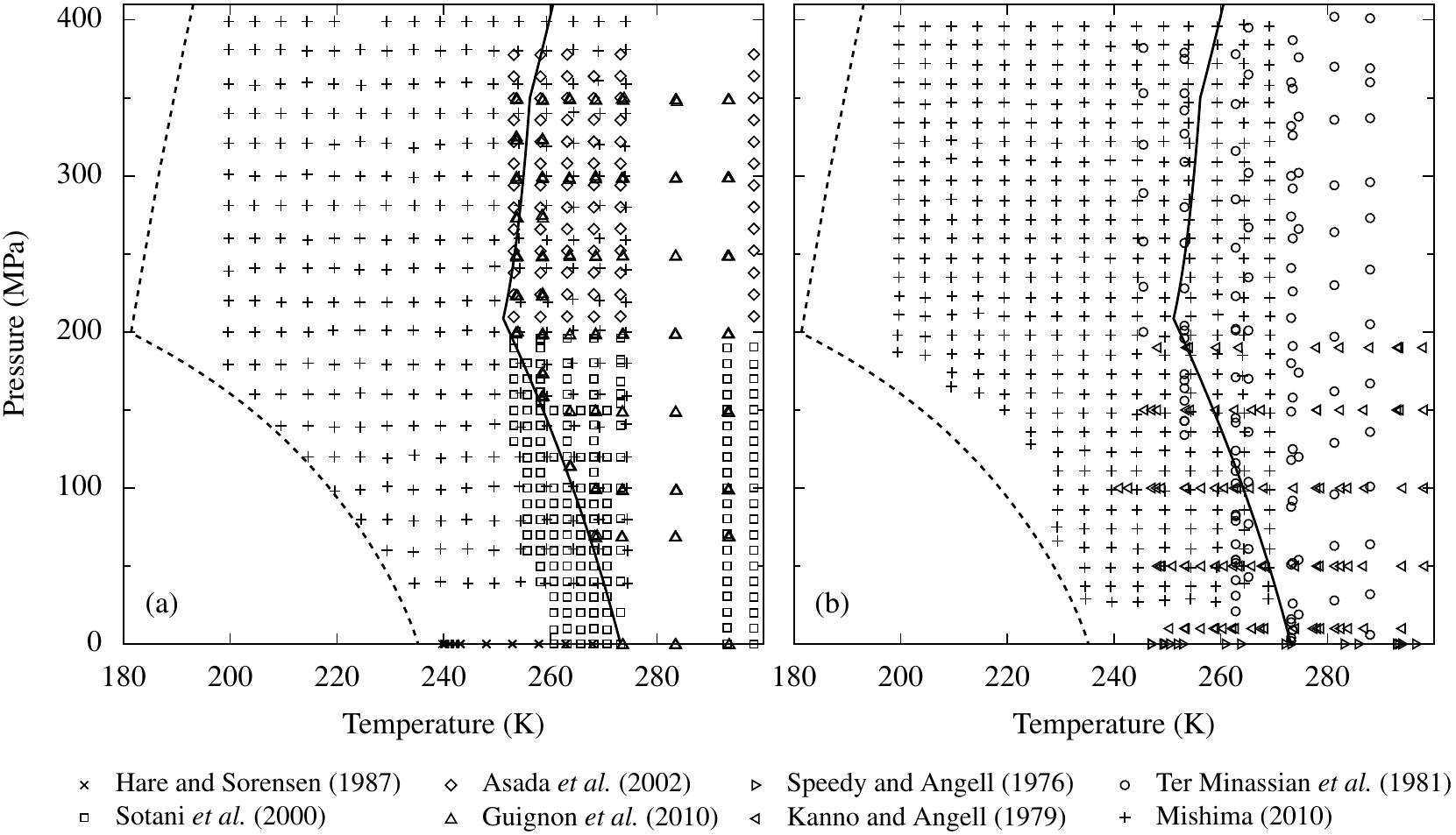}
\caption{\label{fig:ExpPT}(a) Location of the experimental H$_2$O density data
\cite{hare87,sotani2000,asada2002,mishima2010,guignon2010}. The solid curve is the
ice--liquid phase boundary \cite{iapwsmeltsub2011,*wagner2011}; the dashed curve is the
homogeneous ice nucleation limit \cite{kanno1975,kanno2006}. The location of the dashed
curve above 300~MPa is uncertain. At 0.1~MPa and in the stable-liquid region, data from
several older sources have been omitted for clarity. (b) Location of the experimental
H$_2$O density-derivative data. Ter Minassian \ea \cite{terminassian1981} have measured
the expansivity coefficient, other authors \cite{speedy1976,kanno1979,mishima2010} have
measured the isothermal compressibility.}
\end{figure*}

Since the review by Debenedetti,\cite{deben03} new data for the density of supercooled
water have been reported.\cite{asada2002,mishima2010,guignon2010} Most notable is the
recent work of Mishima \cite{mishima2010}, who measured the density and compressibility
down to 200~K and up to 400~MPa; see \figref{fig:ExpPT}(a). More accurate density
measurements have been published by Sotani \ea,\cite{sotani2000} Asada
\ea,\cite{asada2002} and Guignon \ea,\cite{guignon2010} but their lowest temperature is
253~K, so in a larger temperature range Mishima's data are the only data available. The
data of Guignon \ea\ are in the same range as the data of Sotani \ea\ and Asada \ea; the
maximum density difference between the data is 0.25\%, which is within the experimental
uncertainty.

At atmospheric pressure, the density of supercooled water has been measured by several
experimentalists.\cite{deben03} We consider the measurements of Hare and
Sorensen\cite{hare87} of 1987 the most accurate. They showed that their measurements were
not affected by the `excess density' effect, which occurs in thin capillary tubes and
caused too large densities in their 1986 experiments\cite{hare86} and in experiments of
others.

\begin{figure}
\includegraphics{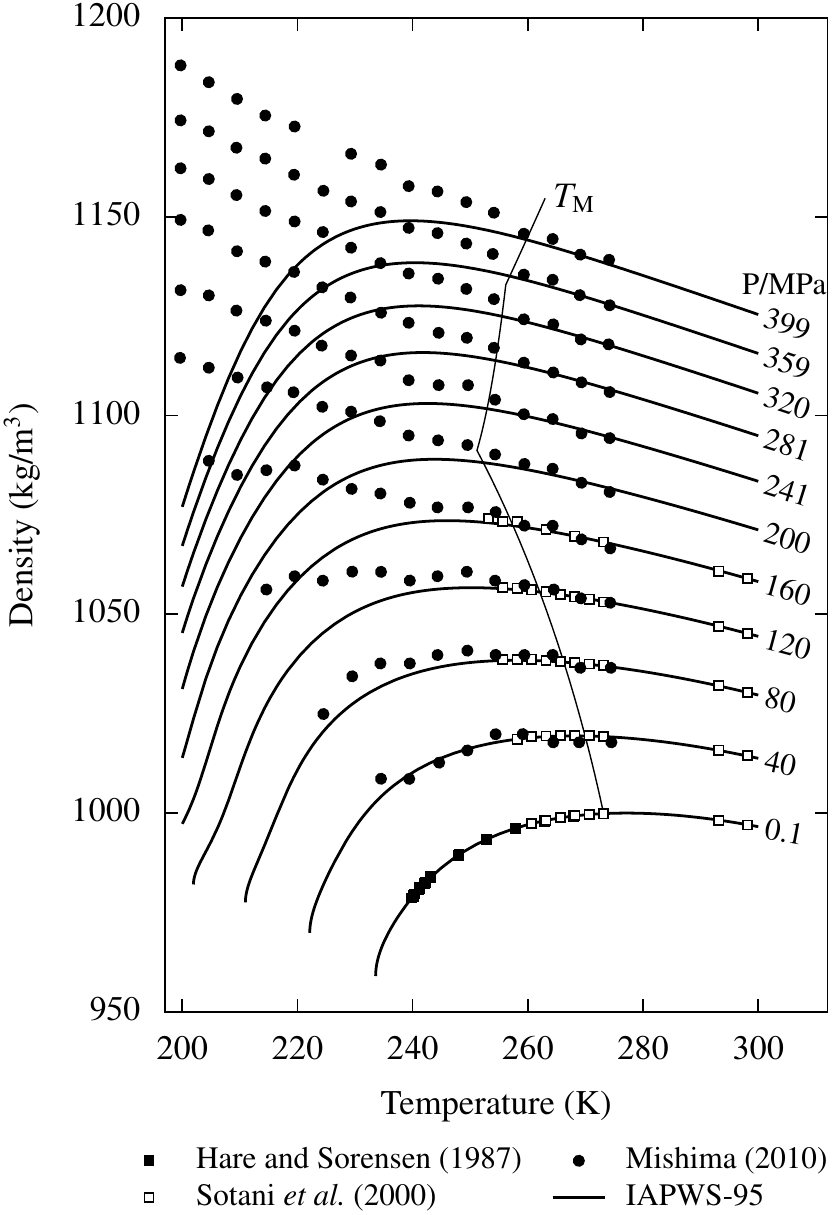}
\caption{\label{fig:densityIAPWST}Densities according to IAPWS-95 (curves).
IAPWS-95 is valid to the right of the melting curve\cite{iapwsmeltsub2011} $T_\text{M}$;
the IAPWS-95 values left of the melting curve are extrapolations.
The symbols represent experimental data of Mishima \cite{mishima2010}, Sotani \ea\ \cite{sotani2000}
and Hare and Sorensen \cite{hare87}. Mishima's data have been
adjusted as described in Appendix~\ref{app:correction}.}
\end{figure}

A comparison of the densities calculated from the IAPWS-95 formulation with the
experimental density data of Hare and Sorensen,\cite{hare87} of Sotani
\ea,\cite{sotani2000} and of Mishima\cite{mishima2010} is shown in
\figref{fig:densityIAPWST}. While the IAPWS-95 formulation reproduces the experimental
density data at ambient pressure, the deviations from the formulation become larger and
larger with increasing pressure. Especially at higher pressures, there is a sizable
discrepancy between the IAPWS-95 formulation and the experimental data; the slope (or the
expansivity) even has a different sign.

\subsection{Derivatives of the density}
Both the isothermal compressibility and the expansivity coefficient of supercooled water
have been measured; see \figref{fig:ExpPT}(b). The most accurate compressibility data are
from Kanno and Angell \cite{kanno1979}, whereas Mishima's \cite{mishima2010} data cover
the largest temperature range. The only expansivity measurements are from Ter Minassian
\ea\cite{terminassian1981} Hare and Sorensen \cite{hare86,hare87} also have published
expansivities (at 0.1~MPa), but these were obtained from the derivative of a fit to their
density data.

\begin{figure}
\includegraphics{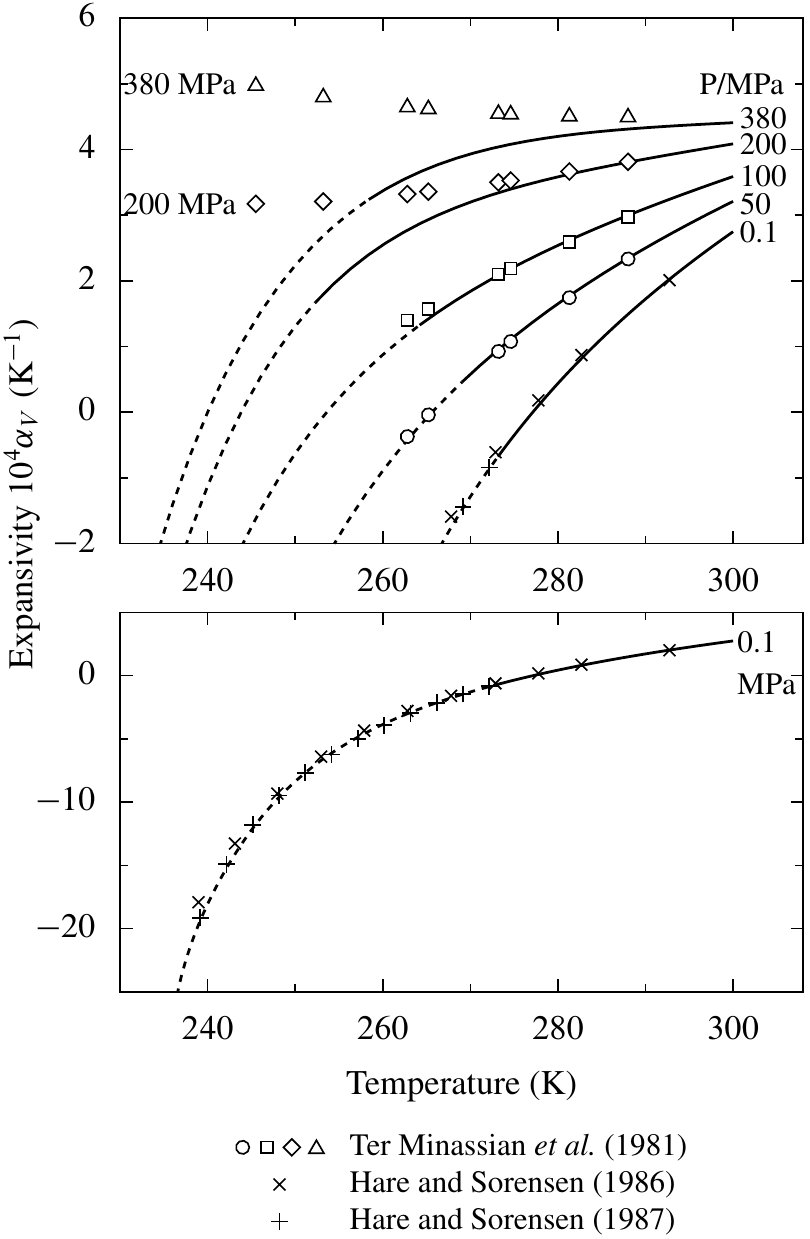}
\caption{\label{fig:expIAPWS}Expansivity coefficient according to IAPWS-95
(solid curves: within region of validity, dashed curves: extrapolation).
Symbols represent experimental data of Ter Minassian \ea\cite{terminassian1981}
and Hare and Sorensen \cite{hare86,hare87}. }
\end{figure}

\begin{figure}
\includegraphics{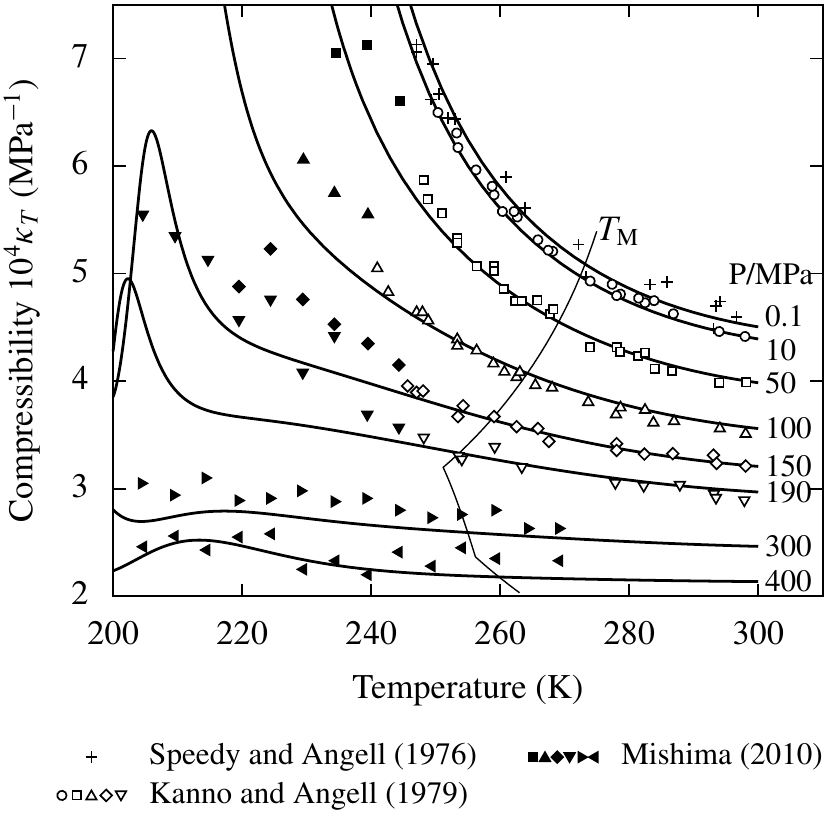}
\caption{\label{fig:compressibility}Isothermal compressibility according to IAPWS-95
(curves).
IAPWS-95 is valid to the right of the melting curve $T_\text{M}$;
the IAPWS-95 values left of the melting curve are extrapolations.
Symbols represent experimental data of Speedy and Angell \cite{speedy1976},
Kanno and Angell \cite{kanno1979}, and Mishima \cite{mishima2010}.
Solid and open symbols with the same shape correspond to the same pressure.}
\end{figure}

According to Wagner and Pruß \cite{wag02nonote}, the behavior of the expansivity
coefficient calculated from the IAPWS-95 formulation should be reasonable in the liquid
region at low temperature. However, from \figref{fig:expIAPWS} we see that the IAPWS-95
expansivity is in error by up to 50\% in the low-temperature region, even at temperatures
above the melting temperature where the IAPWS-95 formulation should be valid. The
isothermal compressibility calculated from the IAPWS-95 formulation agrees with the
experimental data down to about 250~K and up to 400~MPa, as shown in
\figref{fig:compressibility}. However, at lower temperatures, the IAPWS-95
compressibilities do not even qualitatively agree with the data.

\subsection{Heat capacity}\label{sec:cpdata}
\begin{figure}
\includegraphics{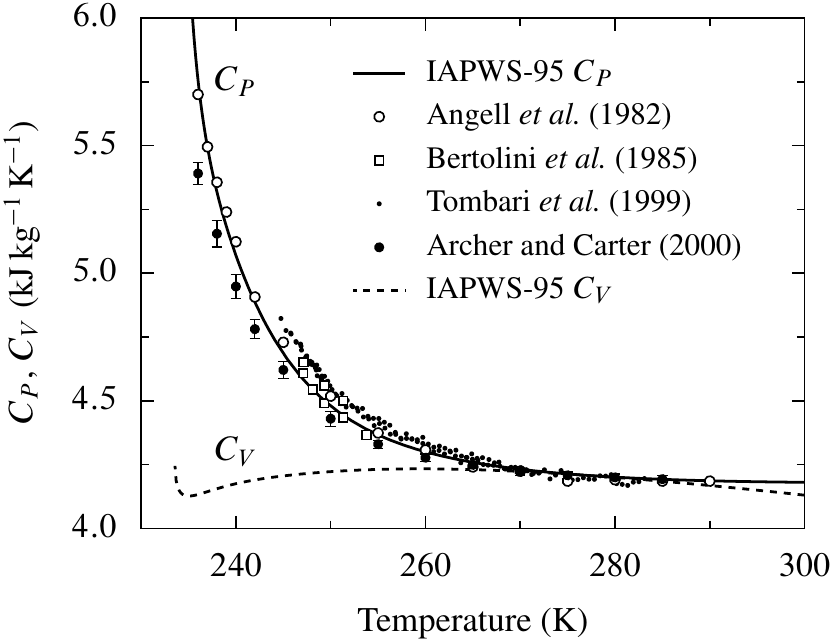}
\caption{\label{fig:IAPWSCpCv}Isobaric and isochoric heat capacity at 0.1~MPa according to IAPWS-95 (curves).
Symbols represent experimental data.\cite{angell1982,bertolini1985,tombari1999,arc00} }
\end{figure}

The isobaric heat capacity $C_P$ of supercooled water has been measured only at
atmospheric pressure.\cite{deben03} Old measurements by Anisimov \ea \cite{anisimov1972}
down to 266~K already showed an anomalous increase of the heat capacity at moderate
supercooling. In breakthrough experiments, Angell \ea
\cite{angell1973,rasmussen1973cp,angell1982} extended the range of measurements down to
236~K, and demonstrated that $C_P$ keeps increasing with decreasing temperature
(\figref{fig:IAPWSCpCv}). More recent measurements by Archer and Carter \cite{arc00},
also down to 236~K, do not perfectly agree with those of Angell \ea\cite{angell1982}
Archer and Carter suggest that the temperature calibration procedure of Angell \ea\ might
cause a systematic error in their measurements. Furthermore, Archer and Carter suspect
that measurements of Tombari \ea \cite{tombari1999} (down to 245~K) could be affected by
even more significant systematic calibration errors. The measurements of Bertolini
\ea\cite{bertolini1985} down to 247~K agree with those of Angell \ea\ (after correction,
as described in Appendix~\ref{app:correction}).

Because there are no high-pressure measurements of $C_P$ in the supercooled region, we
mention here some experiments at high pressure $P$ in the stable region. Sirota \ea
\cite{sirota1970} measured $C_P$ up to 100~MPa and down to 272~K. Recently, Manyà \ea
\cite{manya2011} have measured $C_P$ at 4~MPa from 298~K to 465~K. It turns out that the
results of Manyà \ea\ imply that the derivative $(\pypxl{C_P}{P})_T$ at constant
temperature $T$ is positive for pressures lower than 4~MPa, which contradicts the
thermodynamic relation $(\pypxl{C_P}{P})_T = -T (\partial^2 V / \partial T^2)_P$, with
$V$ being the molar volume. Hence, the data of Manyà \ea\ will not be considered in this
paper.

To our knowledge, the isochoric heat capacity $C_V$ of supercooled water has not been
measured. The values presented by several
investigators\cite{angell1973,rasmussen1973cp,oguni1983} were calculated from other
thermodynamic properties.

The isobaric and isochoric heat capacities predicted by IAPWS-95 at 0.1~MPa are shown in
\figref{fig:IAPWSCpCv}. The $C_P$ curve computed from IAPWS-95 follows the data of Angell
\ea,\cite{angell1982} to which it was fitted. Consequently, IAPWS-95 systematically
deviates from the data of Archer and Carter,\cite{arc00} as discussed by Wagner and
Pruß.\cite{wag02nonote} The isochoric heat capacity $C_V$ does not exhibit a strongly
anomalous behavior according to the IAPWS-95 prediction.

\subsection{Surface tension}
\begin{figure}
\includegraphics{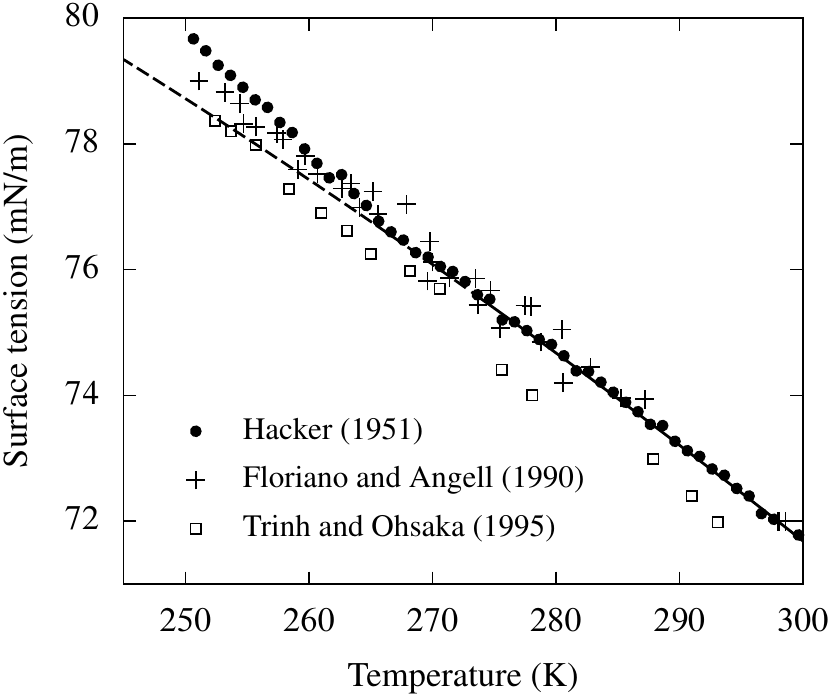}
\caption{\label{fig:IAPWSsurfacetension}
Surface tension according to the IAPWS equation\cite{iapwssurften94}
(solid curve: within region of validity, dashed curve: extrapolation).
Symbols represent experimental data from Hacker,\cite{hac51}
Floriano and Angell,\cite{[][{, data of capillaries 4 to 7 are shown in \figref{fig:IAPWSsurfacetension}.}]flo90}
and Trinh and Ohsaka.\cite{tri95}
}
\end{figure}

As listed by Debenedetti,\cite{deben03} the surface tension of supercooled water with
vapor and with air has been measured by Hacker,\cite{hac51} Floriano and
Angell,\cite{flo90} and Trinh and Ohsaka,\cite{tri95} all down to about 250~K. Hacker's
measurements show an inflection point at about 268 K, below which the surface tension has
a stronger temperature dependence (\figref{fig:IAPWSsurfacetension}). The data of
Floriano and Angell are less accurate, but the authors also noted an inflection. The
measurements of Trinh and Ohsaka show a systematic deviation from the other data, but the
trend agrees with Hacker's data. Furthermore, a molecular dynamics simulation by Lü and
Wei\cite{lu06} shows an inflection point as well.

IAPWS has recommended an equation to represent the surface tension of liquid water in
equilibrium with water vapor.\cite{iapwssurften94} For a comparison with experimental
data, it is necessary to consider the conditions under which the surface tension was
measured. Hacker,\cite{hac51} Floriano and Angell,\cite{flo90} and Trinh and
Ohsaka\cite{tri95} measured the surface tension at atmospheric pressure in air. The
influence of air at atmospheric pressure on the surface tension of water is usually
neglected,\cite{var83} but for supercooled water the effect may be significant since the
solubility of nitrogen and oxygen in water increases with decreasing temperature. From
high-pressure data\cite{mass74} it is known that the surface tension of water with
nitrogen or oxygen is lower than the pure-water surface tension. In
\figref{fig:IAPWSsurfacetension} we show a comparison of the experimental surface-tension
data with the values calculated from the IAPWS equation. As noted above, the experimental
values of the surface tension suggest an inflection point at about 268 K. The
extrapolation of the IAPWS equation does not show an inflection, and the difference with
Hacker's accurate data\cite{hac51} increases with decreasing temperature. The data of
Floriano and Angell\cite{flo90} show more scatter, but below the freezing point most of
their data lie above the IAPWS extrapolation. The measurements of Trinh and
Ohsaka\cite{tri95} lie below all other data; the deviation is about the same as their
experimental uncertainty, which is~1\%.

\subsection{Speed of sound}
\begin{figure}
\includegraphics{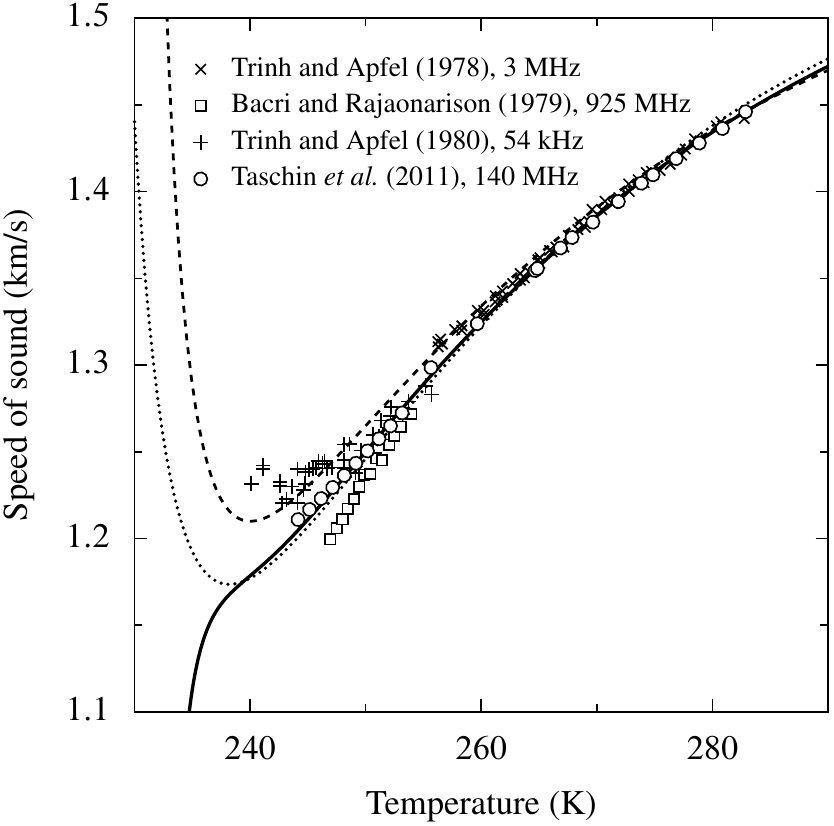}
\caption{\label{fig:IAPWSspeedofsound}
Speed of sound at 0.1~MPa. Solid curve: prediction of the IAPWS-95 formulation.
Dashed curve: model of Sec.~\ref{sec:physmodel}; dotted curve: model of Sec.~\ref{sec:extended}.
Symbols represent experimental data.\cite{trinhapfel1978JASA,trinhapfel1978JCP,%
trinhapfel1980,bacri1979,taschin2011} }
\end{figure}

The speed of sound in supercooled water has been measured in a broad frequency range,
from 54~kHz to about 20~GHz. In stable water at atmospheric pressure, no significant
dependence on the frequency is found in this range, but the speed of sound in supercooled
water shows a dispersion that increases with cooling. In this article we consider the
thermodynamic (zero-frequency) limit of the speed of sound, which is associated with the
adiabatic compressibility.\cite{angell1982book} Taschin \ea\cite{taschin2011} estimate
that dispersion effects become noticeable at frequencies of 1~GHz and higher. We will
therefore not consider the measurements in the 1--10 GHz range obtained with Brillouin
light scattering, which show a speed-of-sound minimum between 250~K and 280~K; the
temperature of the minimum increases with increasing
frequency.\cite{magazu1989,santucci2006} The lower-frequency (ultrasonic) measurements
are shown in \figref{fig:IAPWSspeedofsound}. Below 255~K, there appears to be dispersion;
however, the speed of sound decreases with increasing frequency, which suggests an
anomalous (negative) dispersion. While a negative dispersion of the speed of sound very
close to the vapor--liquid critical point is, in principle, possible due to the
divergence of the thermal conductivity,\cite{anisimov1972thermcond} there is not yet any
experimental indication for an anomaly of the thermal conductivity of supercooled
water.\cite{benchikh1985} Recent simulations of Kumar and Stanley\cite{kumar2011} show a
minimum for this property. Debenedetti\cite{deben03} argues that negative dispersion can
be ruled out because the Brillouin experiments all show positive dispersion. According to
Taschin \ea,\cite{taschin2011} the apparent negative dispersion is a result of systematic
errors in the data of Trinh and Apfel\cite{trinhapfel1980} and Bacri and
Rajaonarison.\cite{bacri1979} The IAPWS-95 formulation agrees with the recent data of
Taschin \ea\cite{taschin2011} to within their uncertainty.

There are no measurements of the speed of sound of supercooled water at high pressure. In
the stable region, Petitet \ea\cite{petitet1983} performed measurements at 10~MHz up to
460~MPa, including the region close to the melting curve, down to 253~K. Petitet \ea\
have also published data down to 253~K at atmospheric pressure, but these data were taken
from Conde \ea\cite{conde1982} Since Conde \ea\ measured at 5~GHz, these data do not
represent the zero-frequency limit of the speed of sound.

\subsection{Spinodal}
\begin{figure}
\includegraphics{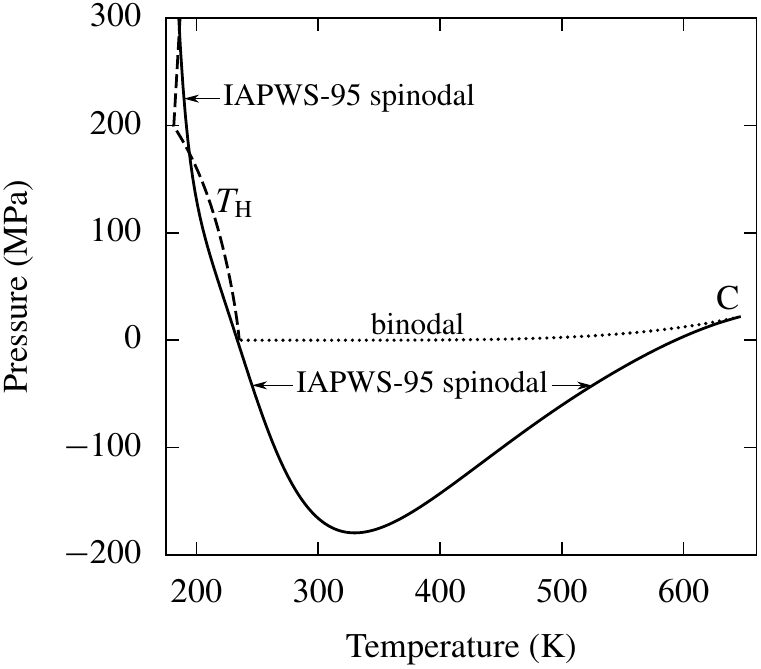}
\caption{\label{fig:IAPWSspinodal}
Location of the liquid spinodal according to the IAPWS-95 formulation.
The curved marked with $T_\text{H}$ is the homogeneous ice nucleation limit.\cite{kanno1975,kanno2006}
Also shown is the location of the binodal, the vapor pressure curve,
and C denotes the vapor--liquid critical point.}
\end{figure}

In the supercooled region, IAPWS-95 predicts a re-entrant liquid spinodal, as shown in
\figref{fig:IAPWSspinodal}. The spinodal pressure becomes positive at 233.6~K, which is a
few degrees below the homogeneous nucleation limit. At about 195~K and 175~MPa, the
spinodal curve crosses the homogeneous nucleation limit and enters the region where
supercooled water can be experimentally observed. Up to about 290~MPa, the spinodal curve
stays in the experimentally accessible range. Since a spinodal has not been observed
there, the spinodal of IAPWS-95 contradicts experimental evidence. While a re-entrant
spinodal is thermodynamically possible, Debenedetti has argued that the re-entrant
spinodal scenario suggested by Speedy\cite{speedy1982a} is implausible for supercooled
water.\cite{deben03,speedy04comment,*deben04reply} Debenedetti's argument may be
summarized as follows. In the pressure--temperature plane (\figref{fig:IAPWSspinodal}), a
re-entrant spinodal must intersect the metastable continuation of the binodal, the
vapor-pressure curve. At the intersection, liquid water is simultaneously in equilibrium
with water vapor and unstable with respect to infinitesimal density fluctuations. Since
these two conditions are mutually exclusive, such an intersection cannot exist. On the
other hand, at the vapor--liquid critical point the binodal and spinodal do coincide,
which is possible because the vapor and liquid phases are indistinguishable there.

\subsection{Liquid--liquid coexistence curve}
Both the existence of a second critical point and its location are still being debated in
the literature. If the second critical point exists, there should be a liquid--liquid
transition (LLT) curve -- separating a hypothetical high-density liquid and low-density
liquid -- which ends at the critical point. At pressures below the critical pressure,
water's response functions exhibit an extremum near the `Widom line,' which is the
extension of the LLT curve into the one-phase region and the locus of maximum
fluctuations of the order parameter.

\begin{figure}
\includegraphics{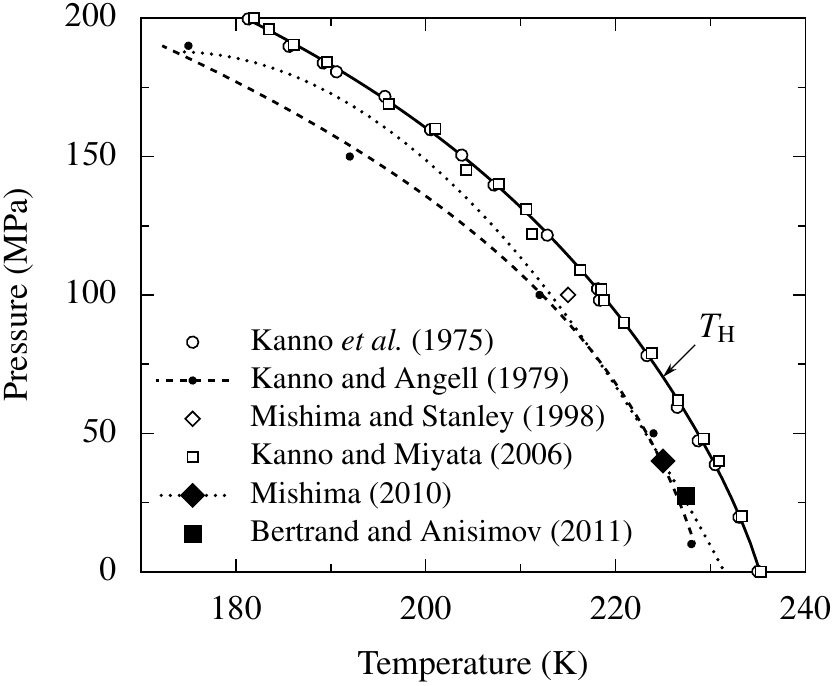}
\caption{\label{fig:LLline}Homogeneous ice nucleation temperatures (open circles \cite{kanno1975}
and squares \cite{kanno2006} and fitted solid curve), Mishima's \cite{mishima2010}
conjectured liquid--liquid coexistence curve (dotted) and liquid--liquid critical point (large solid diamond),
and Kanno and Angell's curve (dashed) connecting the fitted singular
temperatures (solid circles) \cite{kanno1979}.
Open diamond: bend in the melting curve of ice IV \cite{mishima1998};
large solid square: the liquid--liquid critical point suggested by Bertrand and Anisimov.\cite{bertrand2011} }
\end{figure}

While the location of the critical point obtained by different simulations varies
greatly, different attempts to locate the LLT and the Widom line from experimental data
have yielded approximately the same result. Kanno and Angell \cite{kanno1979} fitted
empirical power laws to their compressibility measurements and obtained singular
temperatures located from 5~K to 12~K below the homogeneous nucleation temperature
$T_\text{H}$ (\figref{fig:LLline}), suggesting a LLT that mimics the $T_\text{H}$ curve
but shifted to lower temperature. Mishima measured metastable melting curves of H$_2$O
ice IV \cite{mishima1998} and D$_2$O ices IV and V \cite{mishima2000}, and found that
they suddenly bent at temperatures of 4~K to 7~K below $T_\text{H}$. According to
Mishima, this is indirect evidence for the location of the LLT, but a one-to-one
correspondence between a break in the melting curve and the LLT has been questioned by
Imre and Rzoska \cite{imre2010}.

Mishima \cite{mishima2010} approximated the LLT by a quadratic function of $T$ with
approximately the same shape as the $T_\text{H}$ curve, but allowing a shift to lower
temperature. His final result, shown in \figref{fig:LLline}, is close to Kanno and
Angell's curve for pressures up to 100~MPa, albeit with a different curvature.

\subsection{Heavy water}
\begin{figure}
\includegraphics{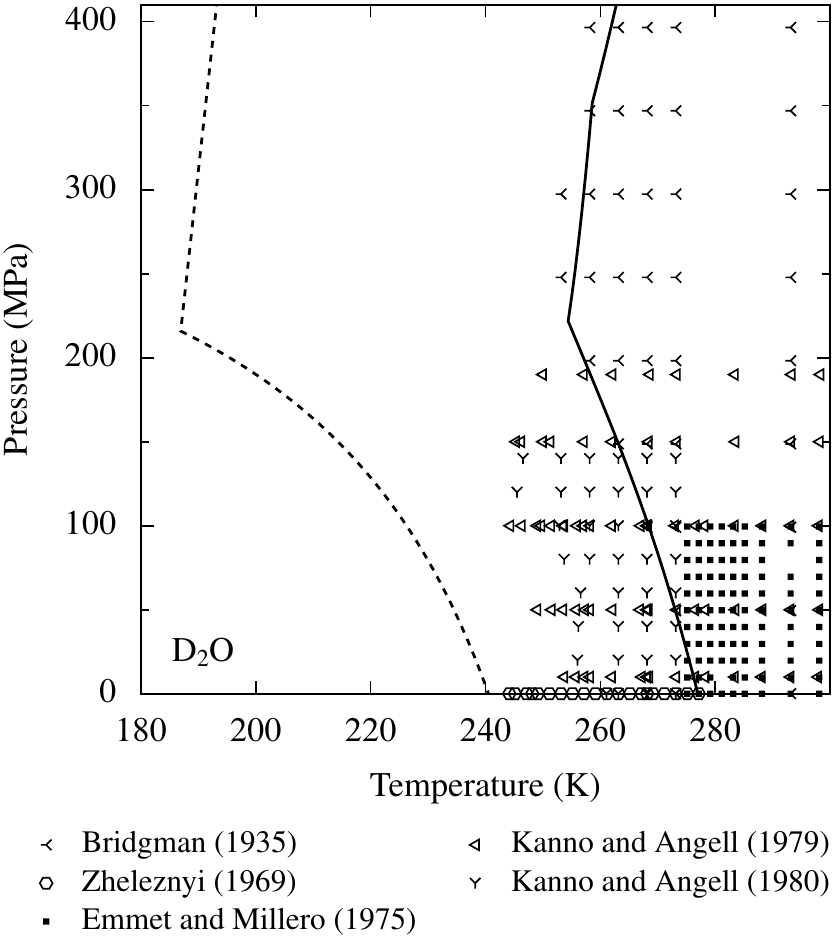}
\caption{\label{fig:ExpPTD2O}(a) Location of the experimental data for the density of D$_2$O and
its derivatives.\cite{bridgman1935,zhe69,emmet1975,kanno1979,kanno1980}
The solid curve is the
ice--liquid phase boundary;\cite{bridgman1935} the dashed curve is the homogeneous
ice nucleation limit \cite{angell1976}. The location of the dashed curve above
300~MPa is uncertain. At 0.1~MPa and in the stable-liquid region, data from several
sources have been omitted for clarity.}
\end{figure}

Experimental data on the properties of supercooled heavy water (D$_2$O) have been
reviewed by Angell\cite{angell1982book,angell83} and Debenedetti.\cite{deben03} The
location of data for the density and its derivatives is shown in \figref{fig:ExpPTD2O}.
At atmospheric pressure, the density has been measured by several
investigators.\cite{zhe69,rassmussen73,kanno1980,hare86} There are differences of up to
0.14\% between the data sets, and it is not clear which is the best set. The isothermal
compressibility and thermal expansion coefficient have been measured in an extensive
temperature and pressure range by Kanno and Angell.\cite{kanno1979,kanno1980} The
isobaric heat capacity has been measured at atmospheric pressure by Angell and
coworkers.\cite{angell1973,rasmussen1973cp,angell1982} The speed of sound of supercooled
D$_2$O was reported by Conde \ea\cite{conde1982} at 5~GHz; at this frequency the speed of
sound likely deviates from the zero-frequency limit, especially at lower temperatures.

\section{Thermodynamic model for supercooled water}\label{sec:theory}
In this section we further develop a scaling model for bulk thermodynamic properties
supercooled water, which was earlier suggested by Fuentevilla and
Anisimov\cite{fuentevilla2006} and, more recently, modified and elaborated by Bertrand
and Anisimov.\cite{bertrand2011}
\subsection{Scaling fields and thermodynamic properties}
Fluids belong to the universality class of Ising-like systems whose critical behavior is
characterized by two independent scaling fields, a ``strong'' scaling field $h_1$
(ordering field) and a ``weak'' scaling field $h_2$, and by a dependent scaling field
$h_3$ which asymptotically close to the critical point becomes a generalized homogeneous
function of $h_1$ and $h_2$ \cite{fisher1974,kadanoff1976,fisher1983}:
\begin{equation}\label{eq:h3fun}
h_{3}(h_1,h_2) \approx \left\vert h_{2}\right\vert ^{2-\alpha }f^{\pm }\left( \frac{h_{1}}
{\left\vert h_{2}\right\vert ^{2-\alpha -\beta }}\right).
\end{equation}%
In this expression $\alpha \simeq 0.110$ and $\beta \simeq 0.326$ are universal critical
exponents \cite{pelissetto2002,sengers2009} and $f^\pm$, with the superscripts $\pm$
referring to $h_2
> 0$ and $h_2 < 0$, is a universal scaling function except for two system-dependent
amplitudes. Associated with these scaling fields are two conjugate scaling densities, a
strongly fluctuating scaling density $\phi_1$ (order parameter) and a weakly fluctuating
scaling density $\phi_2$, such that
\begin{equation}
\di h_{3}=\phi _{1}\,\di h_{1}+\phi _{2}\,\di h_{2}
\end{equation}%
with
\begin{equation}\label{eq:phi1phi2}
    \phi_1  = \pypxd{h_3}{h_1}{h_2},\qquad \phi_2  = \pypxd{h_3}{h_2}{h_1}.
\end{equation}
In addition one can define three susceptibilities, a ``strong'' susceptibility $\chi_1$,
a ``weak'' susceptibility $\chi_2$, and a ``cross'' susceptibility $\chi_{12}$:
\begin{gather}
    \chi_1 = \pypxd{\phi_1}{h_1}{h_2},\qquad \chi_2 = \pypxd{\phi_2}{h_2}{h_1},\\
    \chi_{12} = \pypxd{\phi_1}{h_2}{h_1} = \pypxd{\phi_2}{h_1}{h_2}.
\end{gather}
In fluids and fluid mixtures one encounters a large variety of different types of
critical phenomena \cite{rowlinson1982}. The asymptotic thermodynamic behavior near all
kinds of critical points can be described in terms of \eqref{eq:h3fun}. The differences
arise from the actual relationships between the scaling fields and the physical fields
\cite{anisimov1995}, subject to the condition that at the critical point
\begin{equation}\label{eq:hcrit}
    h_1 = h_2 = h_3 = 0.
\end{equation}
In one-component fluids the relevant physical fields are the chemical potential $\mu$
(Gibbs energy per mole), the temperature $T$, and the pressure $P$. To satisfy condition
(\ref{eq:hcrit}) one defines $\Delta\mu = \mu - \mu_\text{c}$, $\Delta T = T - \Tc$, and
$\Delta P = P - \Pc$. In this article we adopt the usual convention that a subscript c
refers to the value of the property at the critical point. There are two special models
for critical behavior that deserve some attention. The first is the lattice gas in which
the ordering field $h_1$ is asymptotically proportional to $\Delta\mu$ and the weak
scaling field proportional to $\Delta T$ \cite{lee1952,fisher1967,sengerscroxton}. Hence,
in the lattice gas $\phi_1$ is proportional to $\Delta\rho = \rho - \rho_\text{c}$ and
$\phi_2$ proportional to $\Delta s = s - s_\text{c}$, where $\rho$ is the molar density
and $s$ the entropy density. The lattice gas provides a model for the vapor--liquid
critical point where the molar density yields the major contribution to the order
parameter. In practice, the asymptotic critical behavior of a fluid near the
vapor--liquid critical point, including that of H$_2$O \cite{leveltsengers1983}, can be
described by a slight modification of the lattice-gas model to account for some lack of
vapor--liquid symmetry in real fluids. Another special model, introduced by Bertrand and
Anisimov,\cite{bertrand2011} is a ``lattice liquid'' in which the ordering field is
asymptotically proportional to $\Delta T$ and in which the weak scaling field is
proportional to $\Delta\mu$ \cite{bertrand2011}. Near the liquid--liquid critical point
in weakly compressible supercooled water the entropy yields the major contribution to the
order parameter and not the mass density, as first pointed out by Fuentevilla and
Anisimov.\cite{fuentevilla2006} Thus the thermodynamic properties near this
liquid--liquid critical point can be described by a slight modification of the
lattice-liquid model to account for some lack of symmetry in the order
parameter.\cite{bertrand2011}

To implement a scaled thermodynamic representation it is convenient to make all
thermodynamic properties dimensionless in terms of the critical parameters $\Tc$ and
$\rho_\text{c}$ or $\Vc = \rho_\text{c}^{-1}$:
\begin{equation}
    \Th=\frac{T}{\Tc}, \quad
    \mh=\frac{\mu}{R\Tc}, \quad
    \Ph=\frac{P\Vc}{R\Tc},
\label{dim_field}
\end{equation}%
where $R$ is the ideal-gas constant. For the dimensionless physical densities we define
\begin{equation}
    \Vh = \frac{V}{\Vc}, \quad
    \Sh = \frac{S}{R}, \quad
    \Cph = \frac{C_P}{R},
\end{equation}
where $V$ is the molar volume, $S$ the molar entropy, and $C_P$ the isobaric molar heat
capacity. The thermodynamic model of Bertrand and Anisimov was formulated in terms of
$\Ph(\mh,\Th)$ for which
\begin{equation}
    \di\Ph = \Vh^{-1}\di\mh + \Vh^{-1}\Sh\,\di\Th.
\end{equation}
We have found it more convenient, and practically equivalent, to formulate this
thermodynamic model in terms of $\mh(\Ph,\Th)$ for which
\begin{equation}\label{eq:dmu}
    \di \mh = \Vh\, \di \Ph - \Sh\, \di \Th.
\end{equation}
Thus in this formulation, similar to that suggested earlier by Fuentevilla and
Anisimov,\cite{fuentevilla2006} we identify the order parameter with the entropy itself
instead of the entropy density. In our model the scaling fields are related to the
physical fields as
\begin{align}
    h_1 &= \Dt + a'\Dp, \label{eq:h1}\\
    h_2 &= -\Dp + b'\Dt, \label{eq:h2}\\
    h_3 &= \Dp - \Dm + \mr, \label{eq:h3}
\end{align}
with
\begin{equation}
    \Dt = \frac{T-\Tc}{\Tc},\quad \Dp = \frac{(P-\Pc)\Vc}{R\Tc},\quad
    \Dm = \frac{\mu-\mu_\text{c}}{R\Tc}.
\end{equation}
In \eqref{eq:h1} $a'$ represents the slope $-\di\Th/\di\Ph$ of the phase-coexistence or
Widom line at the critical point. In \eqref{eq:h2} $b'$ is a so-called mixing coefficient
which accounts for the fact that the critical phase transition in supercooled water is
not completely symmetric in terms of the entropy order parameter. Introduction of mixing
of this type is also known in the literature as the revised-scaling approximation
\cite{behnejad2010}. \eeqref{eq:h3fun} only represents the asymptotic behavior of the
so-called singular critical contributions to the thermodynamic properties. To obtain a
complete representation of the thermodynamic properties we need to add a regular (i.e.,
analytic) background contribution. As has been common practice in developing scaled
equations of state in fluids near the vapor-liquid critical point
\cite{behnejad2010,leveltsengers1983}, the regular background contribution is represented
by a truncated Taylor-series expansion around the critical point:
\begin{equation}\label{eq:backgr}
    \mr = \sum_{m,\,n} c_{mn} (\Dt)^m (\Dp)^n, \quad \text{with}\quad c_{00} = c_{10} = c_{01} = 0.
\end{equation}
The first two terms in the temperature expansion of $\mr$ depend on the choice of zero
entropy and energy and do not appear in the expressions of any of the physically
observable thermodynamic properties. Hence, these coefficients may be set to zero.
Furthermore, the coefficient $c_{01} = \Vh_\text{c} - 1 = 0$. Strictly speaking, critical
fluctuations also yield an analytic contribution to $h_3$
\cite{anisimov1992,anisimov1998}. In this article we incorporate this contribution into
the linear background contribution as has also been done often in the past.

The current treatment of the background contribution differs from that in earlier
publications.\cite{fuentevilla2006,bertrand2011} Previously, a temperature-dependent
background was added to the critical part of each property. Because the background of
each property was treated separately, the resulting property values were not mutually
consistent. In this work, the background is added to the chemical potential, and the
backgrounds in the derived properties follow, ensuring thermodynamic consistency.

From the fundamental thermodynamic differential relation (\ref{eq:dmu}) it follows that
\begin{alignat}{2}
    \Vh &=  &\pypxd{\mh}{\Ph}{T}   &= 1 - a'\phi_1 + \phi_2 + \mr_{\Ph},\\
    \Sh &= -&\pypxd{\mh}{\Th}{P}   &= \phi_1 + b'\phi_2 - \mr_{\Th}.
\end{alignat}
In this article we adopt the convention that a subscript $\Ph$ indicates a derivative
with respect to $\Ph$ at constant $\Th$ and a subscript $\Th$ a derivative with respect
to $\Th$ at constant $\Ph$. Finally, the dimensionless isothermal compressibility $\kap$,
expansivity coefficient $\alp$, and isobaric heat capacity $\Cph$ can be expressed in
terms of the scaling susceptibilities $\chi_1$, $\chi_{2}$, and $\chi_{12}$:
\begin{gather}\label{eq:compressibility}
    \kap = -\frac{1}{\Vh}\pypxd{\Vh}{\Ph}{T} =
        \frac{1}{\Vh} \left[(a')^2\chi_1 + \chi_2 - 2a'\chi_{12} - \mr_{\Ph\Ph}\right],\\
\begin{split}
    \alp &= \frac{1}{\Vh}\pypxd{\Vh}{\Th}{P} \\&=
        \frac{1}{\Vh} \left[ -a'\chi_1 + b'\chi_2 + (1-a'b')\chi_{12} + \mr_{\Th\Ph}\right],
\end{split}\\
    \Cph = \Th \pypxd{\Sh}{\Th}{P} =
        \Th \left[\chi_1 + (b')^2\chi_2 + 2b'\chi_{12} -\mr_{\Th\Th} \right].
\end{gather}

\subsection{Parametric equation of state}
It is not possible to write the scaled expression (\ref{eq:h3fun}) for $h_3$ as an
explicit function of $h_1$ and $h_2$. Such attempts always cause singular behavior of the
thermodynamic potential in the one-phase region either at $h_1 = 0$ or at $h_2 = 0$. This
problem is solved by replacing the two independent scaling fields, $h_1$ and $h_2$, with
two parametric variables: a variable $r$ which measures a ``distance'' from the critical
point and an angular variable $\theta$ which measures the location on a contour of
constant $r$. A transformation most frequently adopted has the form:
\begin{equation}\label{eq:linmodh1h2}
    h_1 = a r^{2-\alpha-\beta}\theta(1-\theta^2),\qquad h_2 = r(1-b^2\theta^2).
\end{equation}
From \eqsref{eq:h3fun}{eq:phi1phi2} it then follows that the order parameter $\phi_1$
must have the form \cite{fisher1971}:
\begin{equation}
    \phi_1 = k r^\beta M(\theta),
\end{equation}
where $M(\theta)$ is a universal analytic function of $\theta$. In principle, this
function can be calculated from the renormalization-group theory of critical phenomena
\cite{brezin1972}. In practice one adopts an analytic approximant for $M(\theta)$, the
simplest one being $M(\theta) = \theta$ \cite{schofield1969a}:
\begin{equation}\label{eq:linmodphi1}
    \phi_1 = k r^\beta \theta.
\end{equation}
\eeqsref{eq:linmodh1h2}{eq:linmodphi1} define what is known as the ``linear model''
parametric equation of state. In these equations $a$ and $k$ are two system-dependent
amplitudes related to the two system-dependent amplitudes in \eqref{eq:h3fun}, while
$b^2$ is a universal constant which is often approximated by \cite{schofield1969b}
\begin{equation}\label{eq:bsq}
    b^2 = \frac{2-\alpha -4\beta }{(2-\alpha -2\beta )(1-2\beta )} \simeq 1.361.
\end{equation}
\eeqsref{eq:linmodh1h2}{eq:linmodphi1} with the specific choice (\ref{eq:bsq}) for $b^2$
is known as the ``restricted'' linear model \cite{sengerscroxton}. The resulting
parametric equations for the various thermodynamic properties can be found in many
publications
\cite{sengerscroxton,leveltsengers1983,behnejad2010,anisimov1998,hohenberg1972,moldover1979}.
In this paper we are using the ``restricted'' linear model to describe the
supercooled-water anomalies. The parametric equations needed for the analysis in this
article are listed in Appendix~\ref{app:linmodel}.

\section{Comparison with experimental data}\label{sec:physmodel}
The scaling theory, formulated in the preceding section, represents the thermodynamic
behavior asymptotically close to the critical point. Specifically, the liquid--liquid
(LLT) curve that follows from \eqref{eq:h1} is a straight line, while the actual LLT
curve should exhibit curvature (as shown in \figref{fig:LLline}). In this section we
investigate the thermodynamic properties of supercooled liquid water in a range of
pressures and temperatures, where the asymptotic theory appears to be adequate. Issues
related to nonasymptotic features of the scaling theory will be addressed in
Sec.~\ref{sec:extended}. Hence, we restrict the asymptotic theoretical model to pressures
not exceeding 150~MPa, where the LLT can be reasonably approximated by a single straight
line. The slope of the LLT was constrained to values that are close to the slopes of the
curves of Kanno and Angell\cite{kanno1979} and Mishima\cite{mishima2010} in the range of
0~MPa to 150~MPa. Specifically, the value of $a'$ in \eqref{eq:h1} was restricted to the
range of~0.065 to~0.090. Because the position of the LLT is not precisely known, the
critical point was allowed to deviate up to 3~K from Mishima's curve. It was found that
the results of the model were rather insensitive to the critical pressure $\Pc$, so $\Pc$
was constrained to the value of 27.5~MPa obtained by Bertrand and
Anisimov.\cite{bertrand2011} It was also found that a nonzero mixing coefficient $b'$ did
not significantly improve the fit, so $b'$ was set to zero. This means physically that
the liquid-critical behavior in supercooled water exhibits little asymmetry in the order
parameter and is indeed very close to lattice-liquid behavior.

\begin{table}
\caption{\label{tab:physmodel}Parameter values for H$_2$O in the model of
Sec.~\ref{sec:physmodel}}
\begin{ruledtabular}
\begin{tabular}{lD{.}{.}{12}lD{.}{.}{12}}
Parameter & \multicolumn{1}{c}{Value\footnotemark[1]} & Parameter & \multicolumn{1}{c}{Value} \\ \hline
$\Tc$/K	&	224.23	&	$c_{11}$	&	1.536\,3\times 10^{-1}	\\
$\Pc$/MPa	&	27.5	&	$c_{12}$	&	-6.487\,9\times 10^{-3}	\\
$\rho_\text{c}$/(kg m$^{-3}$)	&	948.77	&	$c_{13}$	&	7.709\,0\times 10^{-3}	\\
$a$	&	0.229\,24	&	$c_{20}$	&	-3.888\,8\times 10^{0}	\\
$k$	&	0.377\,04	&	$c_{21}$	&	1.734\,7\times 10^{-1}	\\
$a'$	&	0.090	&	$c_{22}$	&	-6.415\,7\times 10^{-2}	\\
$c_{02}$	&	7.177\,9\times 10^{-2}	&	$c_{23}$	&	-6.985\,0\times 10^{-3}	\\
$c_{03}$	&	-4.093\,6\times 10^{-4}	&	$c_{30}$	&	6.981\,3\times 10^{-1}	\\
$c_{04}$	&	-1.099\,6\times 10^{-3}	&	$c_{31}$	&	-1.145\,9\times 10^{-1}	\\
$c_{05}$	&	2.949\,7\times 10^{-4}	&	$c_{32}$	&	7.500\,6\times 10^{-2}	\\
\end{tabular}
\end{ruledtabular}
\footnotetext[1]{Final digits of parameter values are given to allow reproducing the
values of properties with the model but do not have physical significance.}
\end{table}

\begin{table}
\caption{\label{tab:D2Omodel}Parameter values for D$_2$O in the model of
Sec.~\ref{sec:physmodel}}
\begin{ruledtabular}
\begin{tabular}{lD{.}{.}{12}lD{.}{.}{12}}
Parameter & \multicolumn{1}{c}{Value} & Parameter & \multicolumn{1}{c}{Value} \\ \hline
$\Tc$/K	&	232.65	&	$c_{11}$	&	1.282\,8\times 10^{-1}	\\
$\Pc$/MPa	&	32.29	&	$c_{12}$	&	-1.626\,7\times 10^{-3}	\\
$\rho_\text{c}$/(kg m$^{-3}$)	&	1055.74	&	$c_{13}$	&	9.555\,2\times 10^{-3}	\\
$a$	&	0.229\,24	&	$c_{20}$	&	-4.411\,8\times 10^{0}	\\
$k$	&	0.377\,04	&	$c_{21}$	&	3.000\,2\times 10^{-1}	\\
$a'$	&	0.078\,757	&	$c_{22}$	&	-9.720\,4\times 10^{-2}	\\
$c_{02}$	&	6.907\,2\times 10^{-2}	&	$c_{23}$	&	-1.440\,2\times 10^{-2}	\\
$c_{03}$	&	1.765\,1\times 10^{-4}	&	$c_{30}$	&	8.496\,8\times 10^{-1}	\\
$c_{04}$	&	-1.445\,8\times 10^{-3}	&	$c_{31}$	&	-2.718\,8\times 10^{-1}	\\
$c_{05}$	&	4.333\,5\times 10^{-4}	&	$c_{32}$	&	1.441\,8\times 10^{-1}	\\
\end{tabular}
\end{ruledtabular}
\end{table}

Changes in the third decimal place of the values of the critical exponents $\alpha$ and
$\beta$ result in small density changes that are of the order of 0.1\%. However, some of
the density measurements for water are more accurate than 0.1\%; for example, the
accuracy of the data of Hare and Sorensen\cite{hare87} and Sotani \ea\cite{sotani2000} is
0.01\%. Therefore, the values of the critical exponents must be given with at least four
decimal places. We have adopted the values of Pelissetto and Vicari\cite{pelissetto2002}
and have set $\alpha = 0.1100$ and $\beta = 0.3265$. The value for the molar mass of
H$_2$O (18.015\,268 g/mol) was taken from Wagner and Pruß\cite{wag02nonote} and the
ideal-gas constant $R$ (8.314\,4621 J mol$^{-1}$ K$^{-1}$) was taken from Mohr
\ea\cite{codata2010} The molar mass of D$_2$O (20.027\,508 g/mol) was taken from an IAPWS
guideline.\cite{iapwsfundam}

The number of terms in the background $\mr$ [\eqref{eq:backgr}] was increased step by
step until the experimental data could be accurately represented. The final background
contains fourteen free parameters. The reason for the many background terms of higher
order in temperature and pressure is that the response functions are second derivatives
of the thermodynamic potential. To obtain, for example, a background term in the
compressibility of second order in pressure, it is necessary to have a fourth-order
pressure-dependent term in the potential. The terms in the backgrounds for each property
are at most third order in temperature or in pressure. We want to emphasize that the
observed anomalies are indeed due to the critical part of the equation of state since the
nonlinear contributions to the regular part are needed only when the maximum pressure
considered is higher than about 100~MPa. Besides the background parameters, there are
five additional parameters to be determined: the critical temperature $\Tc$ and volume
$\Vc$, the linear-model amplitudes $a$ and $k$, and the slope of the LLT line $a'$. As
noted, the values of $\Tc$ and $a'$ were constrained to a limited range.

The model was fitted to heat-capacity data of Archer and Carter \cite{arc00} and
IAPWS-95, expansivity data of Hare and Sorensen \cite{hare87}, IAPWS-95 and Ter Minassian
\cite{terminassian1981}, compressibility data of Speedy and Angell \cite{speedy1976},
Kanno and Angell \cite{kanno1979} and Mishima \cite{mishima2010}, density data of Hare
and Sorensen \cite{hare87}, Sotani \ea \cite{sotani2000}, IAPWS-95 and Mishima
\cite{mishima2010}, and speed-of-sound data of Taschin \ea\cite{taschin2011} We have made
small adjustments to the data of Mishima as described in Appendix~\ref{app:correction}.
For all quantities except the heat capacity, values from IAPWS-95 were only used at
0.1~MPa and above 273~K. For the heat capacity, IAPWS-95 values in the range of 273~K to
305~K up to 100~MPa were used. (We considered heat-capacity values calculated from
IAPWS-95 more reliable than high-pressure heat-capacity data of Sirota
\ea\cite{sirota1970}) To reduce the time needed for optimization, not all data points
were used in the fitting process; about 60 points were selected for each of the
quantities heat capacity, expansivity, compressibility, and density. The selected data
are given in the supplemental material.\cite{SCWsupplement}

The model was optimized by minimizing the sum of squared residuals, where the residual is
the difference between model and experiment, divided by the experimental
uncertainty.\cite{lemmon2010} For some data the uncertainty was not given and had to be
estimated. The resulting optimized parameters are listed in Tables \ref{tab:physmodel}
and \ref{tab:D2Omodel}. The value of $a'$ for H$_2$O is exactly 0.09 because $a'$ was
restricted to the range of~0.065 to~0.090, and the optimum is located at the edge of this
range. The fitted models are valid up to 300~K and from 0 to 150~MPa.

\begin{figure}
\centering
\includegraphics{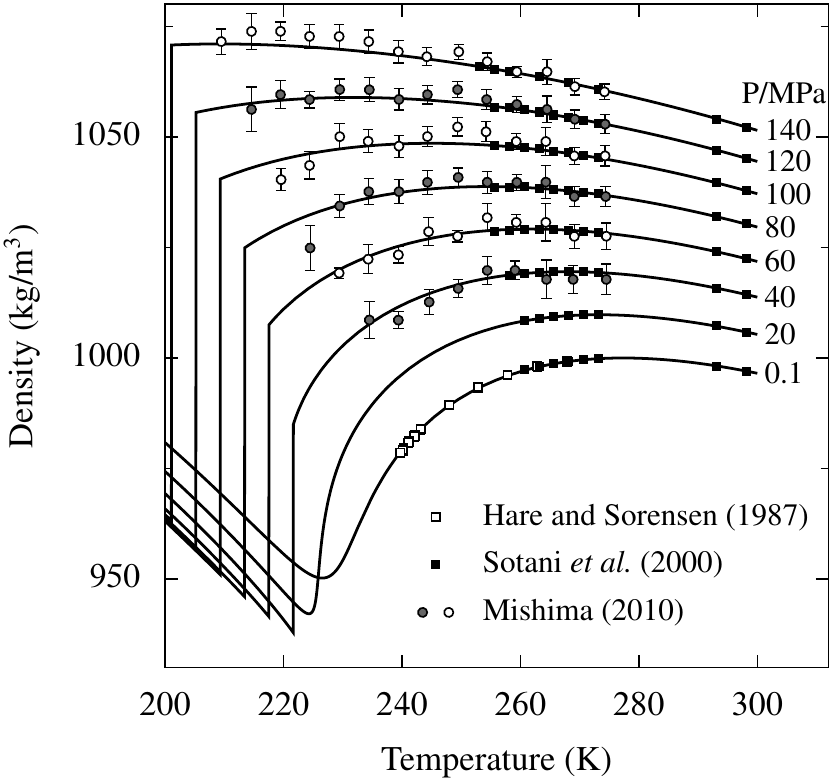}
\caption{\label{fig:densityT150MPa}Densities of H$_2$O according to the model (curves).
The symbols represent experimental data of Mishima \cite{mishima2010}, Sotani \ea \cite{sotani2000}
and Hare and Sorensen \cite{hare87}. The symbols for Mishima's densities on different
isobars are alternatingly open and filled to guide the eye.}
\end{figure}

\begin{figure}
\includegraphics{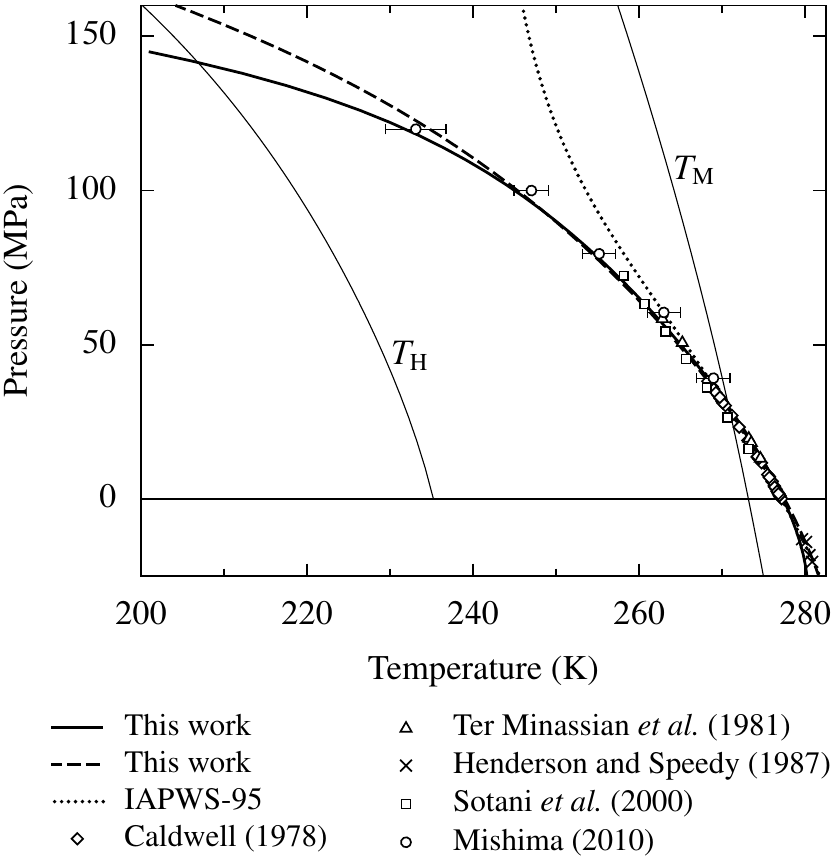}
\caption{\label{fig:TMD} Temperature of maximum density of H$_2$O as a function
of pressure according to the model of Sec.~\ref{sec:physmodel} (thick solid curve),
the model of Sec.~\ref{sec:extended} (dashed) and IAPWS-95 (dotted).
$T_\text{M}$ marks the melting curve \cite{iapwsmeltsub2011} and its extension to negative pressures
\cite{henderson1987b}; $T_\text{H}$ denotes the homogeneous nucleation limit \cite{kanno1975}.
Symbols represent experimental data
\cite{caldwell1978,terminassian1981,henderson1987a,sotani2000,mishima2010}.
The temperatures of maximum density for Mishima's data \cite{mishima2010} were determined by locating
the maxima of fits to his density data.}
\end{figure}

As can be seen in \figref{fig:densityT150MPa}, the model represents the experimental
density data well. The density jumps at low temperature because the isobars cross the LLT
curve there. In \figref{fig:TMD}, the temperature of maximum density is plotted as a
function of pressure, both for the model and for the IAPWS-95 formulation. At pressures
higher than about 60~MPa, the values of IAPWS-95 deviate from the experimental data,
while the current model agrees with the data (except at negative pressures, where the
model was not fitted to any experimental data). The curve of the temperature of maximum
density does not intersect the homogeneous nucleation curve at its break point at about
200~MPa and 180~K. Such an intersection was expected by Dougherty.\cite{dougherty2004}

\begin{figure}
\includegraphics{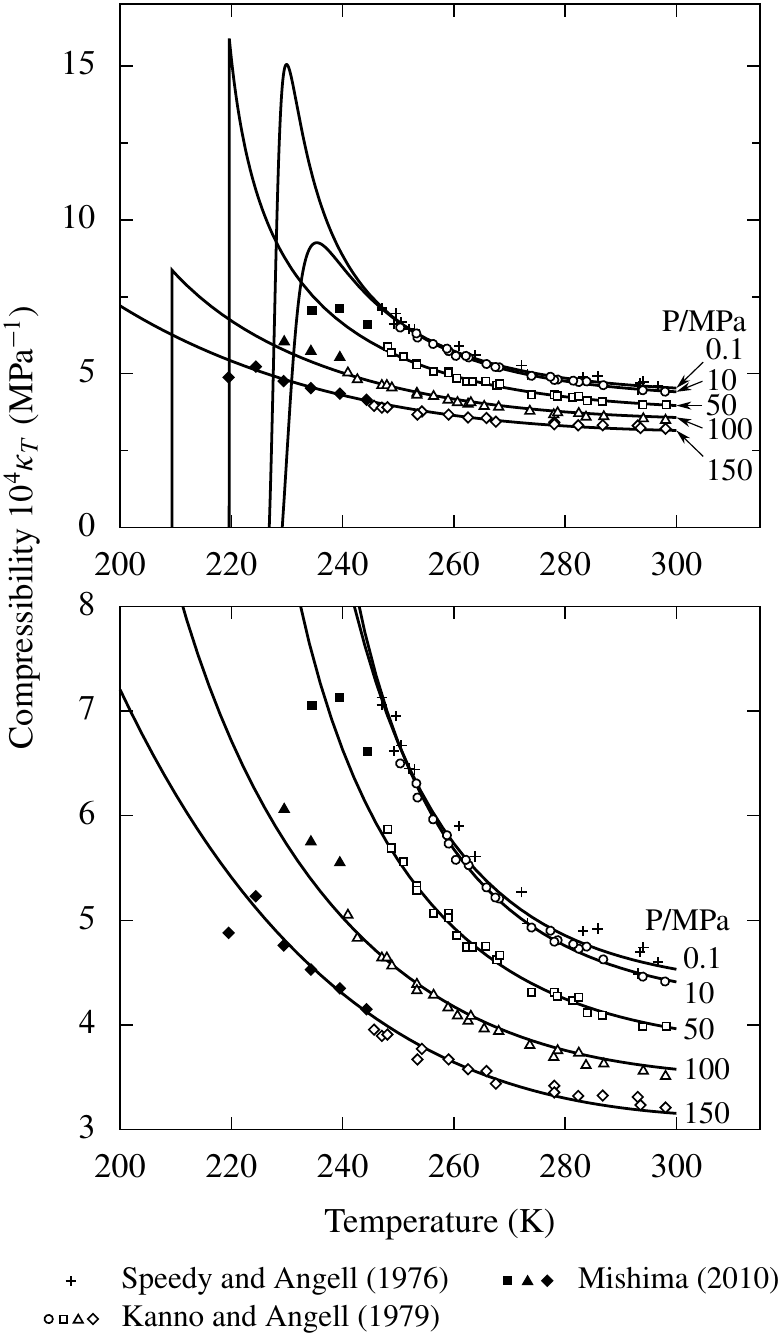}
\caption{\label{fig:comp150MPa}Isothermal compressibility of H$_2$O according to the model
(curves). For clarity, the curves are not shown for temperatures below the LLT line in the bottom graph.
Symbols represent experimental data of Speedy and Angell \cite{speedy1976},
Kanno and Angell \cite{kanno1979}, and Mishima \cite{mishima2010}.
Solid and open symbols with the same shape correspond to the same pressure.}
\end{figure}

The agreement of the compressibility data with values of the model is shown
\figref{fig:comp150MPa}. An interesting feature is the intersection of the isobars of
0.1~MPa and 10~MPa at about 250~K. The experimental data do not confirm or rule out such
an intersection because of the scatter and the lack of data below 245~K. However, the
intersection implies that the pressure derivative of the compressibility,
$(\pypxl{\kappa_T}{P})_T$, is positive at low temperature and ordinary pressures.

\begin{figure}
\includegraphics{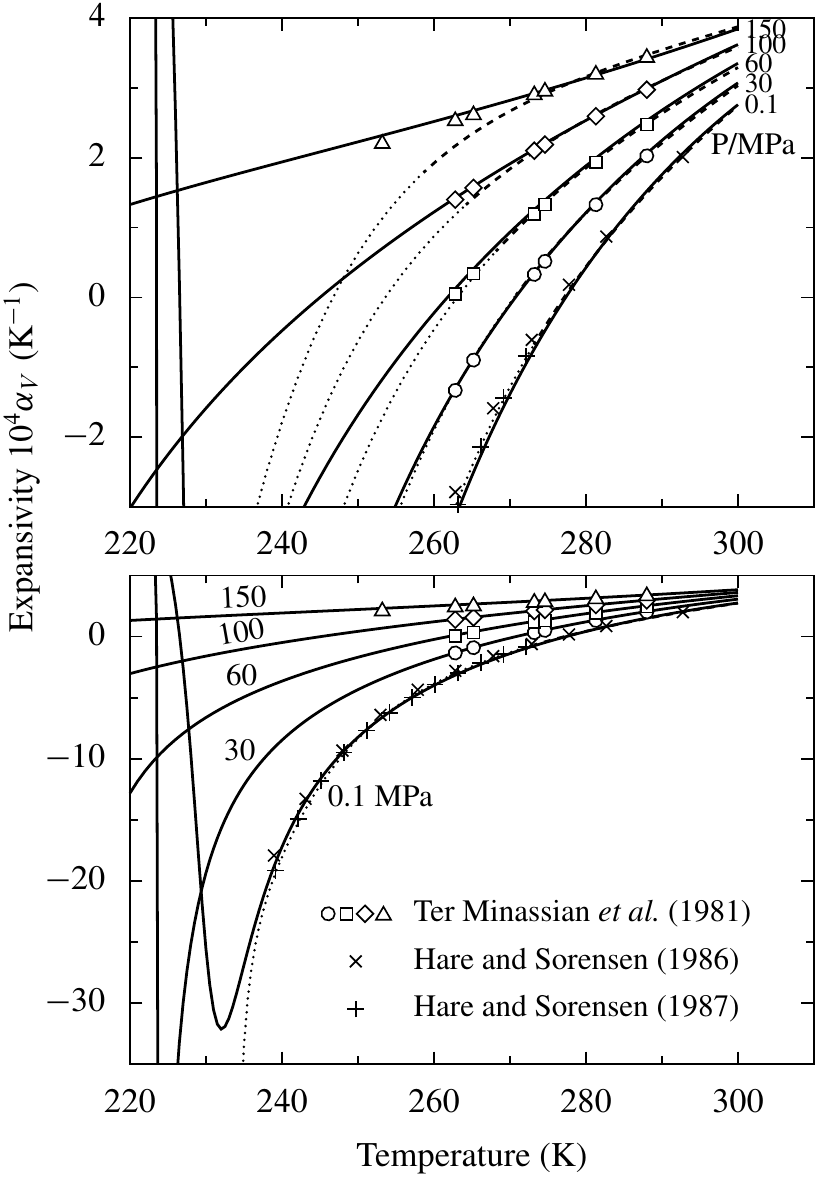}
\caption{\label{fig:exp150MPa}Expansivity coefficient of H$_2$O according to the model
(solid curves) and IAPWS-95 (dashed: within region of validity, dotted: extrapolations).
Symbols represent experimental data of Ter Minassian \ea\cite{terminassian1981}
and Hare and Sorensen.\cite{hare86,hare87} }
\end{figure}

\ffigref{fig:exp150MPa} shows experimental data for the expansivity coefficient and the
values predicted by the model for five pressures. The model follows the experimental
data, contrary to the IAPWS-95 formulation. At 240~K, where the difference between Hare
and Sorensen's data of 1986 and 1987 is largest, the expansivity predicted by the model
lies between them.

\begin{figure}
\includegraphics{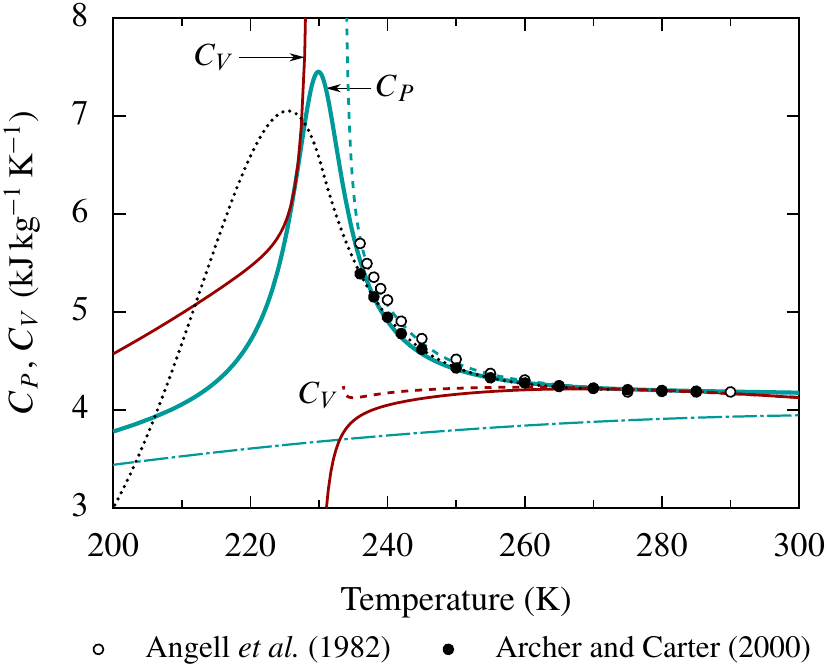}
\caption{\label{fig:Cp150MPa}Isobaric and isochoric heat capacity of H$_2$O versus temperature at 0.1~MPa
according to the model (solid curves), IAPWS-95 (dashed), and the prediction of Murphy and Koop (dotted).
Symbols represent experimental data of Angell \ea\cite{angell1982}
and Archer and Carter.\cite{arc00}
The dash-dotted line is the estimated regular part of $C_P$.
The predicted thermodynamic behavior of $C_V$ is discussed in Sec.~\ref{sec:stability}.}
\end{figure}

\begin{figure}
\includegraphics{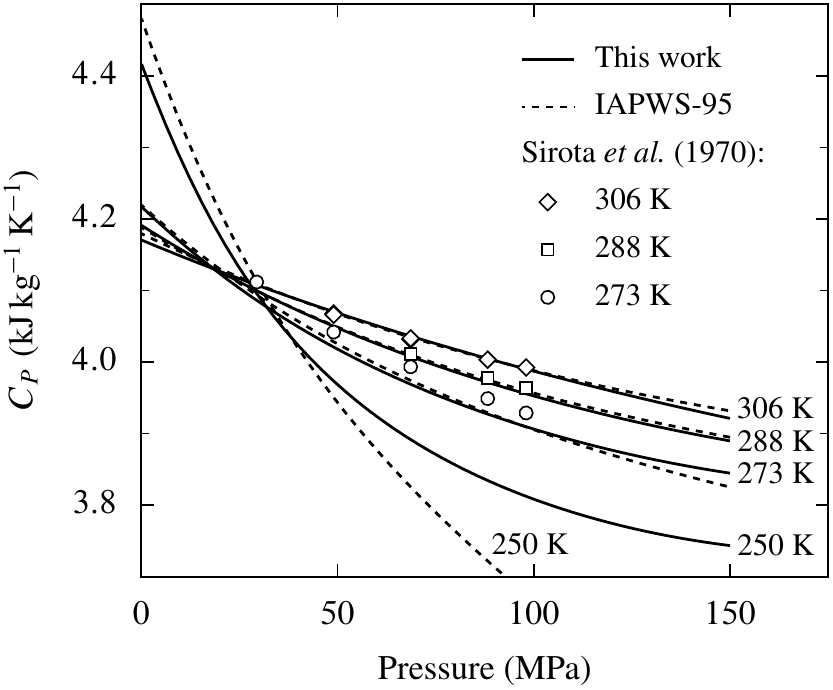}
\caption{\label{fig:CpP150MPa}Isobaric heat capacity of H$_2$O versus pressure according to the model
(solid curves) and IAPWS-95 (dashed).
Symbols represent experimental data of Sirota \ea\cite{sirota1970} }
\end{figure}

Heat-capacity data are compared with the model's predictions in
\figsref{fig:Cp150MPa}{fig:CpP150MPa}. In \figref{fig:Cp150MPa}, it is seen that the
model follows the data of Archer and Carter,\cite{arc00} whereas IAPWS-95 follows the
data of Angell \ea,\cite{angell1982} to which it was fitted. However, the curvature of
the 0.1~MPa isobar of the model is slightly higher than that suggested by the data of
Archer and Carter. Murphy and Koop\cite{mur05} proposed a heat-capacity curve with a
broader peak than that of our model, but with about the same maximum value. The predicted
isochoric heat capacity $C_V$ diverges, as will be discussed in Sec.~\ref{sec:stability}.
\ffigref{fig:CpP150MPa} shows the heat capacity as a function of pressure. There is a
systematic difference between the data of Sirota \ea\cite{sirota1970} in the stable
region and the values of IAPWS-95, and the data of Sirota \ea\ were not selected for the
fit of the current model. At 250~K and pressures above about 50~MPa, the model predicts a
smaller pressure dependence of the heat capacity than IAPWS-95. The pressure dependence
of the heat capacity is thermodynamically related to the expansivity coefficient, and we
have seen that the expansivity coefficient of IAPWS-95 does not agree with experimental
data at low temperature and high pressure (see \figref{fig:exp150MPa}). Therefore,
differences between the heat-capacity values of the current model and IAPWS-95 are to be
expected.

The speed of sound predicted by the model is shown in \figref{fig:IAPWSspeedofsound}. The
prediction agrees fairly well with the experimental data. At 240~K, the model shows a
minimum in the speed of sound whereas IAPWS-95 predicts a monotonically decreasing speed
of sound with decreasing temperature. The divergence of the speed of sound of the model
at low temperature is related to a stability limit as will be discussed in
Sec.~\ref{sec:stability}.

For heavy water, the model was fitted to heat-capacity data of Angell
\ea\cite{angell1982} and Zábranský \ea,\cite{zabransky2010} expansivity data of
Kell\cite{kell67} and Kanno and Angell,\cite{kanno1980} compressibility data of Kanno and
Angell,\cite{kanno1979} density data of Kell,\cite{kell67} Zheleznyi,\cite{zhe69} and
Rasmussen and MacKenzie\cite{rassmussen73}, and speed-of-sound data of Chen and
Millero\cite{chen1977} and Marczak.\cite{marczak1997} It has been found that in the
equation of state for H$_2$O and D$_2$O near the vapor--liquid critical point deviations
from corresponding states only appear in the analytic background contributions, while the
amplitudes of the scaling fields are identical in accordance with corresponding
states.\cite{kamgarparsi1983,kostrow2000} Thus also for the second critical point we
assigned to the linear-model amplitudes $a$ and $k$ the same values for D$_2$O as for
H$_2$O.

\begin{figure}
\centering
\includegraphics{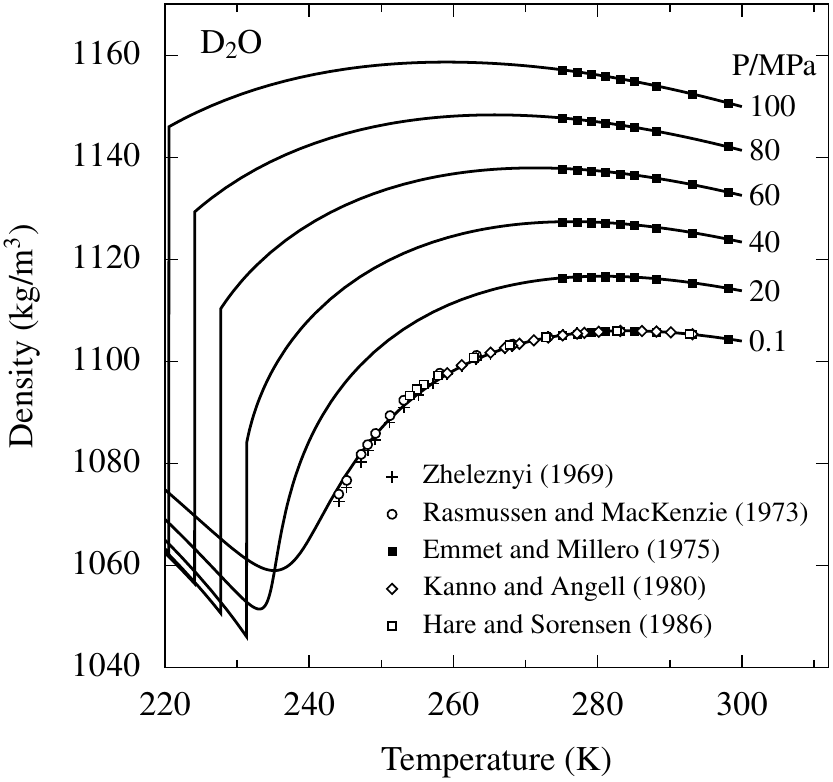}
\caption{\label{fig:densityD2O150MPa}Density of D$_2$O according to the model (curves).
The symbols represent experimental data.\cite{zhe69,rassmussen73,emmet1975,kanno1980,hare86} }
\end{figure}

\begin{figure}
\includegraphics{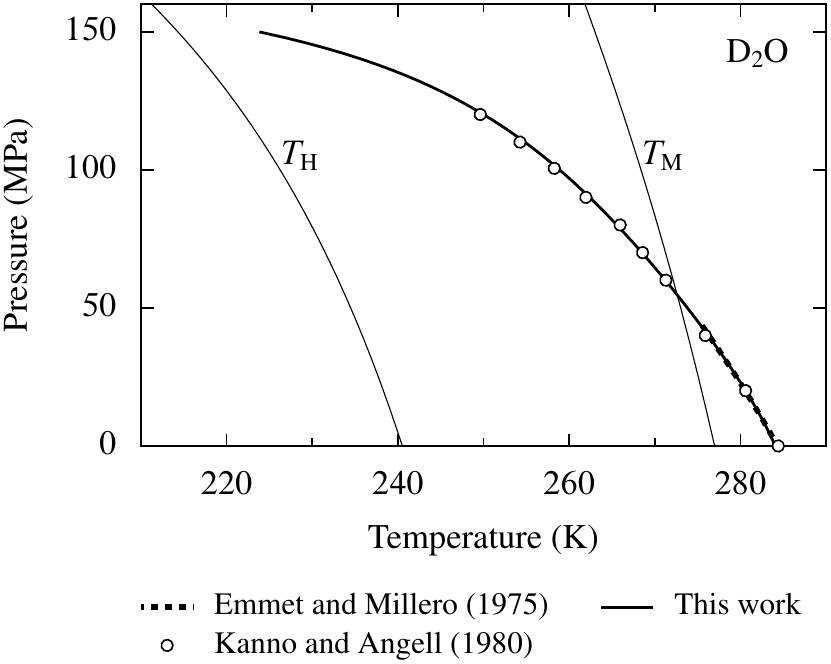}
\caption{\label{fig:TMDD2O} Temperature of maximum density of D$_2$O as a function
of pressure according to the model (thick solid curve).
$T_\text{M}$ marks the melting curve\cite{bridgman1935}
and $T_\text{H}$ denotes the homogeneous nucleation limit.\cite{angell1976}
Symbols represent experimental data.\cite{emmet1975,kanno1980} }
\end{figure}

\begin{figure}
\includegraphics{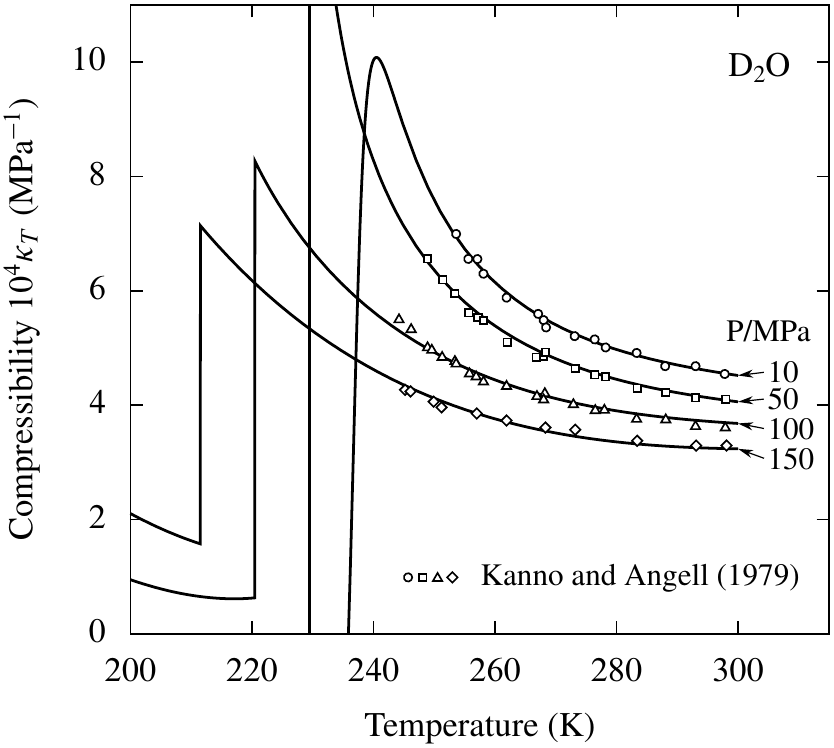}
\caption{\label{fig:compD2O150MPa}Isothermal compressibility of D$_2$O according to the model
(curves).
Symbols represent experimental data of Kanno and Angell.\cite{kanno1979} }
\end{figure}

\begin{figure}
\includegraphics{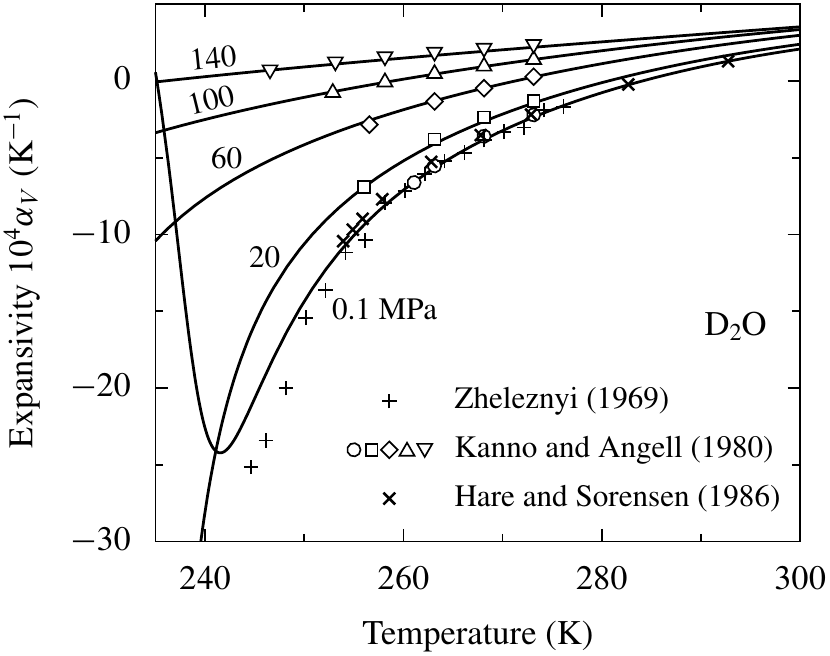}
\caption{\label{fig:expD2O150MPa}Expansivity coefficient of D$_2$O according to the model
(curves). Symbols represent experimental data.\cite{zhe69,kanno1980,hare86}
The data of Zheleznyi\cite{zhe69} (plusses) likely overestimate the anomaly in the expansivity (see the text).}
\end{figure}

\begin{figure}
\includegraphics{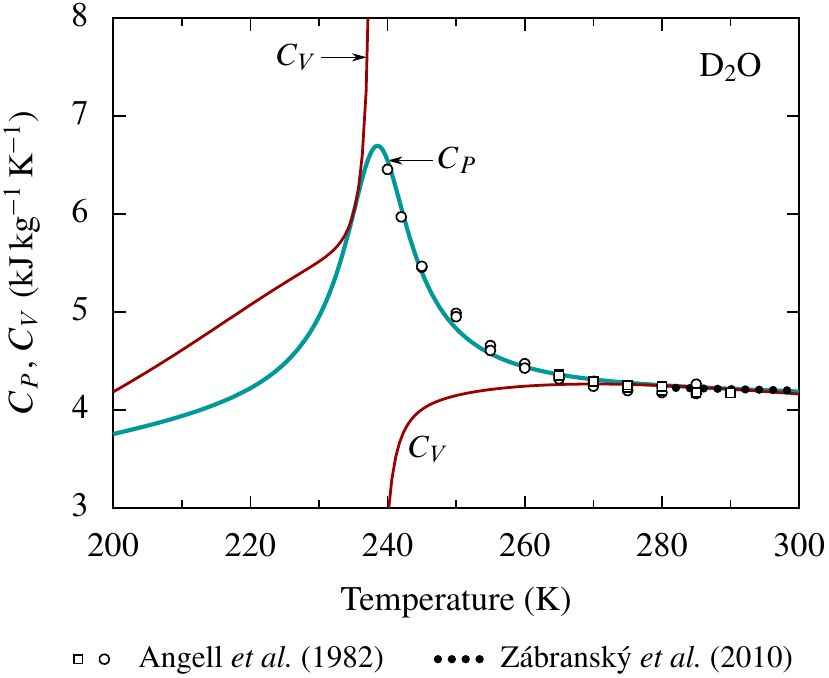}
\caption{\label{fig:CpD2O150MPa}Isobaric and isochoric heat capacity of D$_2$O versus temperature at 0.1~MPa
according to the model (solid curves). Symbols represent experimental $C_P$ data of Angell \ea\cite{angell1982}
(squares: bulk water, circles: emulsion) and the dotted curve shows the
correlation of Zábranský \ea\cite{zabransky2010} for $C_P$.
The predicted thermodynamic behavior of $C_V$ is discussed in Sec.~\ref{sec:stability}.}
\end{figure}

\begin{figure}
\includegraphics{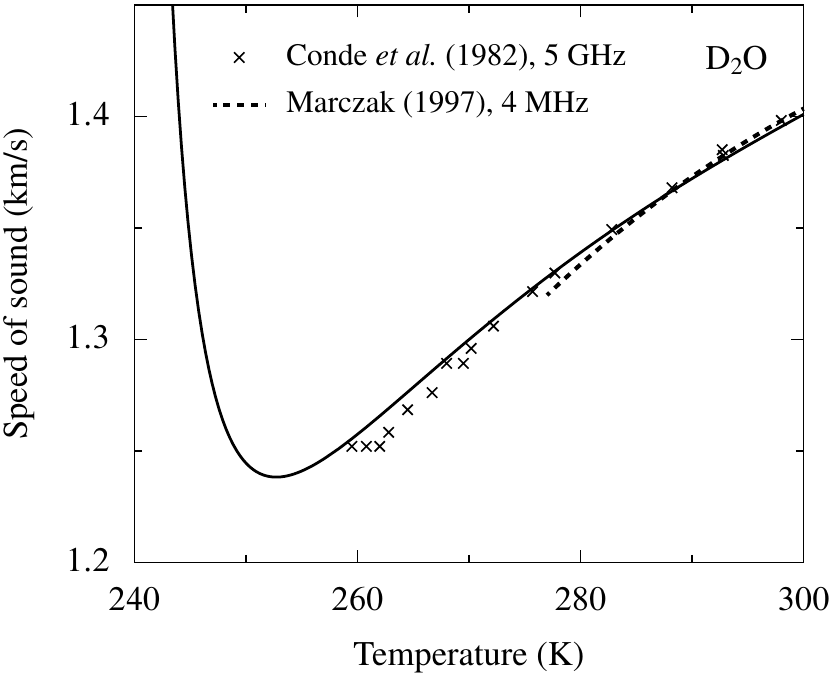}
\caption{\label{fig:speedofsoundD2O}Speed of sound of D$_2$O at 0.1~MPa according
to the model (solid curve).
Symbols represent experimental data of Conde \ea\cite{conde1982} and the dashed line
shows the correlation of Marczak.\cite{marczak1997} }
\end{figure}

Figures \ref{fig:densityD2O150MPa}--\ref{fig:speedofsoundD2O} show the predictions of the
model for heavy water compared with experimental data on density, compressibility,
expansion coefficient, heat capacity, and speed of sound. The results are similar to
those for ordinary water. The quality of the description is almost as good as for
ordinary water, except for the speed of sound; the experimental data for speed of sound
show noticeable (a few percent) deviation from the model. The expansivity data of
Zheleznyi,\cite{zhe69} shown in \figref{fig:expD2O150MPa}, deviate from the prediction of
the model below about 255~K. For ordinary water, Hare and Sorensen\cite{hare87} concluded
that the data of Zheleznyi ``show smaller densities for $T < -25$~°C than our bulk data
and thus overestimate the anomaly in the expansivity.'' For heavy water, the density and
expansivity values of Zheleznyi are also likely too low, since the densities measured by
others\cite{rassmussen73,kanno1980,hare86} are all higher.

While the critical parts of the thermodynamic properties of H$_2$O and D$_2$O follow the
law of corresponding states (the critical amplitudes $a$ and $k$ are the same) the
corresponding regular parts do not follow this law. The critical compressibility factors
$Z_\text{c} = \Pc/(\rho_\text{c} R \Tc)$ for H$_2$O ($Z_\text{c} = 0.28$) and D$_2$O
($Z_\text{c} = 0.32$) LLCP differ by about 13\% while for the vapor--liquid critical
point these factors differ by only 1\%.\cite{kostrow2000} The difference of 13\% should
be taken too seriously since the fit of the model is insensitive to the value of $\Pc$ at
the LLCP.

\section{Semi-empirical extension of scaling model}\label{sec:extended}
As mentioned in the preceding section, the asymptotic theoretical model implies a linear
LLT line. An attempt to include an additional term accounting for curvature of the LLT
curve has been made by Fuentevilla and Anisimov\cite{fuentevilla2006} by introducing a
pressure-dependent coefficient $a'$ in \eqref{eq:h1}. However, such a procedure yields
terms proportional to $\phi_1$ in some response functions which do not vanish far away
from the critical point, where the critical part should not play a role anymore. In
principle, this problem can be solved by including crossover from singular critical
behavior asymptotically close to the critical point to analytic behavior far away from
the critical point, as has been done in the equation of state for H$_2$O near the
vapor--liquid critical point,\cite{kostrow2000} and has also been suggested by
Kiselev.\cite{kiselev2001,kiselev2002} However, it turns out that the theoretical model
can represent all experimental data for H$_2$O up to the maximum available pressure of
400~MPa, if we simply remove any constraints on the slope of the LLT and on the critical
parameters in the unstable region which cannot be measured experimentally. With the
addition of only two background terms, we were then able to fit almost all experimental
data for supercooled water with this semi-empirical extension of the theoretical model.

\begin{table}
\caption{\label{tab:extendedmodel}Parameter values for the extended model of H$_2$O}
\begin{ruledtabular}
\begin{tabular}{lD{.}{.}{12}lD{.}{.}{12}}
Parameter & \multicolumn{1}{c}{Value} & Parameter & \multicolumn{1}{c}{Value} \\ \hline
$\Tc$/K	&	213.89	&	$c_{12}$	&	-4.656\,9\times 10^{-3}	\\
$\Pc$/MPa	&	56.989	&	$c_{13}$	&	2.362\,7\times 10^{-3}	\\
$\rho_\text{c}$/(kg m$^{-3}$)	&	949.87	&	$c_{14}$	&	-2.869\,7\times 10^{-4}	\\
$a$	&	0.116\,24	&	$c_{20}$	&	-3.614\,4\times 10^{0}	\\
$k$	&	0.432\,80	&	$c_{21}$	&	-1.500\,9\times 10^{-2}	\\
$a'$	&	0.108\,98	&	$c_{22}$	&	-2.460\,9\times 10^{-2}	\\
$c_{02}$	&	4.079\,3\times 10^{-2}	&	$c_{23}$	&	9.867\,9\times 10^{-4}	\\
$c_{03}$	&	-6.791\,2\times 10^{-4}	&	$c_{30}$	&	5.426\,7\times 10^{-1}	\\
$c_{04}$	&	-7.566\,9\times 10^{-6}	&	$c_{31}$	&	1.062\,0\times 10^{-1}	\\
$c_{05}$	&	1.092\,2\times 10^{-5}	&	$c_{32}$	&	1.275\,9\times 10^{-2}	\\
$c_{11}$	&	1.954\,7\times 10^{-1}	&	$c_{41}$	&	-7.997\,0\times 10^{-2}	\\
\end{tabular}
\end{ruledtabular}
\end{table}

\begin{figure}[t]
\includegraphics{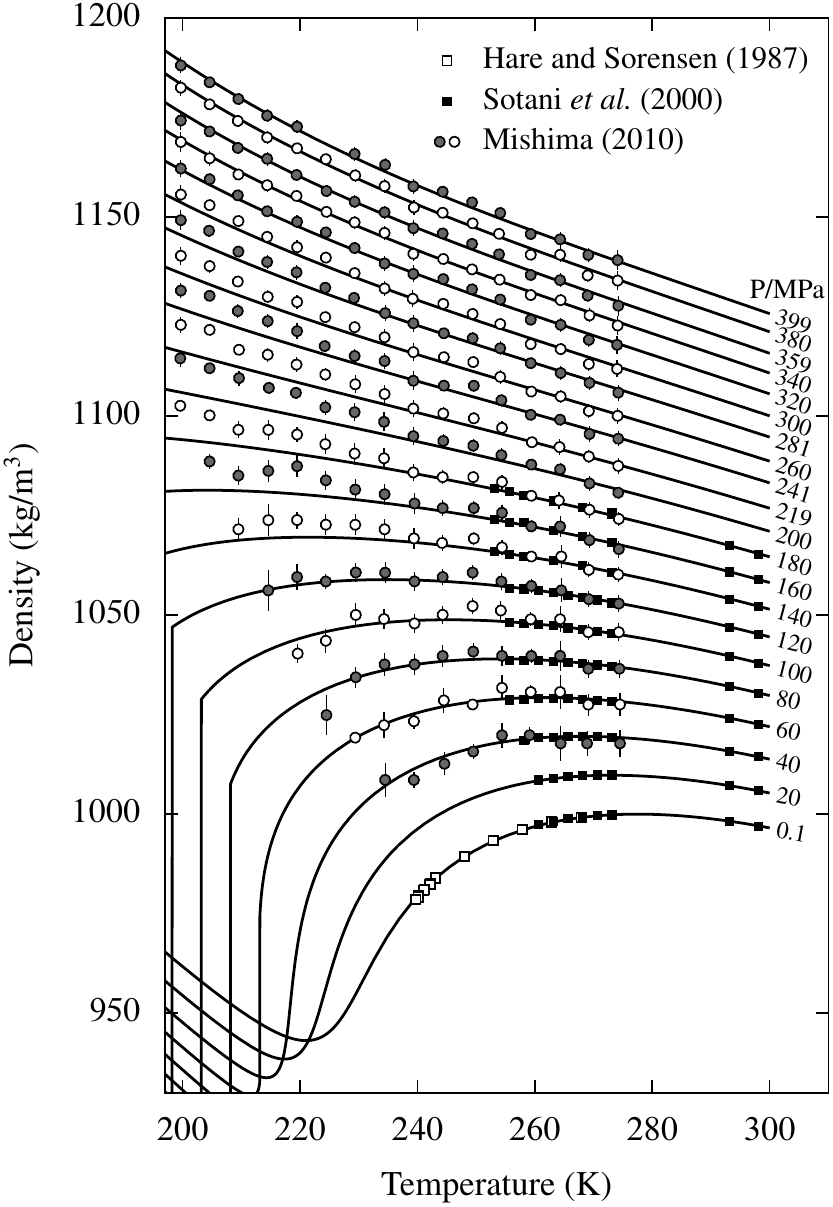}
\caption{\label{fig:densityT400MPa}Densities of H$_2$O according to the extended model (curves).
The symbols represent experimental data of Mishima \cite{mishima2010}, Sotani \ea \cite{sotani2000}
and Hare and Sorensen \cite{hare87}. The symbols for Mishima's densities on different
isobars are alternatingly open and filled to guide the eye.
The vertical lines through Mishima's points are uncertainties given by Mishima.}
\end{figure}

After fitting the model to a selection of experimental data\cite{SCWsupplement} at
pressures up to 400~MPa, the parameters listed in \tabref{tab:extendedmodel} were
obtained. Some parameters are significantly different from the earlier values. In
particular, the critical pressure is about a factor of two higher than in the previous
section, and the critical temperature is 10~K lower. Nevertheless, the extended model
does represent almost all available experimental thermodynamic data for supercooled
water. A comparison between the density values predicted by the model is presented in
\figref{fig:densityT400MPa}. The model reproduces the data, except for Mishima's points
between 160~MPa and 300~MPa below 230~K. The temperature of maximum density is plotted in
\figref{fig:TMD}. The calculated behavior of the extended model is similar to that of the
asymptotic model in the previous section in the range from 0~MPa to 120~MPa. At higher
pressures, the extended model predicts higher temperatures of maximum density than the
previous model, but the results are still within the experimental uncertainty.

\begin{figure}
\includegraphics{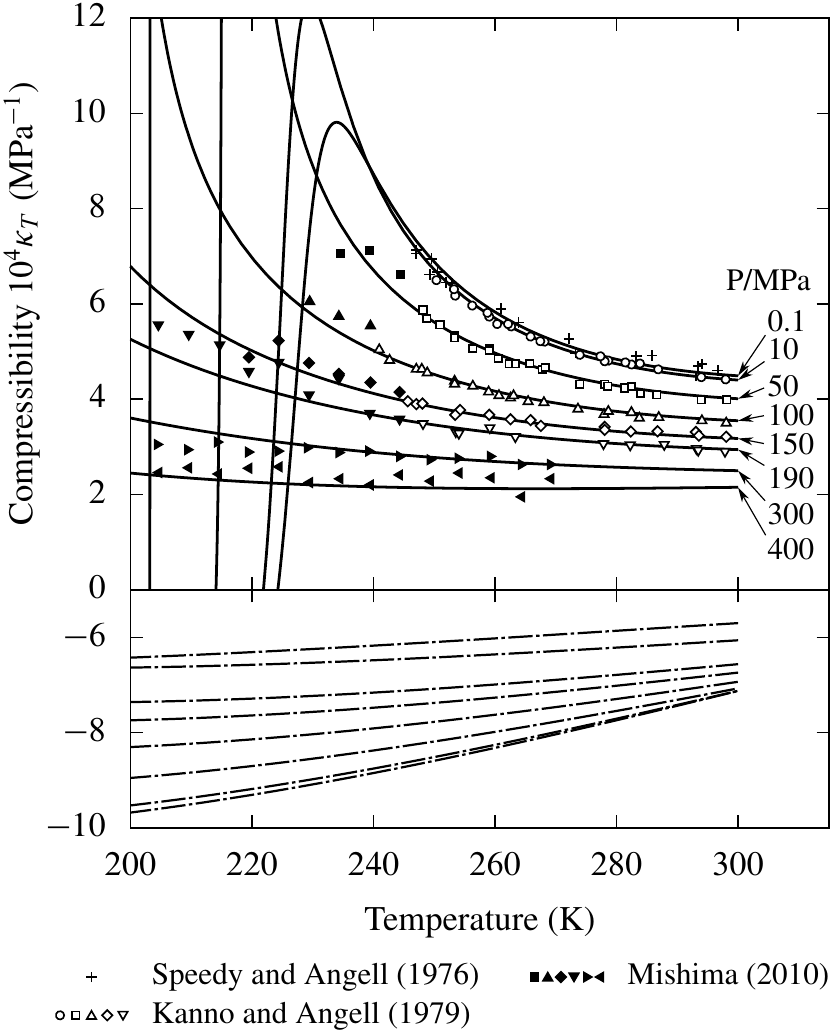}
\caption{\label{fig:comp400MPa}Isothermal compressibility of H$_2$O according to the extended model
(curves). Symbols represent experimental data of Speedy and Angell \cite{speedy1976},
Kanno and Angell \cite{kanno1979}, and Mishima.\cite{mishima2010}
Solid and open symbols with the same shape correspond to the same pressure.
The regular backgrounds for the compressibility are shown by dash-dotted lines.
The negative values of the regular parts are due to incorporating the fluctuation-induced
critical background in $\chi_2$ as noted in Appendix~\ref{app:linmodel}. }
\end{figure}

\begin{figure}
\includegraphics{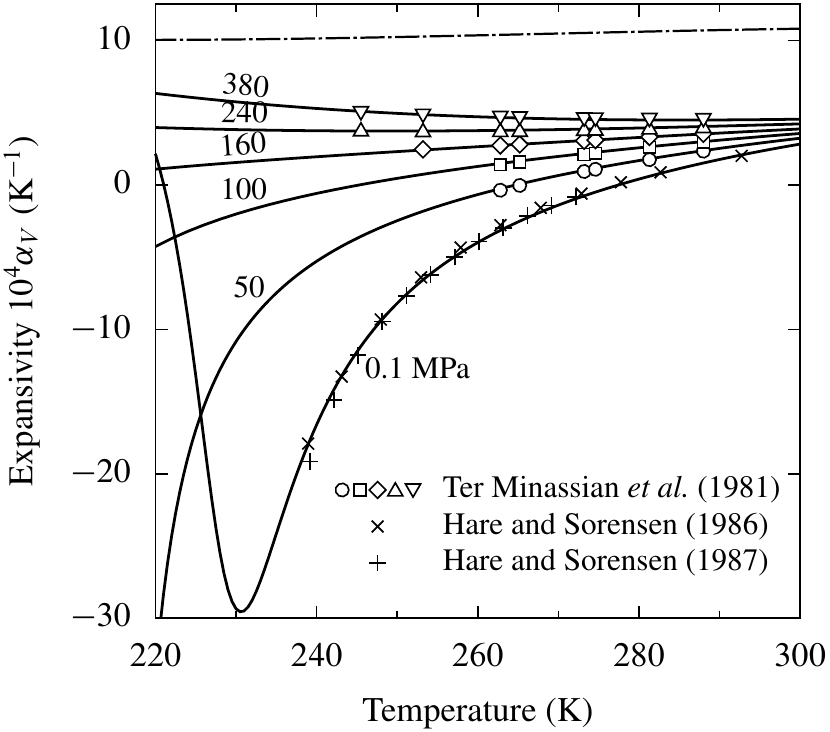}
\caption{\label{fig:exp400MPa}Expansivity coefficient of H$_2$O to the extended model (curves).
Symbols represent experimental data of Ter Minassian \ea\cite{terminassian1981}
and Hare and Sorensen.\cite{hare86,hare87}
The dash-dotted line is the estimated regular part at 0.1~MPa.}
\end{figure}

Compressibility data are compared with values of the extended model in
\figref{fig:comp400MPa}. As in the previous model, the isobars of 0.1~MPa and 10~MPa
intersect, but the intersection is located at a lower temperature. The data of Speedy and
Angell\cite{speedy1976} and Kanno and Angell\cite{kanno1979} are well represented. The
extended model also reproduces most of Mishima's data, which have a lower accuracy than
the data of Angell and coworkers. \ffigref{fig:exp400MPa} shows the expansivity
coefficient predicted by the extended model, which agrees with the data of Ter Minassian
\ea\cite{terminassian1981} Below 250~K, the model agrees better with the data of Hare and
Sorensen of 1986 than with their data of 1987.

\begin{figure}
\includegraphics{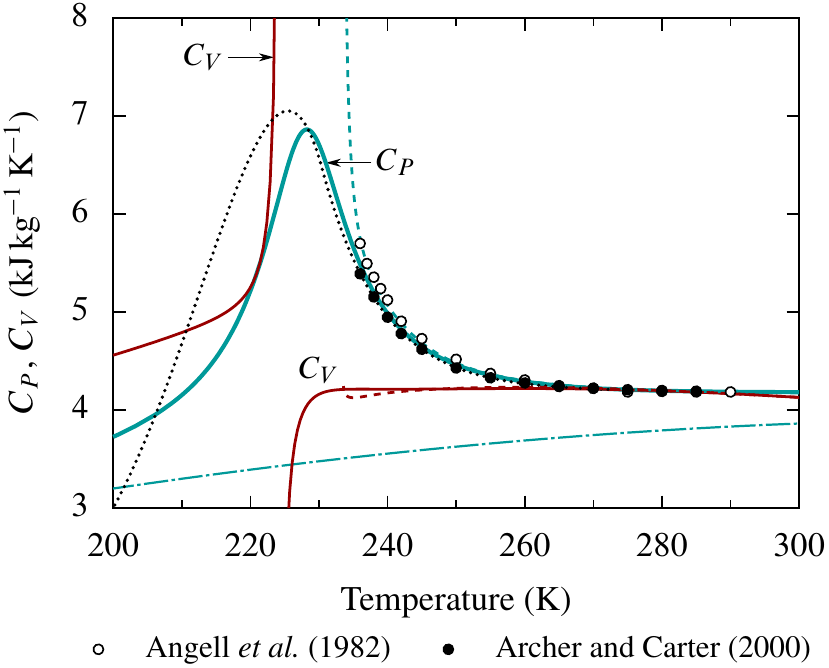}
\caption{\label{fig:Cp400MPa}Isobaric and isochoric heat capacity of H$_2$O versus temperature at 0.1~MPa
according to the extended model (solid curves), IAPWS-95 (dashed),
and the prediction of Murphy and Koop (dotted).
Symbols represent experimental data of Angell \ea\cite{angell1982}
and Archer and Carter.\cite{arc00}
The dash-dotted line is the estimated regular part of $C_P$.
The predicted thermodynamic behavior of $C_V$ is discussed in Sec.~\ref{sec:stability}.}
\end{figure}

\begin{figure}
\includegraphics{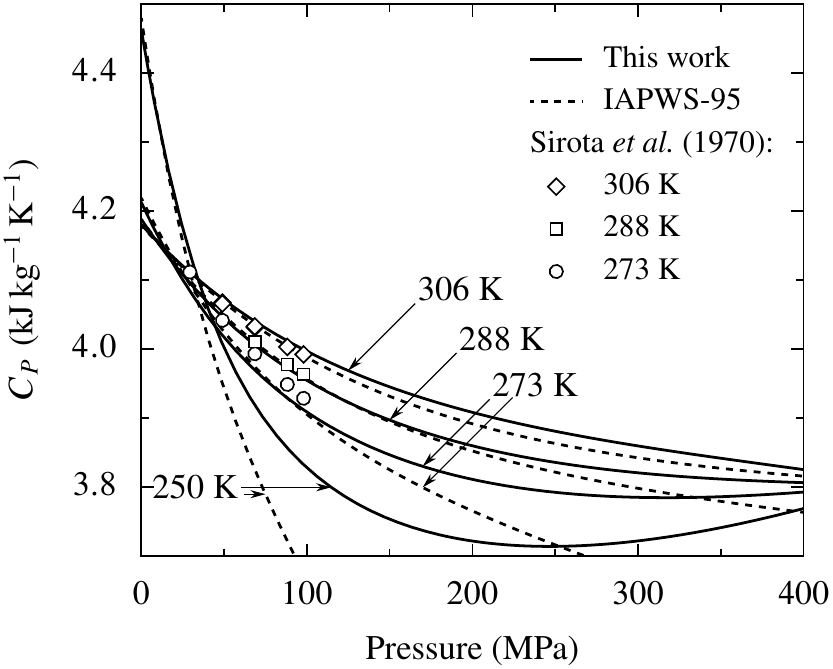}
\caption{\label{fig:CpP400MPa}Isobaric heat capacity of H$_2$O versus pressure according to
the extended model (solid curves) and IAPWS-95 (dashed).
Symbols represent experimental data of Sirota \ea\cite{sirota1970} }
\end{figure}

The calculated heat capacity at 0.1~MPa is compared with experimental data in
\figref{fig:Cp400MPa}. For the extended model, the maximum of the heat capacity is lower
than for the asymptotic model, and the curvature of the data of Archer and
Carter\cite{arc00} is better represented. The predicted heat capacity is shown as a
function of pressure in \figref{fig:CpP400MPa}. There are large differences between the
results of the model and those of IAPWS-95; the model predicts a minimum in the 250~K
heat-capacity isotherm at about 240~MPa. A minimum in the heat capacity at this location
was also predicted by Ter Minassian \ea\cite{terminassian1981} based on their
measurements of the expansivity coefficient.

The speed of sound predicted by the model, shown in \figref{fig:IAPWSspeedofsound},
agrees fairly well with the experimental data of Taschin \ea\cite{taschin2011}

\section{Physical interpretation of the model}\label{sec:interpretation}
\subsection{Two states in supercooled water}
Two features make the second critical point in water phenomenologically different from
the well-known vapor--liquid critical point. The negative slope of the liquid--liquid
phase transition line in the $P$--$T$ plane means that high-density liquid water is the
phase with larger entropy. The relatively large value of this slope at the liquid--liquid
critical point (about 25 times greater than for the vapor--liquid transition at the
critical point) indicates the significance of the entropy change relative to the density
change, and, correspondingly, the importance of the entropy fluctuations. These features
suggest that liquid--liquid phase separation in water is mostly driven by entropy rather
than by energy, thus supporting the ``lattice-liquid'' choice for the scaling fields
given by Eqs.~(\ref{eq:h1})--(\ref{eq:h3}) with $b'=0$.

As the first step to understand a relation between water's polyamorphism and the behavior
of cold aqueous solutions, Bertrand and Anisimov\cite{bertrand2011} have introduced a
mean-field ``two-state'' model that clarifies the nature of the phase separation in a
polyamorphic single-component liquid. Two-fluid-states models trace their lineage back to
a 19th century paper by Röntgen.\cite{roentgen} Relatively recently, Ponyatovski\u{\i}
\ea\cite{ponyatovskii1994} and, more quantitatively, Moynihan\cite{moynihan1997} have
described the emergence of a LLCP in supercooled water as resulting from the effects of
nonideality in a mixture of two ``components,'' with their fraction being controlled by
thermodynamic equilibrium. However, while Moynihan assumed a ``regular-solution'' type of
nonideality, which implies an energy-driven phase separation, such as the vapor--liquid
transition or the conventional liquid--liquid transition in binary solutions, we believe
that a near ``athermal-solution'' type of nonideality is mainly responsible for the
entropy-driven phase separation in metastable water near the LLCP.

It is assumed that liquid water is a ``mixture'' of two states, A and B, of the same
molecular species. For instance, these two states could represent two different
arrangements of the hydrogen-bond network in water and correspond to the low-density and
high-density states of water. The fraction of water molecules, involved in either
structure, denoted $\phi$ for state A and $1-\phi$ for state B, is controlled by
thermodynamic equilibrium between these two structures. Unlike a binary fluid, the
fraction $\phi$ is not an independent variable, but is determined as a function of
pressure $P$ and temperature $T$ from the condition of thermodynamic equilibrium. The
simplest ``athermal solution model''\cite{prigogine1954} predicts a symmetric
liquid--liquid phase separation for any temperature, with the critical fraction
$\phi_\text{c} = 1/2$, if the interaction parameter, which controls the excess
(non-ideal) entropy of mixing is higher than its critical value. However, unlike a purely
athermal non-ideal binary fluid, the entropy-driven phase separation in a polyamorphic
single-component liquid will not be present at any temperature. On the contrary, the
critical temperature $\Tc$ is to be specified through the temperature dependence of the
equilibrium constant $K$ by
\begin{equation}
    \ln K = \lambda \left(\frac{1}{T} - \frac{1}{\Tc}\right),
\end{equation}
where $\lambda$ is the heat (enthalpy change) of ``reaction'' between A and B. A finite
slope of liquid--liquid coexistence in the $P$--$T$ plane can be incorporated into the
two-state lattice liquid model if one assumes that the Gibbs energy change of the
``reaction'' also depends on pressure.

A conceptually similar two-state model suggested by Hrubý and
Holten\cite{hrubyholten2004,*hol04} has explained the inflection point in the surface
tension (\figref{fig:IAPWSsurfacetension}) by a rapid increase of the fraction of water
molecules in the low-density state as water is cooled down.

\subsection{Liquid--liquid phase separation}
\begin{figure}
\includegraphics{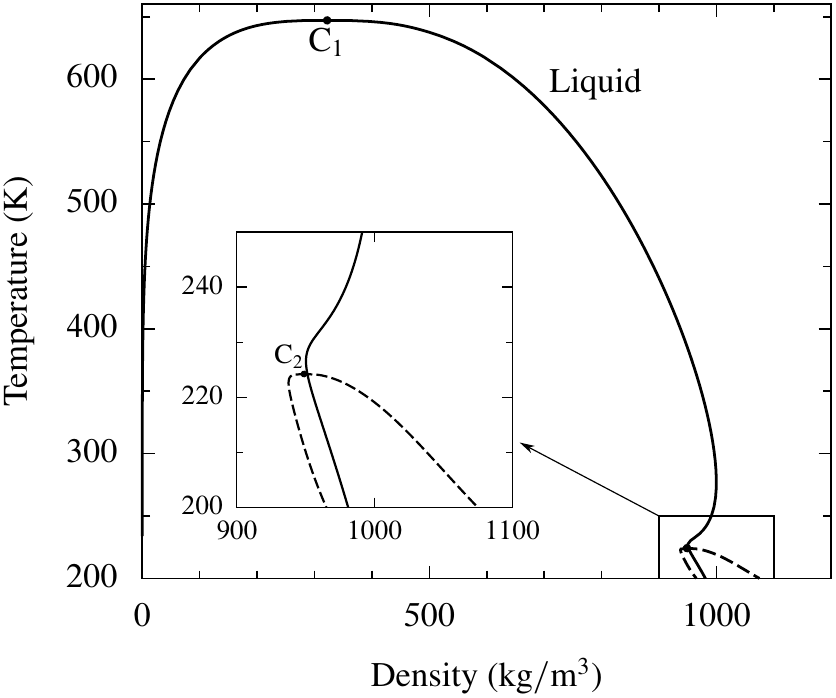}
\caption{\label{fig:Trho}Temperature--density diagram.
C$_1$ and C$_2$ indicate the first and second critical point.
The line left of C$_1$ is the saturated vapor density according to IAPWS-95.
Down to 250~K, the line marked by `Liquid' is the saturated liquid density computed with IAPWS-95; below
250~K, the line represents the liquid density at 0.1~MPa predicted by our model.
The dashed lines show the densities of the two liquid phases in equilibrium on the LLT line.}
\end{figure}

\begin{figure}
\includegraphics{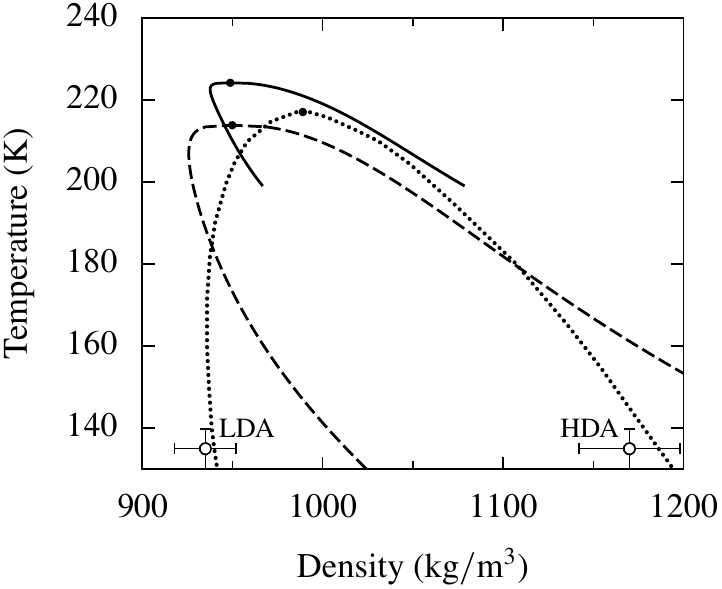}
\caption{\label{fig:Trhocompare}Temperature--density diagram of the two liquid phases in equilibrium on the LLT line.
Solid line: asymptotic model, dashed line: extended model,
dotted line: prediction of the two-state regular-solution model of Moynihan.\cite{moynihan1997}
Solid dots mark the second critical points. Open circles are the densities of the low-density amorphous
(LDA) and high-density amorphous (HDA) phases of water estimated from the experiments
of Mishima.\cite{mishima1994}
}
\end{figure}

A consequence of the second critical point is a phase separation into a high-density and
a low-density liquid at temperatures below the critical temperature. From the
``lattice-liquid'' two-state model with a steep slope of the liquid--liquid transition
(LLT) line in the $P$--$T$ phase diagram it follows that the phase separation is mostly
driven by entropy rather than density. The LLT line is not exactly vertical in the phase
diagram due to a small density difference between the two liquid phases, as pointed out
by Bertrand and Anisimov.\cite{bertrand2011} In \figref{fig:Trho}, a temperature--density
phase diagram is shown including both the vapor--liquid and liquid--liquid critical
points. The region below the second critical point is shown in more detail in
\figref{fig:Trhocompare}, where predictions by several models of the densities of the two
liquid phases are compared.

\begin{figure}
\includegraphics{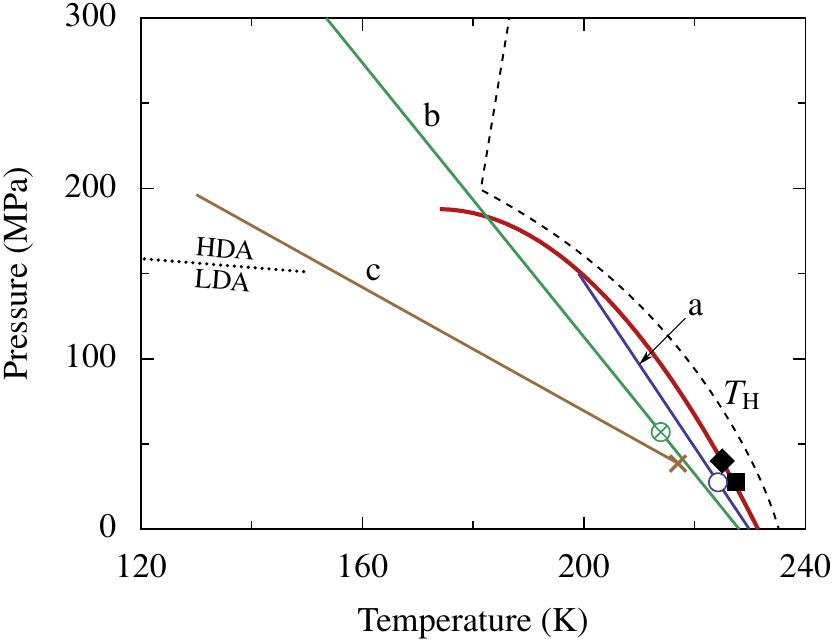}
\caption{\label{fig:LLTline2}Different linearizations of the LLT curve. The thick solid curve
is the LLT curve suggested by Mishima.\cite{mishima2010}
The lines marked a and b are the linearized LLT lines of our models restricted to 150~MPa and 400~MPa, respectively;
the line marked c is the LLT line of Moynihan.\cite{moynihan1997}
Symbols mark the liquid--liquid critical points;
the solid diamond is the critical point of Mishima\cite{mishima2010},
the solid square is that of Bertrand and Anisimov,\cite{bertrand2011}
the open circle is that of our asymptotic model,
the crossed circle is that of our extended model,
and the cross is that of Moynihan.\cite{moynihan1997}
The dotted line is the phase-transition line between the two amorphous phases LDA and HDA
as estimated by Whalley \ea\cite{whalley1989} }
\end{figure}

\begin{figure*}
\includegraphics{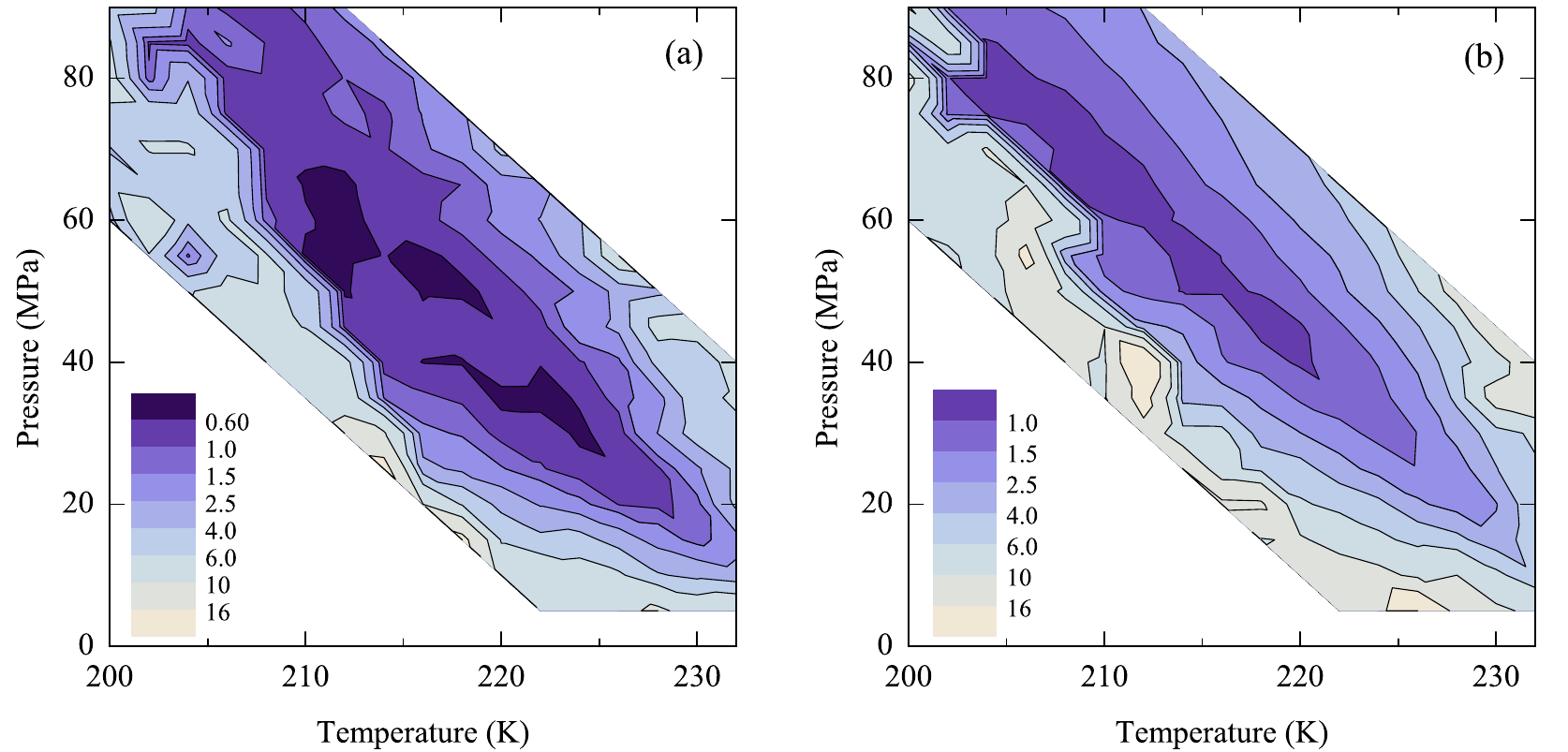}
\caption{\label{fig:contours} Reduced sum of squared residuals as a function of the
location of the liquid--liquid critical point for H$_2$O. (a) Asymptotic model, fitted to
experimental data up to 150~MPa; (b) Extended model, fitted to experimental data up to
400~MPa.}
\end{figure*}

In the second-critical-point scenario it is assumed that the LLT line is connected to the
phase-transition line between two amorphous phases found below about
130~K.\cite{mishima1998review} Since the amorphous phase-transition line is nearly
horizontal in the $P$--$T$ diagram, as shown in \figref{fig:LLTline2}, it follows that
the LLT must be strongly curved to account for such a connection. In our model, we have
linearized the LLT, so the predicted equilibrium densities at 135~K differ from the
experimental densities of the amorphous phases, as seen in \figref{fig:Trhocompare}. The
two-state model of Moynihan\cite{moynihan1997} was fitted to the experimental
amorphous-phase densities, as shown in \figref{fig:Trhocompare}. In Moynihan's model, the
LLT is also a straight line, but its slope, $\di P/\di T$, is less than half of the slope
of our LLT lines (\figref{fig:LLTline2}). Comparing our models for the 0--150~MPa and
0--400~MPa ranges, we see that a lower LLT slope is needed for the larger pressure range.
In conclusion, with a linearized LLT, the larger the desired temperature and pressure
range of the model, the smaller $\di P/\di T$ slope of the linearized LLT line is
required.

Based on the fit of the model to the experimental data, the liquid--liquid critical point
can be located in a certain range of temperatures and pressures, depending on the
pressure limit for experimental data involved in the fit. Without any preliminary idea on
the location of the LLT curve, the position of the liquid--liquid critical point is
uncertain. In principle, we cannot exclude even a negative value for the critical
pressure. Our assumption on the LLT curve is based on the predictions of different
authors\cite{kanno1979,mishima2000,mishima2010,zhang2011} which give approximately the
same result. We have linearized this curve for a certain range of pressures. If the data
are restricted up to 100~MPa, the critical pressure is optimized to 28~MPa and the
critical temperature is found to be 227~K.\cite{bertrand2011} If we include more
high-pressure data and linearize the LLT curve in the pressure range up to 150~MPa, the
optimal value for the critical pressure becomes less certain and moves up. This tendency
continues when we include high-pressure data and linearize the LLT curve up to 400~MPa.
As illustrated in \figref{fig:contours}(a), best fits for the 150~MPa model are obtained
with critical-pressure values between about 25~MPa and 70~MPa. For the extended, 400~MPa,
model, the optimum critical-pressure values move to a higher range from 40~MPa to 85~MPa
[\figref{fig:contours}(b)].

\subsection{Absolute stability limit of the liquid state}\label{sec:stability}
\begin{figure}
\includegraphics{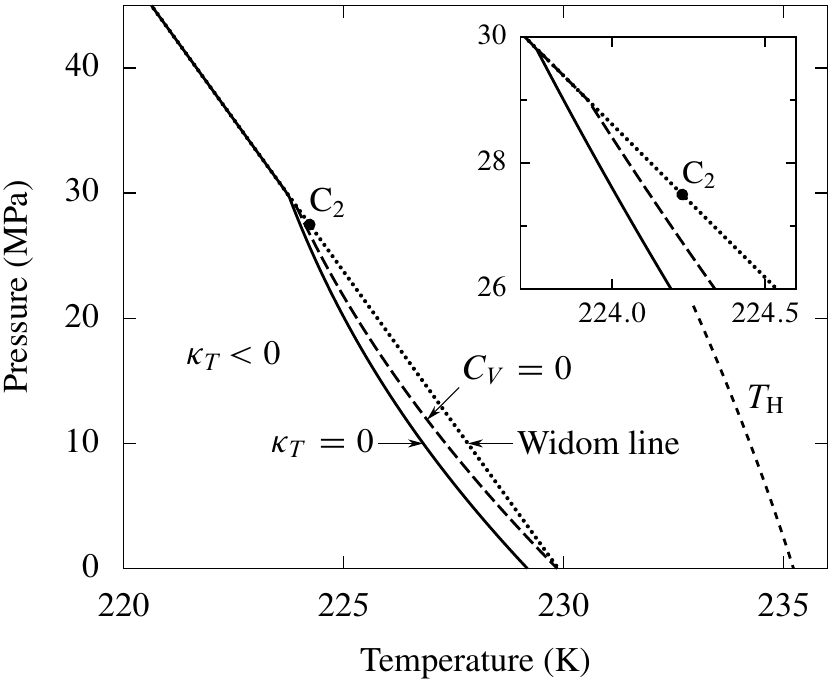}
\caption{\label{fig:stability}
Absolute stability limit of the liquid state, predicted by the model.
Solid line: limit of mechanical stability, where the isothermal compressibility $\kappa_T$ is zero.
Long dashed line: limit of thermal stability, where the isochoric heat capacity $C_V$ is zero.
$T_\text{H}$ marks the experimental homogeneous nucleation limit,
and C$_2$ indicates the second critical point.
The dotted line, at which $h_1=0$ [\eqref{eq:h1}],
indicates the liquid--liquid transition line and its extension below the critical pressure, the Widom line.
The inset shows the stability limits in the vicinity of the critical point.}
\end{figure}

For the supercooled liquid state to be thermally and mechanically stable, both the
isochoric heat capacity $C_V$ and the isothermal compressibility $\kappa_T$ must be
positive. The scaling model predicts regions in the phase diagram where $C_V$ or
$\kappa_T$ is negative, as shown in \figref{fig:stability}. As noted by Bertrand and
Anisimov,\cite{bertrand2011} the stability locus is located close to the Widom line, but
its exact location depends on the parameters of the model. Our model predicts that the
thermal stability condition is violated before the mechanical stability condition as
water is cooled down, contrary to the violation order shown by a cubic equation of state
that describes the vapor--liquid transition.\cite{tmbook,oconnellbook} At pressures
slightly above the critical pressure, the mechanical stability limit coincides with the
LLT line; that is, one of the liquid phases (the low-density liquid) is unstable. We note
that a spinodal for the LLT is not defined in the scaled equation of
state.\cite{fisher1999}

The isochoric heat capacity $C_V$ is related to the isobaric heat capacity
by\cite{deben03}
\begin{equation}
    C_V = C_P - \frac{T\alpha_V^2}{\rho\kappa_T}.
\end{equation}
The critical part of $C_V$ expressed through the scaling susceptibilities is derived in
Appendix~\ref{app:CV}. As the compressibility $\kappa_T$ becomes smaller with decreasing
temperature, $C_V$ first becomes negative and then diverges to negative infinity as
$\kappa_T$ approaches zero. At temperatures below the mechanical stability limit, where
$\kappa_T$ is negative, $C_V$ is larger than $C_P$ (\figref{fig:Cp150MPa}). Another
property that diverges is the speed of sound $c$, which is related to $C_P$, $C_V$ and
$\kappa_T$ according to\cite{deben03}
\begin{equation}
    c = (\rho\kappa_T C_V/C_P)^{-1/2}.
\end{equation}
The speed of sound diverges to positive infinity (\figref{fig:IAPWSspeedofsound}) when
$C_V$ approaches zero (at the thermal stability limit), which occurs at a slightly higher
temperature than the zero-compressibility limit. In the unstable region, the speed of
sound is imaginary.

Kiselev\cite{kiselev2001} has computed a kinetic spinodal of supercooled water and found
that this stability limit was located a few kelvin below the homogeneous nucleation
limit. Kiselev calls the area in the phase diagram below the stability limit a
``nonthermodynamic habitat,'' not because liquid water is necessarily mechanically
unstable there, but because the lifetime of liquid water is smaller than the time needed
for equilibration. Using molecular simulations, Moore and Molinero\cite{moore2011} have
also found a stability limit just below the homogeneous nucleation temperature. They show
that below this limit, the time for crystallization is shorter than the relaxation time
of the liquid, so liquid water cannot not be equilibrated.

\begin{figure}
\includegraphics{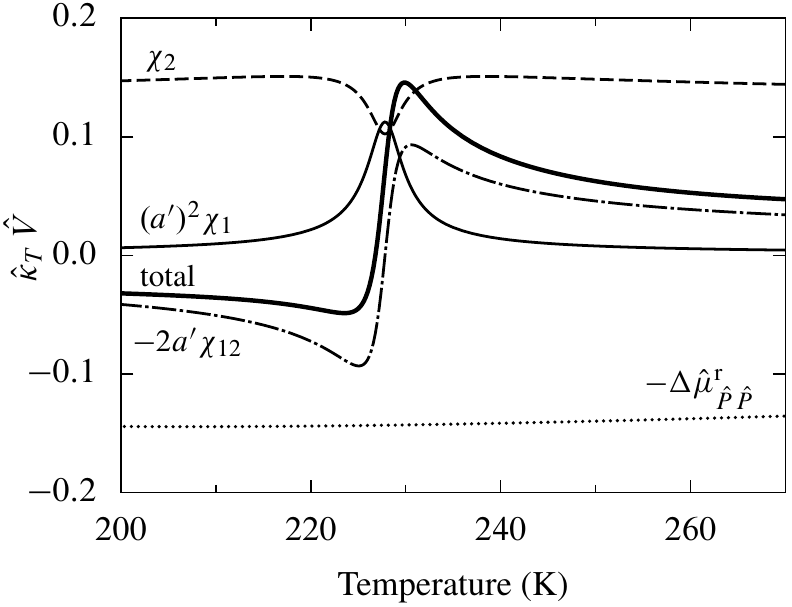}
\caption{\label{fig:susceptibilities}
Contribution of the susceptibilities, $\chi_1$, $\chi_2$, and $\chi_{12}$,
and the regular part to the compressibility,
as given by \eqref{eq:compressibility}, at 10~MPa.
The thick line represents the total compressibility multiplied by the volume, $\kap\Vh$.
}
\end{figure}

An important question arises: What terms in our model are responsible for the instability
of the liquid state? \ffigref{fig:susceptibilities} shows the contributions of the
individual susceptibilities to the compressibility, as given by
\eqref{eq:compressibility}. The strong and weak susceptibilities $\chi_1$ and $\chi_2$
are positive in the entire temperature range, whereas the cross susceptibility
$\chi_{12}$ changes sign on the Widom line. The sum of the contributions of the three
susceptibilities is positive in the entire range; it is the contribution of the regular
part, which is negative, that results in a negative compressibility below a certain
temperature.

\subsection{Confined water}
\begin{figure}
\includegraphics{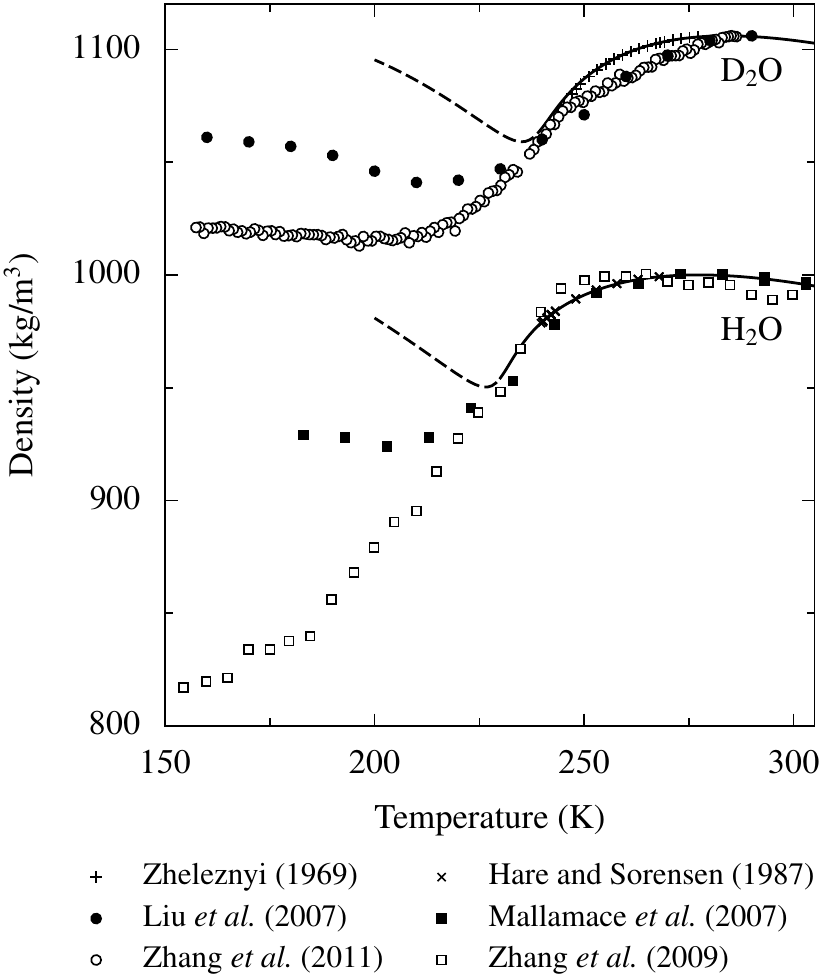}
\caption{\label{fig:densityconfined}Experimental densities of D$_2$O (open and filled circles,
both in hydrophilic confinement),\cite{liu07,zhang2011} and H$_2$O (open squares: hydrophobic confinement, filled
squares: hydrophilic confinement),\cite{mallamace07b,zhang2009}
together with bulk densities\cite{zhe69,hare87} (plus and cross symbols).
The predictions of the asymptotic scaling models for bulk D$_2$O and H$_2$O are shown by
solid curves, and the dashed curves represent the extrapolation in the unstable region.
}
\end{figure}

\begin{figure}
\includegraphics{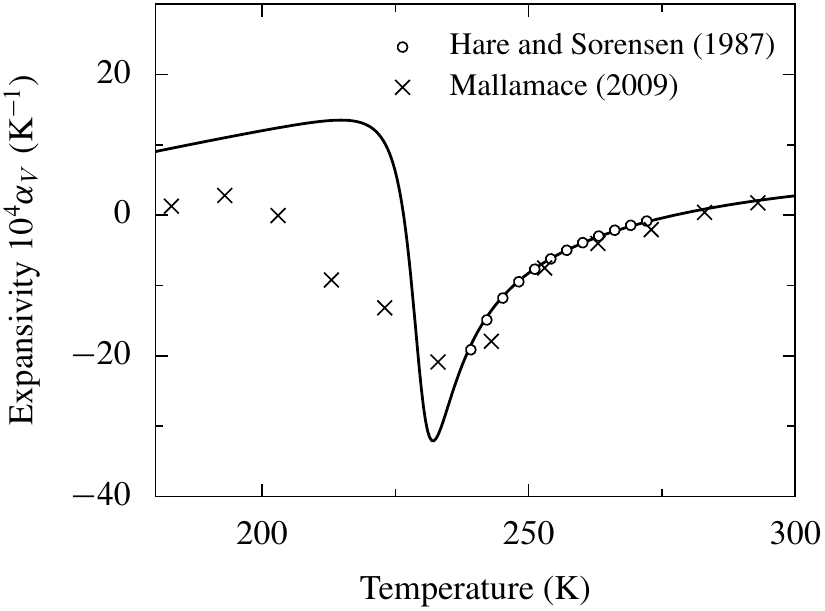}
\caption{\label{fig:expanconfined}Thermal expansion coefficient in confined water as measured
by Mallamace (crosses).\cite{mallamace2009} For comparison, the expansion coefficient of bulk water is also
shown (circles). The curve is the prediction of the asymptotic model.}
\end{figure}

\begin{figure}
\includegraphics{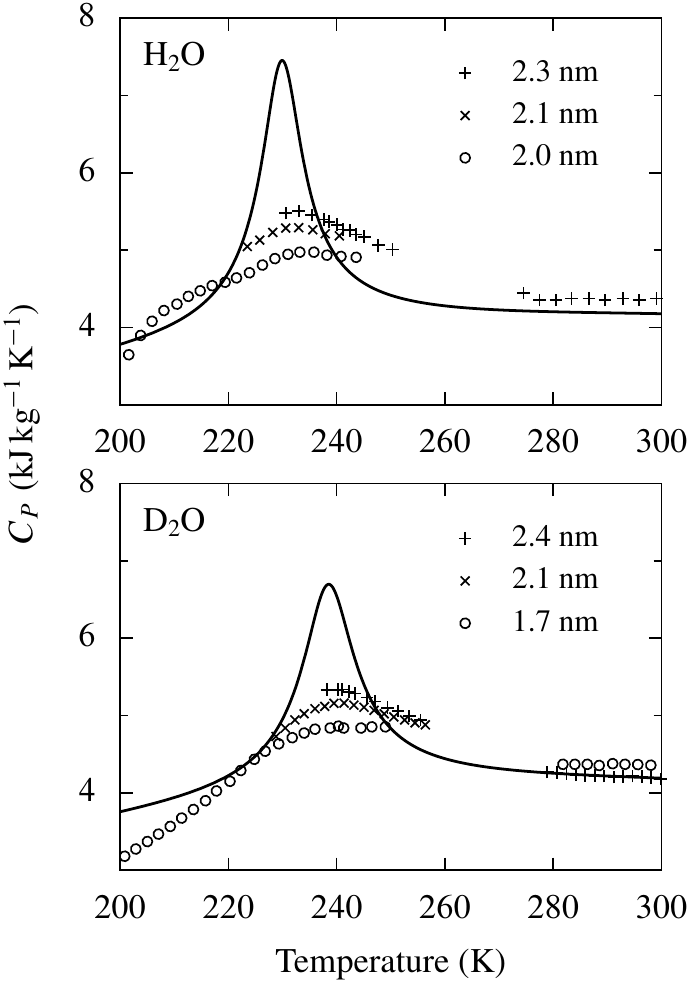}
\caption{\label{fig:cpconfined}Heat capacity of H$_2$O and D$_2$O in silica nanopores
as measured by Nagoe \ea\cite{nagoe2010} Different symbols correspond to different pore diameters.
The curves are the predictions of the asymptotic model.}
\end{figure}

New insights into the thermodynamics of supercooled water have been provided by
measurements of water confined in mesoporous structures. The silica material MCM-41,
which is comprised of a ``honeycomb'' of long cylindrical nanopores, has received
significant attention as a confining medium in part because the diameter of the pores can
be precisely controlled. As the pore diameter is decreased, the freezing point of water
is reduced. Spontaneous crystallization is completely suppressed below a critical pore
diameter, which allows for measurements to be made below the homogeneous-nucleation
temperature. Examples of such measurements for hydrated MCM-41 are presented for the
density in \figref{fig:densityconfined}, for the expansion coefficient in
\figref{fig:expanconfined}, and for the heat capacity in \figref{fig:cpconfined}, to be
compared with the prediction for bulk water.

Many of the results of experiments on confined water can be interpreted in the framework
of the second critical point hypothesis. Remarkably, the heat capacities exhibit maxima
upon crossing the Widom line predicted for bulk water. However, the height of the maximum
and the shape of the heat capacity in porous media differ significantly from those
predicted in bulk water (\figref{fig:cpconfined}). It has recently been reported that the
density of confined water shows significant hysteresis with temperature variation at
elevated pressures.\cite{zhang2011} This can be interpreted in terms of the crossing of
the first-order liquid-liquid coexistence curve. The exact connection between the bulk
thermodynamics of supercooled water and the results found in confinement is an active
area of research. Surface effects, both in the form of finite-size effects and
surface-water interactions, are thought to be contributing factors to the observed
differences. In computer simulations of water confined in MCM-41, investigators have
found that water found in the immediate vicinity of the surface behaves differently from
water in the center of the pore.\cite{gallo2010} This distinction raises the possibility
that wetting-type phenomena may be responsible for some of the observed behavior. That
the surface interactions play a significant role can be seen from the fact that the
measured density for hydrophobic and hydrophilic confinement are qualitatively
different.\cite{zhang2009} However, even in similar hydrophilic environments, the
measurements for D$_2$O show considerable difference (\figref{fig:densityconfined}). The
difference could be caused by the uncertainty in the background of the neutron
diffraction intensity, which must be subtracted from the measured intensity in
neutron-scattering experiments.\cite{zhang2011,zhang2009} In addition, the relative
densities that are obtained must be scaled to obtain absolute densities, which introduces
an additional uncertainty. The shape of the response functions, the expansivity
coefficient and the heat capacity, as shown in
\figsref{fig:expanconfined}{fig:cpconfined}, is more certain. The rounded anomalies of
the heat capacity around the Widom line in the confined mW water model are demonstrated
by recent simulations of Xu and Molinero.\cite{xu2011} An initial qualitative analysis of
the finite-size effects based on the LLCP hypothesis has been made by Bertrand and
Anisimov.\cite{bertrand2011} This work indicates that finite-size effects can account for
some of the differences between confined and bulk results.

\section{Discussion}\label{sec:discussion}
We have evaluated the IAPWS-95 formulation for ordinary water when it is extrapolated
into the supercooled region. At 0.1~MPa, it was found that IAPWS-95 reproduces the
experimental data of all properties except the isobaric heat capacity. With increasing
pressure, however, the deviations of the IAPWS-95 values from the data become larger and
larger. A property that is particularly poorly described is the expansion coefficient,
for which the IAPWS-95 extrapolation has a wrong sign in a large region of the phase
diagram. Even in the stable region close to the melting line, where IAPWS-95 should be
valid, the expansion coefficient predicted by IAPWS-95 is a factor of two lower than the
experimental value. The IAPWS equation for the surface tension performs well when
extrapolated down to about 266~K; below that, the experimentally observed increase of the
temperature dependence is not reproduced and the difference between the equation and the
data increases with decreasing temperature.

We have demonstrated that a theoretical model based on the assumption of a liquid--liquid
critical point in supercooled water can represent the thermodynamic properties of both
supercooled H$_2$O and D$_2$O to pressures of 150~MPa. Moreover, by allowing the slope of
the liquid--liquid transition (LLT) line and the critical pressure to be freely
adjustable parameters the model can represent almost all available thermodynamic property
data for supercooled water. Nevertheless, there are still a number of issues that need to
be considered.

A principal issue is that the existence of a liquid--liquid critical point is not the
only possible explanation for the anomalous behavior of the thermodynamic properties of
supercooled water. Scenarios for a singularity-free or critical-point-free interpretation
have also been proposed \cite{sastry1996,rebelo1998,angell2008}. Recently, Stokely
\ea\cite{stokely2010} have shown, for a water-like lattice model, how all such scenarios
can be described by varying two quantities, the strength of the hydrogen bonds and the
cooperativity of the hydrogen bonds. An intriguing possibility of the existence of
multiple critical points in supercooled water, as predicted by some
simulations,\cite{henriques2005,brovchenko2005} is discussed by Brovchenko and
Oleinikova.\cite{brovchenkobook} Another possibility is that response functions like the
compressibility do not diverge at a single temperature corresponding to a critical
temperature but at a range of pressure-dependent temperatures $T_\text{s}(P)$
corresponding to a crystallization spinodal, the absolute limit of stability of the
liquid phase. Most recently, the discussion on the nature of the anomalies observed in
supercooled water received an additional impetus after Limmer and Chandler reported new
simulation results~\cite{limmer2011} for two atomistic models of water,
mW~\cite{molinero2009} and mST2~\cite{liu2009}. They did not find two liquid states in
the supercooled region and excluded the possibility of a critical point for the models
studied. However, the most recent simulations by Sciortino \ea\cite{sciortino2011} of the
original ST2 model confirm the existence of a liquid--liquid critical point for that
particular model, with two distinct liquid states. They found no evidence of
crystallization during their simulation time, while there was enough time for the liquid
to equilibrate. It would be important to compare the anomalies obtained for various
models with those exhibited by real water.

Recent simulations of Moore and Molinero\cite{moore2011} indicate that for simulation
cells larger than the critical ice nucleus, spontaneous crystallization occurs before
liquid--liquid separation can equilibrate. An approach to consider a coupling between
spontaneous crystallization near the absolute stability limit of the liquid state and
liquid--liquid separation is the theory of weak crystallization pioneered by
Brazovski\u{\i},\cite{brazovskii1975} further developed by Brazovski\u{\i}
\ea,\cite{brazovskii1987} and reviewed by Kats \ea\cite{kats1993} According to this
theory, the fluctuations of the translational (short-wavelength) order parameter $\psi$
renormalize the mean-field distance $\Delta_0 = (T - T_0)/T$ between the temperature $T$
and the mean-field absolute stability limit $T_0$ of the liquid phase:
\begin{equation}\label{eq:gap}
    \Delta = \Delta_0 + \kappa\Delta^{-1/2},
\end{equation}
where $\Delta = (T - T_\text{s})/T$ where $T_\text{s}$ is the fluctuation-affected
stability-limit temperature, and $\kappa$ is a molecular parameter, similar to the
Ginzburg number that defines the validity of the mean-field approximation in the theory
of critical phenomena.\cite{anisimov1992,anisimovbook} The theory of weak crystallization
requires $\kappa\Delta^{-1/2} \ll \Delta_0$ and $\Delta_0 \ll 1$. There are two
non-trivial consequences of the effects of translational-order fluctuations: (1) The heat
capacity would contain a fluctuation correction $\delta C_P \sim \Delta^{-3/2}$; (2) The
renormalized gap $\Delta$ may be positive even at negative $\Delta_0$, meaning that
translational-order fluctuations stabilize the liquid phase.

The theory of weak crystallization was used to describe a coupling between
one-dimensional density modulation and orientational fluctuations in liquid crystals; it
was supported by accurate light-scattering and heat-capacity
measurements.\cite{gohin1983,anisimov1987a,*anisimov1987b,anisimovbook}

Similarly, we could consider a coupling between the translational-order fluctuations and
the critical fluctuations of the liquid--liquid order parameter $\phi_1$. A relationship
between the structural transformation in liquid water and its crystallization rate is
suggested by recent simulations by Moore and Molinero.\cite{moore2011} The lowest-order
term associated with such a coupling in the free-energy Landau expansion would be
$\sim\psi^2\phi_1$. Such a coupling would be possible even if the virtual critical point
is below the stability limit of the liquid phase.

Formation of ice is not ``weak crystallization.'' This is why application of weak
crystallization theory to supercooled water is questionable. However, it may not be
hopeless. The gap $\Delta$ between the melting temperature $T_\text{m}$ and the
temperature of spontaneous crystallization $T_\text{H}$ is of the order of 0.1. Hence, a
further investigation of the possibility of such coupling would be worthwhile. The actual
question is how the experimental observations can be explained by a theory that accounts
for both crystallization and liquid--liquid separation. The final conclusion concerning
the existence of the liquid--liquid critical point in water should be based on the
ability to quantitatively describe the experimental data.

Assuming the existence of the LLCP in supercooled water, we confirm the finding of
Fuentevilla and Anisimov\cite{fuentevilla2006} and Bertrand and
Anisimov\cite{bertrand2011} that the critical point is located at much lower pressure
than predicted by most simulations, definitely below 100~MPa, while the precise value of
the critical pressure is uncertain. However, regardless of all open questions, we have
shown that a critical-point parametric equation of state describes the available
thermodynamic data for supercooled water within experimental accuracy, thus establishing
a benchmark for any further developments in this area.

\begin{acknowledgments}
We acknowledge collaboration with D. Fuentevilla and J.~Kalová at an early stage of this
research. We thank R.~Torre for sending us the experimental data on the sound velocity,
C.~A. Angell, A.~H. Harvey, S. Sastry, and H.~E. Stanley for useful comments and
suggestions, and S.-H. Chen and Y. Zhang for discussing their results on confined
supercooled water. We also thank J.~P. O'Connell for clarifying the independence of the
mechanical and thermal instability criteria. The research of V.~Holten at the University
of Maryland has been supported by the Burgers Program of the University of Maryland (four
months). The research of C.~E.~Bertrand and current research of V.~Holten has been
supported by the Division of Chemistry of the US National Science Foundation under Grant
No. CHE-1012052. Travel support for V.~Holten was provided by the J.~M. Burgerscentrum in
the Netherlands.
\end{acknowledgments}

\appendix

\section{Linear-model parametric equations}\label{app:linmodel}
The formulas for the linear model in this appendix are taken from Behnejad \ea
\cite{behnejad2010} [In Behnejad \ea \cite{behnejad2010}, the formula corresponding to
our \eqref{eq:q2} erroneously does not contain the factor~$L_0$.]

The $r$ and $\theta$ coordinates are related to $h_1$ and $h_2$ by
\begin{align}
    h_1 &= a r^{\beta +\gamma} \theta(1-\theta^2),\\
    h_2 &= r (1-b^2 \theta^2),
\end{align}
with $\gamma = 2 - \alpha - 2\beta$ and
\begin{equation}
    b^2 = \frac{\gamma -2\beta}{\gamma(1-2\beta)}.
\end{equation}
The scaling densities are given by
\begin{align}
    \phi_1 &= k r^\beta \theta\\
    \phi_2 &= a k r^{1-\alpha} s(\theta),
\end{align}
with
\begin{align}
    s(\theta) &= L_0(s_0+s_2\theta^2),\\
    L_0 &= 1/[2 b^4 (1 - \alpha) \alpha],\\
    s_0 &= (\gamma - 2 \beta) - b^2 \alpha\gamma,\\
    s_2 &= (\alpha - 1) (\gamma - 2 \beta) b^2.
\end{align}
The scaling susceptibilities are given by
\begin{align}
    \chi_1 &= \frac{k}{a}r^{-\gamma}q_1(\theta),\\
    \chi_{12} &= k r^{\beta-1}q_{12}(\theta),\\
    \chi_2 &= a k r^{-\alpha}q_2(\theta),
\end{align}
with
\begin{align}
    q_1(\theta) &=&&(1 - b^2\theta^2 + 2\beta b^2 \theta^2)/q_0(\theta),\\
    q_{12}(\theta) &=&&\theta [-\gamma +(\gamma -2\beta )\theta ^2]/q_0(\theta),\\
    q_2(\theta) &=&&\bigl[(1-\alpha )(1-3\theta ^2) s(\theta) \notag\\
        &&& - (\beta +\gamma )2 s_2 L_0 \theta ^2(1-\theta ^2)\bigr] / q_0(\theta),\label{eq:q2}\\
    q_0(\theta) &=&&(1-3\theta ^2)(1-b^2\theta ^2)+2b^2(\beta +\gamma )\theta ^2(1-\theta ^2).
\end{align}

The weakly divergent scaling susceptibility $\chi_2$ also contains a negative
fluctuation-induced analytical background, which, as suggested by Bertrand and
Anisimov,\cite{bertrand2011} was incorporated in the regular part of the chemical
potential in our approach.

\section{Isochoric heat capacity in scaling models}\label{app:CV}
Equations (\ref{eq:h1})--(\ref{eq:h3}) can be generalized as
\begin{align}
    h_1 &= c\Dt + a\Dp, \label{eq:h1gen}\\
    h_2 &= -d\Dp + b\Dt, \label{eq:h2gen}\\
    h_3 &= \Dp - \Dm + \mr, \label{eq:h3gen}
\end{align}
The primes of the parameters $a'$ and $b'$ have been omitted for brevity. With $a = b =
0$ and $c = d = 1$ we recover the pure lattice-liquid model, whereas with $a = b = 1$ and
$c = d = 0$ we obtain the pure lattice-gas model. The case $c=0$ for the lattice-gas
model corresponds to $h_1 = \Dp = \Dm$, as discussed in Ref.~\onlinecite{bertrand2011}.
The expression for the isochoric heat capacity $\Cvh$,
\begin{equation}
    \Cvh = \Cph - \Vh\Th\frac{\alp^2}{\kap},
\end{equation}
leads to the critical part of $\Cvh$,
\begin{equation}
     \frac{\Cvh}{\Th} = A \frac{\chi _1 \chi _2 -\chi _{12}^2}{a^2 \chi _1-2 a d \chi _{12}+d^2 \chi _2},
\end{equation}
with $A = (a b+c d)^2$. For the pure lattice liquid, we obtain
\begin{equation}
    \frac{\Cvh}{\Th} = \chi _1-\frac{\chi _{12}^2}{\chi _2},
\end{equation}
and for the pure lattice gas,
\begin{equation}
    \frac{\Cvh}{\Th} = \chi _2-\frac{\chi _{12}^2}{\chi _1},
\end{equation}
as obtained by Behnejad \ea\cite{behnejad2010} In our model, the coefficient $a$ is small
but finite and $\Cvh$ is given by
\begin{equation}
    \frac{\Cvh}{\Th} = \frac{\chi _1 \chi _2 -\chi _{12}^2 }{a^2 \chi _1-2 a \chi _{12}+\chi _2}.
\end{equation}

\section{Adjustment of experimental data}\label{app:correction}
\begin{figure}
\includegraphics{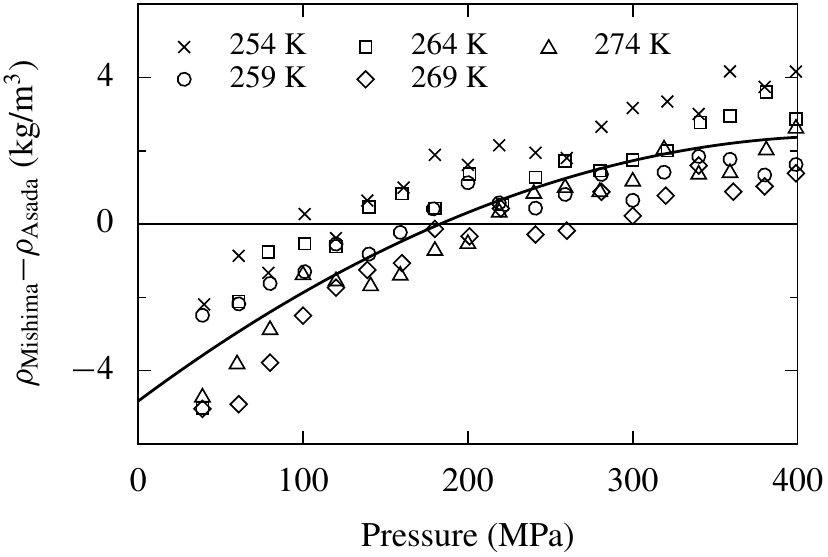}
\caption{\label{fig:mishimacorrection}Difference of Mishima's \cite{mishima2010} density
and the correlation of Asada \ea\cite{asada2002} The curve is a quadratic fit,
$\Delta\rho = -4.83 + 0.03348\, P - 3.87\times 10^{-5} P^2$, with $\Delta\rho$ in kg/m$^3$
and pressure $P$ in MPa.}
\end{figure}

The density data of Mishima show systematic differences of up to 5~kg/m$^3$ with the
densities reported by Asada \ea,\cite{asada2002} with IAPWS-95 (in the region where it is
reliable), and with the equation of state of Saul and Wagner.\cite{saulwagner89} The
differences appear to be mostly pressure dependent and only slightly temperature
dependent; see \figref{fig:mishimacorrection}. In the range from 273~K to 373~K and 0~MPa
to 380~MPa, where the three other data sets (Asada \ea, IAPWS-95, Saul and Wagner)
overlap and can be considered reliable, their maximum mutual difference is 0.8~kg/m$^3$.
Therefore, we decided that a correction of Mishima's data is justified. A
pressure-dependent density correction was determined by fitting a quadratic function of
the pressure to the difference of Mishima's densities and the correlation\footnote{There
is a sign error in the correlation of Asada \ea \cite{asada2002}: the coefficient $a_6$
in their Table~1 should have a positive sign.} of Asada \ea, between 245~K and 274~K
(\figref{fig:mishimacorrection}). This density correction was then subtracted from all
densities measured by Mishima.

Because of an incorrect graph, the heat-capacity data shown in the article of Bertolini
\ea\cite{bertolini1985} are about 4\% too low. The discrepancy was noted by Duan
\ea,\cite{duan2008} but they did not apply a correction. Tombari \ea\cite{tombari1999}
and Mishima\cite{mishima2010} corrected the data without comment. A quantitative
correction can be obtained because the graph of Bertolini \ea\ also contains a plot of an
equation of Leyendekkers and Hunter,\cite{ley85a} and the data in the graph can be scaled
until the curve matches the equation. The corrected heat capacity $C_P^\text{corr}$ may
be calculated from the original $C_P^\text{org}$ by $C_P^\text{corr} = 13.74 + 0.863\,
C_P^\text{org}$, with $C_P$ in J\,mol$^{-1}$K$^{-1}$. This correction is slightly smaller
than that of Tombari \ea\ and of Mishima.

\bibliography{vincentnew,supercooled}

\begin{thebibliography}{147}%
\makeatletter
\providecommand \@ifxundefined [1]{%
 \@ifx{#1\undefined}
}%
\providecommand \@ifnum [1]{%
 \ifnum #1\expandafter \@firstoftwo
 \else \expandafter \@secondoftwo
 \fi
}%
\providecommand \@ifx [1]{%
 \ifx #1\expandafter \@firstoftwo
 \else \expandafter \@secondoftwo
 \fi
}%
\providecommand \natexlab [1]{#1}%
\providecommand \enquote  [1]{``#1''}%
\providecommand \bibnamefont  [1]{#1}%
\providecommand \bibfnamefont [1]{#1}%
\providecommand \citenamefont [1]{#1}%
\providecommand \href@noop [0]{\@secondoftwo}%
\providecommand \href [0]{\begingroup \@sanitize@url \@href}%
\providecommand \@href[1]{\@@startlink{#1}\@@href}%
\providecommand \@@href[1]{\endgroup#1\@@endlink}%
\providecommand \@sanitize@url [0]{\catcode `\\12\catcode `\$12\catcode
  `\&12\catcode `\#12\catcode `\^12\catcode `\_12\catcode `\%12\relax}%
\providecommand \@@startlink[1]{}%
\providecommand \@@endlink[0]{}%
\providecommand \url  [0]{\begingroup\@sanitize@url \@url }%
\providecommand \@url [1]{\endgroup\@href {#1}{\urlprefix }}%
\providecommand \urlprefix  [0]{URL }%
\providecommand \Eprint [0]{\href }%
\providecommand \doibase [0]{http://dx.doi.org/}%
\providecommand \selectlanguage [0]{\@gobble}%
\providecommand \bibinfo  [0]{\@secondoftwo}%
\providecommand \bibfield  [0]{\@secondoftwo}%
\providecommand \translation [1]{[#1]}%
\providecommand \BibitemOpen [0]{}%
\providecommand \bibitemStop [0]{}%
\providecommand \bibitemNoStop [0]{.\EOS\space}%
\providecommand \EOS [0]{\spacefactor3000\relax}%
\providecommand \BibitemShut  [1]{\csname bibitem#1\endcsname}%
\let\auto@bib@innerbib\@empty
\bibitem [{\citenamefont {Debenedetti}(2003)}]{deben03}%
  \BibitemOpen
  \bibfield  {author} {\bibinfo {author} {\bibfnamefont {P.~G.}\ \bibnamefont
  {Debenedetti}},\ }\href@noop {} {\bibfield  {journal} {\bibinfo  {journal}
  {J.~Phys.: Condens. Matter}\ }\textbf {\bibinfo {volume} {15}},\ \bibinfo
  {pages} {R1669} (\bibinfo {year} {2003})}\BibitemShut {NoStop}%
\bibitem [{\citenamefont {Poole}\ \emph {et~al.}(1992)\citenamefont {Poole},
  \citenamefont {Sciortino}, \citenamefont {Essmann},\ and\ \citenamefont
  {Stanley}}]{poole1992}%
  \BibitemOpen
  \bibfield  {author} {\bibinfo {author} {\bibfnamefont {P.~H.}\ \bibnamefont
  {Poole}}, \bibinfo {author} {\bibfnamefont {F.}~\bibnamefont {Sciortino}},
  \bibinfo {author} {\bibfnamefont {U.}~\bibnamefont {Essmann}}, \ and\
  \bibinfo {author} {\bibfnamefont {H.~E.}\ \bibnamefont {Stanley}},\
  }\href@noop {} {\bibfield  {journal} {\bibinfo  {journal} {Nature (London)}\
  }\textbf {\bibinfo {volume} {360}},\ \bibinfo {pages} {324} (\bibinfo {year}
  {1992})}\BibitemShut {NoStop}%
\bibitem [{\citenamefont {Bertrand}\ and\ \citenamefont
  {Anisimov}(2011)}]{bertrand2011}%
  \BibitemOpen
  \bibfield  {author} {\bibinfo {author} {\bibfnamefont {C.~E.}\ \bibnamefont
  {Bertrand}}\ and\ \bibinfo {author} {\bibfnamefont {M.~A.}\ \bibnamefont
  {Anisimov}},\ }\href@noop {} {\bibfield  {journal} {\bibinfo  {journal}
  {J.~Phys. Chem.~B}\ }\textbf {\bibinfo {volume} {115}},\ \bibinfo {pages}
  {14099} (\bibinfo {year} {2011})}\BibitemShut {NoStop}%
\bibitem [{\citenamefont {Jeffery}\ and\ \citenamefont
  {Austin}(1999)}]{jeffery1999}%
  \BibitemOpen
  \bibfield  {author} {\bibinfo {author} {\bibfnamefont {C.~A.}\ \bibnamefont
  {Jeffery}}\ and\ \bibinfo {author} {\bibfnamefont {P.~H.}\ \bibnamefont
  {Austin}},\ }\href@noop {} {\bibfield  {journal} {\bibinfo  {journal}
  {J.~Chem. Phys.}\ }\textbf {\bibinfo {volume} {110}},\ \bibinfo {pages} {484}
  (\bibinfo {year} {1999})}\BibitemShut {NoStop}%
\bibitem [{\citenamefont {Kiselev}(2001)}]{kiselev2001}%
  \BibitemOpen
  \bibfield  {author} {\bibinfo {author} {\bibfnamefont {S.~B.}\ \bibnamefont
  {Kiselev}},\ }\href@noop {} {\bibfield  {journal} {\bibinfo  {journal} {Int.
  J. Thermophys.}\ }\textbf {\bibinfo {volume} {22}},\ \bibinfo {pages} {1421}
  (\bibinfo {year} {2001})}\BibitemShut {NoStop}%
\bibitem [{\citenamefont {Kiselev}\ and\ \citenamefont
  {Ely}(2002)}]{kiselev2002}%
  \BibitemOpen
  \bibfield  {author} {\bibinfo {author} {\bibfnamefont {S.~B.}\ \bibnamefont
  {Kiselev}}\ and\ \bibinfo {author} {\bibfnamefont {J.~F.}\ \bibnamefont
  {Ely}},\ }\href@noop {} {\bibfield  {journal} {\bibinfo  {journal} {J.~Chem.
  Phys.}\ }\textbf {\bibinfo {volume} {116}},\ \bibinfo {pages} {5657}
  (\bibinfo {year} {2002})}\BibitemShut {NoStop}%
\bibitem [{\citenamefont {Fuentevilla}\ and\ \citenamefont
  {Anisimov}(2006)}]{fuentevilla2006}%
  \BibitemOpen
  \bibfield  {author} {\bibinfo {author} {\bibfnamefont {D.~A.}\ \bibnamefont
  {Fuentevilla}}\ and\ \bibinfo {author} {\bibfnamefont {M.~A.}\ \bibnamefont
  {Anisimov}},\ }\href@noop {} {\bibfield  {journal} {\bibinfo  {journal}
  {Phys. Rev. Lett.}\ }\textbf {\bibinfo {volume} {97}},\ \bibinfo {pages}
  {195702} (\bibinfo {year} {2006})},\ \bibinfo {note} {erratum \textit{ibid.}
  \textbf{98}, 149904 (2007)}\BibitemShut {NoStop}%
\bibitem [{\citenamefont {Kalová}\ and\ \citenamefont
  {Mareš}(2010)}]{kalova2010}%
  \BibitemOpen
  \bibfield  {author} {\bibinfo {author} {\bibfnamefont {J.}~\bibnamefont
  {Kalová}}\ and\ \bibinfo {author} {\bibfnamefont {R.}~\bibnamefont {Mareš}},\
  }\href@noop {} {\bibfield  {journal} {\bibinfo  {journal} {Int. J.
  Thermophys.}\ }\textbf {\bibinfo {volume} {31}},\ \bibinfo {pages} {756}
  (\bibinfo {year} {2010})}\BibitemShut {NoStop}%
\bibitem [{\citenamefont {Limmer}\ and\ \citenamefont
  {Chandler}(2011)}]{limmer2011}%
  \BibitemOpen
  \bibfield  {author} {\bibinfo {author} {\bibfnamefont {D.~T.}\ \bibnamefont
  {Limmer}}\ and\ \bibinfo {author} {\bibfnamefont {D.}~\bibnamefont
  {Chandler}},\ }\href@noop {} {\bibfield  {journal} {\bibinfo  {journal}
  {J.~Chem. Phys.}\ }\textbf {\bibinfo {volume} {135}},\ \bibinfo {pages}
  {134503} (\bibinfo {year} {2011})}\BibitemShut {NoStop}%
\bibitem [{\citenamefont {Wikfeldt}, \citenamefont {Nilsson},\ and\
  \citenamefont {Pettersson}(2011)}]{wikfeldt2011}%
  \BibitemOpen
  \bibfield  {author} {\bibinfo {author} {\bibfnamefont {K.~T.}\ \bibnamefont
  {Wikfeldt}}, \bibinfo {author} {\bibfnamefont {A.}~\bibnamefont {Nilsson}}, \
  and\ \bibinfo {author} {\bibfnamefont {L.~G.~M.}\ \bibnamefont
  {Pettersson}},\ }\href@noop {} {\bibfield  {journal} {\bibinfo  {journal}
  {Phys. Chem. Chem. Phys.}\ }\textbf {\bibinfo {volume} {13}},\ \bibinfo
  {pages} {19918} (\bibinfo {year} {2011})}\BibitemShut {NoStop}%
\bibitem [{iap(2009)}]{iapws95}%
  \BibitemOpen
  \href {http://www.iapws.org/relguide/IAPWS95-Rev.pdf} {\emph {\bibinfo
  {title} {Revised Release on the IAPWS Formulation 1995 for the Thermodynamic
  Properties of Ordinary Water Substance for General and Scientific Use}}},\
  \bibinfo {organization} {International Association for the Properties of
  Water and Steam} (\bibinfo {year} {2009}),\ \url
  {http://www.iapws.org/relguide/IAPWS95-Rev.pdf}\BibitemShut {NoStop}%
\bibitem [{\citenamefont {Wagner}\ and\ \citenamefont
  {Pruß}(2002)}]{wag02nonote}%
  \BibitemOpen
  \bibfield  {author} {\bibinfo {author} {\bibfnamefont {W.}~\bibnamefont
  {Wagner}}\ and\ \bibinfo {author} {\bibfnamefont {A.}~\bibnamefont {Pruß}},\
  }\href@noop {} {\bibfield  {journal} {\bibinfo  {journal} {J.~Phys. Chem.
  Ref. Data}\ }\textbf {\bibinfo {volume} {31}},\ \bibinfo {pages} {387}
  (\bibinfo {year} {2002})}\BibitemShut {NoStop}%
\bibitem [{\citenamefont {Angell}(1982)}]{angell1982book}%
  \BibitemOpen
  \bibfield  {author} {\bibinfo {author} {\bibfnamefont {C.~A.}\ \bibnamefont
  {Angell}},\ }in\ \href@noop {} {\emph {\bibinfo {booktitle} {Water and
  Aqueous Solutions at Subzero Temperatures}}},\ \bibinfo {series} {Water: A
  Comprehensive Treatise}, Vol.~\bibinfo {volume} {7},\ \bibinfo {editor}
  {edited by\ \bibinfo {editor} {\bibfnamefont {F.}~\bibnamefont {Franks}}}\
  (\bibinfo  {publisher} {Plenum},\ \bibinfo {address} {New York},\ \bibinfo
  {year} {1982})\ Chap.~\bibinfo {chapter} {1}, pp.\ \bibinfo {pages}
  {1--81}\BibitemShut {NoStop}%
\bibitem [{\citenamefont {Angell}(1983)}]{angell83}%
  \BibitemOpen
  \bibfield  {author} {\bibinfo {author} {\bibfnamefont {C.~A.}\ \bibnamefont
  {Angell}},\ }\href@noop {} {\bibfield  {journal} {\bibinfo  {journal} {Ann.
  Rev. Phys. Chem.}\ }\textbf {\bibinfo {volume} {34}},\ \bibinfo {pages} {593}
  (\bibinfo {year} {1983})}\BibitemShut {NoStop}%
\bibitem [{\citenamefont {Sato}\ \emph {et~al.}(1991)\citenamefont {Sato},
  \citenamefont {Watanabe}, \citenamefont {Levelt~Sengers}, \citenamefont
  {Gallagher}, \citenamefont {Hill}, \citenamefont {Straub},\ and\
  \citenamefont {Wagner}}]{sato1991}%
  \BibitemOpen
  \bibfield  {author} {\bibinfo {author} {\bibfnamefont {H.}~\bibnamefont
  {Sato}}, \bibinfo {author} {\bibfnamefont {K.}~\bibnamefont {Watanabe}},
  \bibinfo {author} {\bibfnamefont {J.~M.~H.}\ \bibnamefont {Levelt~Sengers}},
  \bibinfo {author} {\bibfnamefont {J.~S.}\ \bibnamefont {Gallagher}}, \bibinfo
  {author} {\bibfnamefont {P.~G.}\ \bibnamefont {Hill}}, \bibinfo {author}
  {\bibfnamefont {J.}~\bibnamefont {Straub}}, \ and\ \bibinfo {author}
  {\bibfnamefont {W.}~\bibnamefont {Wagner}},\ }\href@noop {} {\bibfield
  {journal} {\bibinfo  {journal} {J.~Phys. Chem. Ref. Data}\ }\textbf {\bibinfo
  {volume} {20}},\ \bibinfo {pages} {1023} (\bibinfo {year}
  {1991})}\BibitemShut {NoStop}%
\bibitem [{\citenamefont {Feistel}\ \emph {et~al.}(2008)\citenamefont
  {Feistel}, \citenamefont {Wright}, \citenamefont {Miyagawa}, \citenamefont
  {Harvey}, \citenamefont {Hruby}, \citenamefont {Jackett}, \citenamefont
  {McDougall},\ and\ \citenamefont {Wagner}}]{feistel2008}%
  \BibitemOpen
  \bibfield  {author} {\bibinfo {author} {\bibfnamefont {R.}~\bibnamefont
  {Feistel}}, \bibinfo {author} {\bibfnamefont {D.~G.}\ \bibnamefont {Wright}},
  \bibinfo {author} {\bibfnamefont {K.}~\bibnamefont {Miyagawa}}, \bibinfo
  {author} {\bibfnamefont {A.~H.}\ \bibnamefont {Harvey}}, \bibinfo {author}
  {\bibfnamefont {J.}~\bibnamefont {Hruby}}, \bibinfo {author} {\bibfnamefont
  {D.~R.}\ \bibnamefont {Jackett}}, \bibinfo {author} {\bibfnamefont {T.~J.}\
  \bibnamefont {McDougall}}, \ and\ \bibinfo {author} {\bibfnamefont
  {W.}~\bibnamefont {Wagner}},\ }\href@noop {} {\bibfield  {journal} {\bibinfo
  {journal} {Ocean Sci.}\ }\textbf {\bibinfo {volume} {4}},\ \bibinfo {pages}
  {275} (\bibinfo {year} {2008})}\BibitemShut {NoStop}%
\bibitem [{SCW()}]{SCWsupplement}%
  \BibitemOpen
  \href@noop {} {}\bibinfo {note} {See supplementary material at [URL will be
  inserted by AIP] for tables with property data for supercooled \chem{H_2O}
  and \chem{D_2O}.}\BibitemShut {Stop}%
\bibitem [{\citenamefont {Hare}\ and\ \citenamefont {Sorensen}(1987)}]{hare87}%
  \BibitemOpen
  \bibfield  {author} {\bibinfo {author} {\bibfnamefont {D.~E.}\ \bibnamefont
  {Hare}}\ and\ \bibinfo {author} {\bibfnamefont {C.~M.}\ \bibnamefont
  {Sorensen}},\ }\href@noop {} {\bibfield  {journal} {\bibinfo  {journal}
  {J.~Chem. Phys.}\ }\textbf {\bibinfo {volume} {87}},\ \bibinfo {pages} {4840}
  (\bibinfo {year} {1987})}\BibitemShut {NoStop}%
\bibitem [{\citenamefont {Sotani}\ \emph {et~al.}(2000)\citenamefont {Sotani},
  \citenamefont {Arabas}, \citenamefont {Kubota},\ and\ \citenamefont
  {Kijima}}]{sotani2000}%
  \BibitemOpen
  \bibfield  {author} {\bibinfo {author} {\bibfnamefont {T.}~\bibnamefont
  {Sotani}}, \bibinfo {author} {\bibfnamefont {J.}~\bibnamefont {Arabas}},
  \bibinfo {author} {\bibfnamefont {H.}~\bibnamefont {Kubota}}, \ and\ \bibinfo
  {author} {\bibfnamefont {M.}~\bibnamefont {Kijima}},\ }\href@noop {}
  {\bibfield  {journal} {\bibinfo  {journal} {High Temp. High Pressures}\
  }\textbf {\bibinfo {volume} {32}},\ \bibinfo {pages} {433} (\bibinfo {year}
  {2000})}\BibitemShut {NoStop}%
\bibitem [{\citenamefont {Asada}\ \emph {et~al.}(2002)\citenamefont {Asada},
  \citenamefont {Sotani}, \citenamefont {Arabas}, \citenamefont {Kubota},
  \citenamefont {Matsuo},\ and\ \citenamefont {Tanaka}}]{asada2002}%
  \BibitemOpen
  \bibfield  {author} {\bibinfo {author} {\bibfnamefont {S.}~\bibnamefont
  {Asada}}, \bibinfo {author} {\bibfnamefont {T.}~\bibnamefont {Sotani}},
  \bibinfo {author} {\bibfnamefont {J.}~\bibnamefont {Arabas}}, \bibinfo
  {author} {\bibfnamefont {H.}~\bibnamefont {Kubota}}, \bibinfo {author}
  {\bibfnamefont {S.}~\bibnamefont {Matsuo}}, \ and\ \bibinfo {author}
  {\bibfnamefont {Y.}~\bibnamefont {Tanaka}},\ }\href@noop {} {\bibfield
  {journal} {\bibinfo  {journal} {J. Phys.: Condens. Matter}\ }\textbf
  {\bibinfo {volume} {14}},\ \bibinfo {pages} {11447} (\bibinfo {year}
  {2002})}\BibitemShut {NoStop}%
\bibitem [{\citenamefont {Mishima}(2010)}]{mishima2010}%
  \BibitemOpen
  \bibfield  {author} {\bibinfo {author} {\bibfnamefont {O.}~\bibnamefont
  {Mishima}},\ }\href@noop {} {\bibfield  {journal} {\bibinfo  {journal}
  {J.~Chem. Phys.}\ }\textbf {\bibinfo {volume} {133}},\ \bibinfo {pages}
  {144503} (\bibinfo {year} {2010})}\BibitemShut {NoStop}%
\bibitem [{\citenamefont {Guignon}, \citenamefont {Aparicio},\ and\
  \citenamefont {Sanz}(2010)}]{guignon2010}%
  \BibitemOpen
  \bibfield  {author} {\bibinfo {author} {\bibfnamefont {B.}~\bibnamefont
  {Guignon}}, \bibinfo {author} {\bibfnamefont {C.}~\bibnamefont {Aparicio}}, \
  and\ \bibinfo {author} {\bibfnamefont {P.~D.}\ \bibnamefont {Sanz}},\
  }\href@noop {} {\bibfield  {journal} {\bibinfo  {journal} {J.~Chem. Eng.
  Data}\ }\textbf {\bibinfo {volume} {55}},\ \bibinfo {pages} {3338} (\bibinfo
  {year} {2010})}\BibitemShut {NoStop}%
\bibitem [{iap(2011)}]{iapwsmeltsub2011}%
  \BibitemOpen
  \href {http://www.iapws.org/relguide/MeltSub2011.pdf} {\emph {\bibinfo
  {title} {Revised Release on the Pressure along the Melting and Sublimation
  Curves of Ordinary Water Substance}}},\ \bibinfo {organization} {IAPWS}
  (\bibinfo {year} {2011}),\ \url
  {http://www.iapws.org/relguide/MeltSub2011.pdf}\BibitemShut {NoStop}%
\bibitem [{\citenamefont {Wagner}\ \emph {et~al.}(2011)\citenamefont {Wagner},
  \citenamefont {Riethmann}, \citenamefont {Feistel},\ and\ \citenamefont
  {Harvey}}]{wagner2011}%
  \BibitemOpen
  \bibfield  {author} {\bibinfo {author} {\bibfnamefont {W.}~\bibnamefont
  {Wagner}}, \bibinfo {author} {\bibfnamefont {T.}~\bibnamefont {Riethmann}},
  \bibinfo {author} {\bibfnamefont {R.}~\bibnamefont {Feistel}}, \ and\
  \bibinfo {author} {\bibfnamefont {A.~H.}\ \bibnamefont {Harvey}},\
  }\href@noop {} {\bibfield  {journal} {\bibinfo  {journal} {J.~Phys. Chem.
  Ref. Data}\ }\textbf {\bibinfo {volume} {40}},\ \bibinfo {pages} {043103}
  (\bibinfo {year} {2011})}\BibitemShut {NoStop}%
\bibitem [{\citenamefont {Kanno}, \citenamefont {Speedy},\ and\ \citenamefont
  {Angell}(1975)}]{kanno1975}%
  \BibitemOpen
  \bibfield  {author} {\bibinfo {author} {\bibfnamefont {H.}~\bibnamefont
  {Kanno}}, \bibinfo {author} {\bibfnamefont {R.~J.}\ \bibnamefont {Speedy}}, \
  and\ \bibinfo {author} {\bibfnamefont {C.~A.}\ \bibnamefont {Angell}},\
  }\href@noop {} {\bibfield  {journal} {\bibinfo  {journal} {Science}\ }\textbf
  {\bibinfo {volume} {189}},\ \bibinfo {pages} {880} (\bibinfo {year}
  {1975})}\BibitemShut {NoStop}%
\bibitem [{\citenamefont {Kanno}\ and\ \citenamefont
  {Miyata}(2006)}]{kanno2006}%
  \BibitemOpen
  \bibfield  {author} {\bibinfo {author} {\bibfnamefont {H.}~\bibnamefont
  {Kanno}}\ and\ \bibinfo {author} {\bibfnamefont {K.}~\bibnamefont {Miyata}},\
  }\href@noop {} {\bibfield  {journal} {\bibinfo  {journal} {Chem. Phys.
  Lett.}\ }\textbf {\bibinfo {volume} {422}},\ \bibinfo {pages} {507} (\bibinfo
  {year} {2006})}\BibitemShut {NoStop}%
\bibitem [{\citenamefont {Ter~Minassian}, \citenamefont {Pruzan},\ and\
  \citenamefont {Soulard}(1981)}]{terminassian1981}%
  \BibitemOpen
  \bibfield  {author} {\bibinfo {author} {\bibfnamefont {L.}~\bibnamefont
  {Ter~Minassian}}, \bibinfo {author} {\bibfnamefont {P.}~\bibnamefont
  {Pruzan}}, \ and\ \bibinfo {author} {\bibfnamefont {A.}~\bibnamefont
  {Soulard}},\ }\href@noop {} {\bibfield  {journal} {\bibinfo  {journal}
  {J.~Chem. Phys.}\ }\textbf {\bibinfo {volume} {75}},\ \bibinfo {pages} {3064}
  (\bibinfo {year} {1981})}\BibitemShut {NoStop}%
\bibitem [{\citenamefont {Speedy}\ and\ \citenamefont
  {Angell}(1976)}]{speedy1976}%
  \BibitemOpen
  \bibfield  {author} {\bibinfo {author} {\bibfnamefont {R.~J.}\ \bibnamefont
  {Speedy}}\ and\ \bibinfo {author} {\bibfnamefont {C.~A.}\ \bibnamefont
  {Angell}},\ }\href@noop {} {\bibfield  {journal} {\bibinfo  {journal}
  {J.~Chem. Phys.}\ }\textbf {\bibinfo {volume} {65}},\ \bibinfo {pages} {851}
  (\bibinfo {year} {1976})}\BibitemShut {NoStop}%
\bibitem [{\citenamefont {Kanno}\ and\ \citenamefont
  {Angell}(1979)}]{kanno1979}%
  \BibitemOpen
  \bibfield  {author} {\bibinfo {author} {\bibfnamefont {H.}~\bibnamefont
  {Kanno}}\ and\ \bibinfo {author} {\bibfnamefont {C.~A.}\ \bibnamefont
  {Angell}},\ }\href@noop {} {\bibfield  {journal} {\bibinfo  {journal}
  {J.~Chem. Phys.}\ }\textbf {\bibinfo {volume} {70}},\ \bibinfo {pages} {4008}
  (\bibinfo {year} {1979})}\BibitemShut {NoStop}%
\bibitem [{\citenamefont {Hare}\ and\ \citenamefont {Sorensen}(1986)}]{hare86}%
  \BibitemOpen
  \bibfield  {author} {\bibinfo {author} {\bibfnamefont {D.~E.}\ \bibnamefont
  {Hare}}\ and\ \bibinfo {author} {\bibfnamefont {C.~M.}\ \bibnamefont
  {Sorensen}},\ }\href@noop {} {\bibfield  {journal} {\bibinfo  {journal}
  {J.~Chem. Phys.}\ }\textbf {\bibinfo {volume} {84}},\ \bibinfo {pages} {5085}
  (\bibinfo {year} {1986})}\BibitemShut {NoStop}%
\bibitem [{\citenamefont {Anisimov}\ \emph {et~al.}(1972)\citenamefont
  {Anisimov}, \citenamefont {Voronel'}, \citenamefont {Zaugol'nikova},\ and\
  \citenamefont {Ovodov}}]{anisimov1972}%
  \BibitemOpen
  \bibfield  {author} {\bibinfo {author} {\bibfnamefont {M.~A.}\ \bibnamefont
  {Anisimov}}, \bibinfo {author} {\bibfnamefont {A.~V.}\ \bibnamefont
  {Voronel'}}, \bibinfo {author} {\bibfnamefont {N.~S.}\ \bibnamefont
  {Zaugol'nikova}}, \ and\ \bibinfo {author} {\bibfnamefont {G.~I.}\
  \bibnamefont {Ovodov}},\ }\href@noop {} {\bibfield  {journal} {\bibinfo
  {journal} {JETP Lett.}\ }\textbf {\bibinfo {volume} {15}},\ \bibinfo {pages}
  {317} (\bibinfo {year} {1972})}\BibitemShut {NoStop}%
\bibitem [{\citenamefont {Angell}, \citenamefont {Shuppert},\ and\
  \citenamefont {Tucker}(1973)}]{angell1973}%
  \BibitemOpen
  \bibfield  {author} {\bibinfo {author} {\bibfnamefont {C.~A.}\ \bibnamefont
  {Angell}}, \bibinfo {author} {\bibfnamefont {J.}~\bibnamefont {Shuppert}}, \
  and\ \bibinfo {author} {\bibfnamefont {J.~C.}\ \bibnamefont {Tucker}},\
  }\href@noop {} {\bibfield  {journal} {\bibinfo  {journal} {J. Phys. Chem.}\
  }\textbf {\bibinfo {volume} {77}},\ \bibinfo {pages} {3092} (\bibinfo {year}
  {1973})}\BibitemShut {NoStop}%
\bibitem [{\citenamefont {Rasmussen}\ \emph {et~al.}(1973)\citenamefont
  {Rasmussen}, \citenamefont {MacKenzie}, \citenamefont {Angell},\ and\
  \citenamefont {Tucker}}]{rasmussen1973cp}%
  \BibitemOpen
  \bibfield  {author} {\bibinfo {author} {\bibfnamefont {D.~H.}\ \bibnamefont
  {Rasmussen}}, \bibinfo {author} {\bibfnamefont {A.~P.}\ \bibnamefont
  {MacKenzie}}, \bibinfo {author} {\bibfnamefont {C.~A.}\ \bibnamefont
  {Angell}}, \ and\ \bibinfo {author} {\bibfnamefont {J.~C.}\ \bibnamefont
  {Tucker}},\ }\href@noop {} {\bibfield  {journal} {\bibinfo  {journal}
  {Science}\ }\textbf {\bibinfo {volume} {181}},\ \bibinfo {pages} {342}
  (\bibinfo {year} {1973})}\BibitemShut {NoStop}%
\bibitem [{\citenamefont {Angell}, \citenamefont {Oguni},\ and\ \citenamefont
  {Sichina}(1982)}]{angell1982}%
  \BibitemOpen
  \bibfield  {author} {\bibinfo {author} {\bibfnamefont {C.~A.}\ \bibnamefont
  {Angell}}, \bibinfo {author} {\bibfnamefont {M.}~\bibnamefont {Oguni}}, \
  and\ \bibinfo {author} {\bibfnamefont {W.~J.}\ \bibnamefont {Sichina}},\
  }\href@noop {} {\bibfield  {journal} {\bibinfo  {journal} {J. Phys. Chem.}\
  }\textbf {\bibinfo {volume} {86}},\ \bibinfo {pages} {998} (\bibinfo {year}
  {1982})}\BibitemShut {NoStop}%
\bibitem [{\citenamefont {Archer}\ and\ \citenamefont {Carter}(2000)}]{arc00}%
  \BibitemOpen
  \bibfield  {author} {\bibinfo {author} {\bibfnamefont {D.~G.}\ \bibnamefont
  {Archer}}\ and\ \bibinfo {author} {\bibfnamefont {R.~W.}\ \bibnamefont
  {Carter}},\ }\href@noop {} {\bibfield  {journal} {\bibinfo  {journal}
  {J.~Phys. Chem.~B}\ }\textbf {\bibinfo {volume} {104}},\ \bibinfo {pages}
  {8563} (\bibinfo {year} {2000})}\BibitemShut {NoStop}%
\bibitem [{\citenamefont {Tombari}, \citenamefont {Ferrari},\ and\
  \citenamefont {Salvetti}(1999)}]{tombari1999}%
  \BibitemOpen
  \bibfield  {author} {\bibinfo {author} {\bibfnamefont {E.}~\bibnamefont
  {Tombari}}, \bibinfo {author} {\bibfnamefont {C.}~\bibnamefont {Ferrari}}, \
  and\ \bibinfo {author} {\bibfnamefont {G.}~\bibnamefont {Salvetti}},\
  }\href@noop {} {\bibfield  {journal} {\bibinfo  {journal} {Chem. Phys.
  Lett.}\ }\textbf {\bibinfo {volume} {300}},\ \bibinfo {pages} {749} (\bibinfo
  {year} {1999})}\BibitemShut {NoStop}%
\bibitem [{\citenamefont {Bertolini}, \citenamefont {Cassettari},\ and\
  \citenamefont {Salvetti}(1985)}]{bertolini1985}%
  \BibitemOpen
  \bibfield  {author} {\bibinfo {author} {\bibfnamefont {D.}~\bibnamefont
  {Bertolini}}, \bibinfo {author} {\bibfnamefont {M.}~\bibnamefont
  {Cassettari}}, \ and\ \bibinfo {author} {\bibfnamefont {G.}~\bibnamefont
  {Salvetti}},\ }\href@noop {} {\bibfield  {journal} {\bibinfo  {journal}
  {Chem. Phys. Lett.}\ }\textbf {\bibinfo {volume} {199}},\ \bibinfo {pages}
  {553} (\bibinfo {year} {1985})}\BibitemShut {NoStop}%
\bibitem [{\citenamefont {Sirota}, \citenamefont {Grishkov},\ and\
  \citenamefont {Tomishko}(1970)}]{sirota1970}%
  \BibitemOpen
  \bibfield  {author} {\bibinfo {author} {\bibfnamefont {A.~M.}\ \bibnamefont
  {Sirota}}, \bibinfo {author} {\bibfnamefont {A.~J.}\ \bibnamefont
  {Grishkov}}, \ and\ \bibinfo {author} {\bibfnamefont {A.~G.}\ \bibnamefont
  {Tomishko}},\ }\href@noop {} {\bibfield  {journal} {\bibinfo  {journal}
  {Thermal Eng.}\ }\textbf {\bibinfo {volume} {17}},\ \bibinfo {pages} {90}
  (\bibinfo {year} {1970})}\BibitemShut {NoStop}%
\bibitem [{\citenamefont {Manyà}\ \emph {et~al.}(2011)\citenamefont {Manyà},
  \citenamefont {Antal}, \citenamefont {Kinoshita},\ and\ \citenamefont
  {Masutani}}]{manya2011}%
  \BibitemOpen
  \bibfield  {author} {\bibinfo {author} {\bibfnamefont {J.~J.}\ \bibnamefont
  {Manyà}}, \bibinfo {author} {\bibfnamefont {M.~J.}\ \bibnamefont {Antal},
  \bibfnamefont {Jr.}}, \bibinfo {author} {\bibfnamefont {C.~K.}\ \bibnamefont
  {Kinoshita}}, \ and\ \bibinfo {author} {\bibfnamefont {S.~M.}\ \bibnamefont
  {Masutani}},\ }\href@noop {} {\bibfield  {journal} {\bibinfo  {journal} {Ind.
  Eng. Chem. Res.}\ }\textbf {\bibinfo {volume} {50}},\ \bibinfo {pages} {6470}
  (\bibinfo {year} {2011})}\BibitemShut {NoStop}%
\bibitem [{\citenamefont {Oguni}\ and\ \citenamefont
  {Angell}(1983)}]{oguni1983}%
  \BibitemOpen
  \bibfield  {author} {\bibinfo {author} {\bibfnamefont {M.}~\bibnamefont
  {Oguni}}\ and\ \bibinfo {author} {\bibfnamefont {C.~A.}\ \bibnamefont
  {Angell}},\ }\href@noop {} {\bibfield  {journal} {\bibinfo  {journal}
  {J.~Chem. Phys.}\ }\textbf {\bibinfo {volume} {78}},\ \bibinfo {pages} {7334}
  (\bibinfo {year} {1983})}\BibitemShut {NoStop}%
\bibitem [{\citenamefont {Hacker}(1951)}]{hac51}%
  \BibitemOpen
  \bibfield  {author} {\bibinfo {author} {\bibfnamefont {P.~T.}\ \bibnamefont
  {Hacker}},\ }\href
  {http://naca.central.cranfield.ac.uk/reports/1951/naca-tn-2510.pdf} {\enquote
  {\bibinfo {title} {Experimental values of the surface tension of supercooled
  water},}\ }\bibinfo {type} {Technical Note}\ \bibinfo {number} {2510}\
  (\bibinfo  {institution} {National Advisory Committee for Aeronautics},\
  \bibinfo {year} {1951})\ \url
  {http://naca.central.cranfield.ac.uk/reports/1951/naca-tn-2510.pdf}\BibitemShut
  {NoStop}%
\bibitem [{\citenamefont {Floriano}\ and\ \citenamefont
  {Angell}(1990)}]{flo90}%
  \BibitemOpen
  \bibfield  {author} {\bibinfo {author} {\bibfnamefont {M.~A.}\ \bibnamefont
  {Floriano}}\ and\ \bibinfo {author} {\bibfnamefont {C.~A.}\ \bibnamefont
  {Angell}},\ }\href@noop {} {\bibfield  {journal} {\bibinfo  {journal} {J.
  Phys. Chem.}\ }\textbf {\bibinfo {volume} {94}},\ \bibinfo {pages} {4199}
  (\bibinfo {year} {1990})}\BibitemShut {NoStop}%
\bibitem [{\citenamefont {Trinh}\ and\ \citenamefont {Ohsaka}(1995)}]{tri95}%
  \BibitemOpen
  \bibfield  {author} {\bibinfo {author} {\bibfnamefont {E.~H.}\ \bibnamefont
  {Trinh}}\ and\ \bibinfo {author} {\bibfnamefont {K.}~\bibnamefont {Ohsaka}},\
  }\href@noop {} {\bibfield  {journal} {\bibinfo  {journal} {Int. J.
  Thermophys.}\ }\textbf {\bibinfo {volume} {16}},\ \bibinfo {pages} {545}
  (\bibinfo {year} {1995})}\BibitemShut {NoStop}%
\bibitem [{\citenamefont {Lü}\ and\ \citenamefont {Wei}(2006)}]{lu06}%
  \BibitemOpen
  \bibfield  {author} {\bibinfo {author} {\bibfnamefont {Y.~J.}\ \bibnamefont
  {Lü}}\ and\ \bibinfo {author} {\bibfnamefont {B.}~\bibnamefont {Wei}},\
  }\href@noop {} {\bibfield  {journal} {\bibinfo  {journal} {Appl. Phys.
  Lett.}\ }\textbf {\bibinfo {volume} {89}},\ \bibinfo {pages} {164106}
  (\bibinfo {year} {2006})}\BibitemShut {NoStop}%
\bibitem [{iap(1994)}]{iapwssurften94}%
  \BibitemOpen
  \href {http://www.iapws.org/relguide/surf.pdf} {\emph {\bibinfo {title}
  {IAPWS Release on Surface Tension of Ordinary Water Substance}}},\ \bibinfo
  {organization} {IAPWS} (\bibinfo {year} {1994}),\ \url
  {http://www.iapws.org/relguide/surf.pdf}\BibitemShut {NoStop}%
\bibitem [{\citenamefont {Vargaftik}, \citenamefont {Volkov},\ and\
  \citenamefont {Voljak}(1983)}]{var83}%
  \BibitemOpen
  \bibfield  {author} {\bibinfo {author} {\bibfnamefont {N.~B.}\ \bibnamefont
  {Vargaftik}}, \bibinfo {author} {\bibfnamefont {B.~N.}\ \bibnamefont
  {Volkov}}, \ and\ \bibinfo {author} {\bibfnamefont {L.~D.}\ \bibnamefont
  {Voljak}},\ }\href@noop {} {\bibfield  {journal} {\bibinfo  {journal}
  {J.~Phys. Chem. Ref. Data}\ }\textbf {\bibinfo {volume} {12}},\ \bibinfo
  {pages} {817} (\bibinfo {year} {1983})}\BibitemShut {NoStop}%
\bibitem [{\citenamefont {Massoudi}\ and\ \citenamefont {King}(1974)}]{mass74}%
  \BibitemOpen
  \bibfield  {author} {\bibinfo {author} {\bibfnamefont {R.}~\bibnamefont
  {Massoudi}}\ and\ \bibinfo {author} {\bibfnamefont {A.~D.}\ \bibnamefont
  {King}, \bibfnamefont {Jr.}},\ }\href@noop {} {\bibfield  {journal} {\bibinfo
   {journal} {J. Phys. Chem.}\ }\textbf {\bibinfo {volume} {78}},\ \bibinfo
  {pages} {2262} (\bibinfo {year} {1974})}\BibitemShut {NoStop}%
\bibitem [{\citenamefont {Taschin}\ \emph {et~al.}(2011)\citenamefont
  {Taschin}, \citenamefont {Cucini}, \citenamefont {Bartolini},\ and\
  \citenamefont {Torre}}]{taschin2011}%
  \BibitemOpen
  \bibfield  {author} {\bibinfo {author} {\bibfnamefont {A.}~\bibnamefont
  {Taschin}}, \bibinfo {author} {\bibfnamefont {R.}~\bibnamefont {Cucini}},
  \bibinfo {author} {\bibfnamefont {P.}~\bibnamefont {Bartolini}}, \ and\
  \bibinfo {author} {\bibfnamefont {R.}~\bibnamefont {Torre}},\ }\href@noop {}
  {\bibfield  {journal} {\bibinfo  {journal} {Phil. Mag.}\ }\textbf {\bibinfo
  {volume} {91}},\ \bibinfo {pages} {1796} (\bibinfo {year}
  {2011})}\BibitemShut {NoStop}%
\bibitem [{\citenamefont {Magazú}\ \emph {et~al.}(1989)\citenamefont {Magazú},
  \citenamefont {Maisano}, \citenamefont {Majolino}, \citenamefont
  {Mallamace},\ and\ \citenamefont {Migliardo}}]{magazu1989}%
  \BibitemOpen
  \bibfield  {author} {\bibinfo {author} {\bibfnamefont {S.}~\bibnamefont
  {Magazú}}, \bibinfo {author} {\bibfnamefont {G.}~\bibnamefont {Maisano}},
  \bibinfo {author} {\bibfnamefont {D.}~\bibnamefont {Majolino}}, \bibinfo
  {author} {\bibfnamefont {F.}~\bibnamefont {Mallamace}}, \ and\ \bibinfo
  {author} {\bibfnamefont {P.}~\bibnamefont {Migliardo}},\ }\href@noop {}
  {\bibfield  {journal} {\bibinfo  {journal} {J. Phys. Chem.}\ }\textbf
  {\bibinfo {volume} {93}},\ \bibinfo {pages} {942} (\bibinfo {year}
  {1989})}\BibitemShut {NoStop}%
\bibitem [{\citenamefont {Santucci}\ \emph {et~al.}(2006)\citenamefont
  {Santucci}, \citenamefont {Fioretto}, \citenamefont {Comez}, \citenamefont
  {Gessin},\ and\ \citenamefont {Masciovecchio}}]{santucci2006}%
  \BibitemOpen
  \bibfield  {author} {\bibinfo {author} {\bibfnamefont {S.~C.}\ \bibnamefont
  {Santucci}}, \bibinfo {author} {\bibfnamefont {D.}~\bibnamefont {Fioretto}},
  \bibinfo {author} {\bibfnamefont {L.}~\bibnamefont {Comez}}, \bibinfo
  {author} {\bibfnamefont {A.}~\bibnamefont {Gessin}}, \ and\ \bibinfo {author}
  {\bibfnamefont {C.}~\bibnamefont {Masciovecchio}},\ }\href@noop {} {\bibfield
   {journal} {\bibinfo  {journal} {Phys. Rev. Lett.}\ }\textbf {\bibinfo
  {volume} {97}},\ \bibinfo {pages} {225701} (\bibinfo {year}
  {2006})}\BibitemShut {NoStop}%
\bibitem [{\citenamefont {Anisimov}, \citenamefont {Gorodetskii},\ and\
  \citenamefont {Kiyachenko}(1972)}]{anisimov1972thermcond}%
  \BibitemOpen
  \bibfield  {author} {\bibinfo {author} {\bibfnamefont {M.~A.}\ \bibnamefont
  {Anisimov}}, \bibinfo {author} {\bibfnamefont {E.~E.}\ \bibnamefont
  {Gorodetskii}}, \ and\ \bibinfo {author} {\bibfnamefont {{\relax Yu}.~F.}\
  \bibnamefont {Kiyachenko}},\ }\href@noop {} {\bibfield  {journal} {\bibinfo
  {journal} {Sov. Phys.--JETP}\ }\textbf {\bibinfo {volume} {35}},\ \bibinfo
  {pages} {1014} (\bibinfo {year} {1972})}\BibitemShut {NoStop}%
\bibitem [{\citenamefont {Benchikh}, \citenamefont {Fournier},\ and\
  \citenamefont {Boccara}(1985)}]{benchikh1985}%
  \BibitemOpen
  \bibfield  {author} {\bibinfo {author} {\bibfnamefont {O.}~\bibnamefont
  {Benchikh}}, \bibinfo {author} {\bibfnamefont {D.}~\bibnamefont {Fournier}},
  \ and\ \bibinfo {author} {\bibfnamefont {A.~C.}\ \bibnamefont {Boccara}},\
  }\href@noop {} {\bibfield  {journal} {\bibinfo  {journal} {J. Physique}\
  }\textbf {\bibinfo {volume} {46}},\ \bibinfo {pages} {727} (\bibinfo {year}
  {1985})}\BibitemShut {NoStop}%
\bibitem [{\citenamefont {Kumar}\ and\ \citenamefont
  {Stanley}(2011)}]{kumar2011}%
  \BibitemOpen
  \bibfield  {author} {\bibinfo {author} {\bibfnamefont {P.}~\bibnamefont
  {Kumar}}\ and\ \bibinfo {author} {\bibfnamefont {H.~E.}\ \bibnamefont
  {Stanley}},\ }\href@noop {} {\bibfield  {journal} {\bibinfo  {journal}
  {J.~Phys. Chem.~B}\ }\textbf {\bibinfo {volume} {115}},\ \bibinfo {pages}
  {14269} (\bibinfo {year} {2011})}\BibitemShut {NoStop}%
\bibitem [{\citenamefont {Trinh}\ and\ \citenamefont
  {Apfel}(1980)}]{trinhapfel1980}%
  \BibitemOpen
  \bibfield  {author} {\bibinfo {author} {\bibfnamefont {E.}~\bibnamefont
  {Trinh}}\ and\ \bibinfo {author} {\bibfnamefont {R.~E.}\ \bibnamefont
  {Apfel}},\ }\href@noop {} {\bibfield  {journal} {\bibinfo  {journal}
  {J.~Chem. Phys.}\ }\textbf {\bibinfo {volume} {72}},\ \bibinfo {pages} {6731}
  (\bibinfo {year} {1980})}\BibitemShut {NoStop}%
\bibitem [{\citenamefont {Bacri}\ and\ \citenamefont
  {Rajaonarison}(1979)}]{bacri1979}%
  \BibitemOpen
  \bibfield  {author} {\bibinfo {author} {\bibfnamefont {J.-C.}\ \bibnamefont
  {Bacri}}\ and\ \bibinfo {author} {\bibfnamefont {R.}~\bibnamefont
  {Rajaonarison}},\ }\href@noop {} {\bibfield  {journal} {\bibinfo  {journal}
  {J. Physique Lett.}\ }\textbf {\bibinfo {volume} {40}},\ \bibinfo {pages}
  {L403} (\bibinfo {year} {1979})}\BibitemShut {NoStop}%
\bibitem [{\citenamefont {Petitet}, \citenamefont {Tufeu},\ and\ \citenamefont
  {Le~Neindre}(1983)}]{petitet1983}%
  \BibitemOpen
  \bibfield  {author} {\bibinfo {author} {\bibfnamefont {J.~P.}\ \bibnamefont
  {Petitet}}, \bibinfo {author} {\bibfnamefont {R.}~\bibnamefont {Tufeu}}, \
  and\ \bibinfo {author} {\bibfnamefont {B.}~\bibnamefont {Le~Neindre}},\
  }\href@noop {} {\bibfield  {journal} {\bibinfo  {journal} {Int. J.
  Thermophys.}\ }\textbf {\bibinfo {volume} {4}},\ \bibinfo {pages} {35}
  (\bibinfo {year} {1983})}\BibitemShut {NoStop}%
\bibitem [{\citenamefont {Conde}, \citenamefont {Teixeira},\ and\ \citenamefont
  {Papon}(1982)}]{conde1982}%
  \BibitemOpen
  \bibfield  {author} {\bibinfo {author} {\bibfnamefont {O.}~\bibnamefont
  {Conde}}, \bibinfo {author} {\bibfnamefont {J.}~\bibnamefont {Teixeira}}, \
  and\ \bibinfo {author} {\bibfnamefont {P.}~\bibnamefont {Papon}},\
  }\href@noop {} {\bibfield  {journal} {\bibinfo  {journal} {J.~Chem. Phys.}\
  }\textbf {\bibinfo {volume} {76}},\ \bibinfo {pages} {3747} (\bibinfo {year}
  {1982})}\BibitemShut {NoStop}%
\bibitem [{\citenamefont {Trinh}\ and\ \citenamefont
  {Apfel}(1978{\natexlab{a}})}]{trinhapfel1978JASA}%
  \BibitemOpen
  \bibfield  {author} {\bibinfo {author} {\bibfnamefont {E.}~\bibnamefont
  {Trinh}}\ and\ \bibinfo {author} {\bibfnamefont {R.~E.}\ \bibnamefont
  {Apfel}},\ }\href@noop {} {\bibfield  {journal} {\bibinfo  {journal} {J.
  Acoust. Soc. Am.}\ }\textbf {\bibinfo {volume} {63}},\ \bibinfo {pages} {777}
  (\bibinfo {year} {1978}{\natexlab{a}})}\BibitemShut {NoStop}%
\bibitem [{\citenamefont {Trinh}\ and\ \citenamefont
  {Apfel}(1978{\natexlab{b}})}]{trinhapfel1978JCP}%
  \BibitemOpen
  \bibfield  {author} {\bibinfo {author} {\bibfnamefont {E.}~\bibnamefont
  {Trinh}}\ and\ \bibinfo {author} {\bibfnamefont {R.~E.}\ \bibnamefont
  {Apfel}},\ }\href@noop {} {\bibfield  {journal} {\bibinfo  {journal}
  {J.~Chem. Phys.}\ }\textbf {\bibinfo {volume} {69}},\ \bibinfo {pages} {4245}
  (\bibinfo {year} {1978}{\natexlab{b}})}\BibitemShut {NoStop}%
\bibitem [{\citenamefont {Speedy}(1982)}]{speedy1982a}%
  \BibitemOpen
  \bibfield  {author} {\bibinfo {author} {\bibfnamefont {R.~J.}\ \bibnamefont
  {Speedy}},\ }\href@noop {} {\bibfield  {journal} {\bibinfo  {journal} {J.
  Phys. Chem.}\ }\textbf {\bibinfo {volume} {86}},\ \bibinfo {pages} {982}
  (\bibinfo {year} {1982})}\BibitemShut {NoStop}%
\bibitem [{\citenamefont {Speedy}(2004)}]{speedy04comment}%
  \BibitemOpen
  \bibfield  {author} {\bibinfo {author} {\bibfnamefont {R.~J.}\ \bibnamefont
  {Speedy}},\ }\href@noop {} {\bibfield  {journal} {\bibinfo  {journal}
  {J.~Phys.: Condens. Matter}\ }\textbf {\bibinfo {volume} {16}},\ \bibinfo
  {pages} {6811} (\bibinfo {year} {2004})}\BibitemShut {NoStop}%
\bibitem [{\citenamefont {Debenedetti}(2004)}]{deben04reply}%
  \BibitemOpen
  \bibfield  {author} {\bibinfo {author} {\bibfnamefont {P.~G.}\ \bibnamefont
  {Debenedetti}},\ }\href@noop {} {\bibfield  {journal} {\bibinfo  {journal}
  {J.~Phys.: Condens. Matter}\ }\textbf {\bibinfo {volume} {16}},\ \bibinfo
  {pages} {6815} (\bibinfo {year} {2004})}\BibitemShut {NoStop}%
\bibitem [{\citenamefont {Mishima}\ and\ \citenamefont
  {Stanley}(1998{\natexlab{a}})}]{mishima1998}%
  \BibitemOpen
  \bibfield  {author} {\bibinfo {author} {\bibfnamefont {O.}~\bibnamefont
  {Mishima}}\ and\ \bibinfo {author} {\bibfnamefont {H.~E.}\ \bibnamefont
  {Stanley}},\ }\href@noop {} {\bibfield  {journal} {\bibinfo  {journal}
  {Nature}\ }\textbf {\bibinfo {volume} {392}},\ \bibinfo {pages} {164}
  (\bibinfo {year} {1998}{\natexlab{a}})}\BibitemShut {NoStop}%
\bibitem [{\citenamefont {Mishima}(2000)}]{mishima2000}%
  \BibitemOpen
  \bibfield  {author} {\bibinfo {author} {\bibfnamefont {O.}~\bibnamefont
  {Mishima}},\ }\href@noop {} {\bibfield  {journal} {\bibinfo  {journal} {Phys.
  Rev. Lett.}\ }\textbf {\bibinfo {volume} {85}},\ \bibinfo {pages} {334}
  (\bibinfo {year} {2000})}\BibitemShut {NoStop}%
\bibitem [{\citenamefont {Imre}\ and\ \citenamefont {Rzoska}(2010)}]{imre2010}%
  \BibitemOpen
  \bibfield  {author} {\bibinfo {author} {\bibfnamefont {A.~R.}\ \bibnamefont
  {Imre}}\ and\ \bibinfo {author} {\bibfnamefont {S.~J.}\ \bibnamefont
  {Rzoska}},\ }\href@noop {} {\bibfield  {journal} {\bibinfo  {journal} {Adv.
  Sci. Lett.}\ }\textbf {\bibinfo {volume} {3}},\ \bibinfo {pages} {527}
  (\bibinfo {year} {2010})}\BibitemShut {NoStop}%
\bibitem [{\citenamefont {Bridgman}(1935)}]{bridgman1935}%
  \BibitemOpen
  \bibfield  {author} {\bibinfo {author} {\bibfnamefont {P.~W.}\ \bibnamefont
  {Bridgman}},\ }\href@noop {} {\bibfield  {journal} {\bibinfo  {journal}
  {J.~Chem. Phys.}\ }\textbf {\bibinfo {volume} {3}},\ \bibinfo {pages} {597}
  (\bibinfo {year} {1935})}\BibitemShut {NoStop}%
\bibitem [{\citenamefont {Zheleznyi}(1969)}]{zhe69}%
  \BibitemOpen
  \bibfield  {author} {\bibinfo {author} {\bibfnamefont {B.~V.}\ \bibnamefont
  {Zheleznyi}},\ }\href@noop {} {\bibfield  {journal} {\bibinfo  {journal}
  {Russ. J. Phys. Chem.}\ }\textbf {\bibinfo {volume} {43}},\ \bibinfo {pages}
  {1311} (\bibinfo {year} {1969})}\BibitemShut {NoStop}%
\bibitem [{\citenamefont {Emmet}\ and\ \citenamefont
  {Millero}(1975)}]{emmet1975}%
  \BibitemOpen
  \bibfield  {author} {\bibinfo {author} {\bibfnamefont {R.~T.}\ \bibnamefont
  {Emmet}}\ and\ \bibinfo {author} {\bibfnamefont {F.~J.}\ \bibnamefont
  {Millero}},\ }\href@noop {} {\bibfield  {journal} {\bibinfo  {journal}
  {J.~Chem. Eng. Data}\ }\textbf {\bibinfo {volume} {20}},\ \bibinfo {pages}
  {351} (\bibinfo {year} {1975})}\BibitemShut {NoStop}%
\bibitem [{\citenamefont {Kanno}\ and\ \citenamefont
  {Angell}(1980)}]{kanno1980}%
  \BibitemOpen
  \bibfield  {author} {\bibinfo {author} {\bibfnamefont {H.}~\bibnamefont
  {Kanno}}\ and\ \bibinfo {author} {\bibfnamefont {C.~A.}\ \bibnamefont
  {Angell}},\ }\href@noop {} {\bibfield  {journal} {\bibinfo  {journal}
  {J.~Chem. Phys.}\ }\textbf {\bibinfo {volume} {73}},\ \bibinfo {pages} {1940}
  (\bibinfo {year} {1980})}\BibitemShut {NoStop}%
\bibitem [{\citenamefont {Angell}\ and\ \citenamefont
  {Kanno}(1976)}]{angell1976}%
  \BibitemOpen
  \bibfield  {author} {\bibinfo {author} {\bibfnamefont {C.~A.}\ \bibnamefont
  {Angell}}\ and\ \bibinfo {author} {\bibfnamefont {H.}~\bibnamefont {Kanno}},\
  }\href@noop {} {\bibfield  {journal} {\bibinfo  {journal} {Science}\ }\textbf
  {\bibinfo {volume} {193}},\ \bibinfo {pages} {1121} (\bibinfo {year}
  {1976})}\BibitemShut {NoStop}%
\bibitem [{\citenamefont {Rasmussen}\ and\ \citenamefont
  {MacKenzie}(1973)}]{rassmussen73}%
  \BibitemOpen
  \bibfield  {author} {\bibinfo {author} {\bibfnamefont {D.~H.}\ \bibnamefont
  {Rasmussen}}\ and\ \bibinfo {author} {\bibfnamefont {A.~P.}\ \bibnamefont
  {MacKenzie}},\ }\href@noop {} {\bibfield  {journal} {\bibinfo  {journal}
  {J.~Chem. Phys.}\ }\textbf {\bibinfo {volume} {59}},\ \bibinfo {pages} {5003}
  (\bibinfo {year} {1973})}\BibitemShut {NoStop}%
\bibitem [{\citenamefont {Fisher}(1974)}]{fisher1974}%
  \BibitemOpen
  \bibfield  {author} {\bibinfo {author} {\bibfnamefont {M.~E.}\ \bibnamefont
  {Fisher}},\ }\href@noop {} {\bibfield  {journal} {\bibinfo  {journal} {Rev.
  Mod. Phys.}\ }\textbf {\bibinfo {volume} {46}},\ \bibinfo {pages} {597}
  (\bibinfo {year} {1974})}\BibitemShut {NoStop}%
\bibitem [{\citenamefont {Kadanoff}(1976)}]{kadanoff1976}%
  \BibitemOpen
  \bibfield  {author} {\bibinfo {author} {\bibfnamefont {L.~P.}\ \bibnamefont
  {Kadanoff}},\ }in\ \href@noop {} {\emph {\bibinfo {booktitle} {Phase
  Transitions and Critical Phenomena}}},\ Vol.~\bibinfo {volume} {5a},\
  \bibinfo {editor} {edited by\ \bibinfo {editor} {\bibfnamefont
  {C.}~\bibnamefont {Domb}}\ and\ \bibinfo {editor} {\bibfnamefont {M.~S.}\
  \bibnamefont {Green}}}\ (\bibinfo  {publisher} {Academic Press},\ \bibinfo
  {address} {New York},\ \bibinfo {year} {1976})\ pp.\ \bibinfo {pages}
  {1--34}\BibitemShut {NoStop}%
\bibitem [{\citenamefont {Fisher}(1983)}]{fisher1983}%
  \BibitemOpen
  \bibfield  {author} {\bibinfo {author} {\bibfnamefont {M.~E.}\ \bibnamefont
  {Fisher}},\ }in\ \href@noop {} {\emph {\bibinfo {booktitle} {Critical
  Phenomena, Lecture Notes in Physics}}},\ Vol.\ \bibinfo {volume} {186},\
  \bibinfo {editor} {edited by\ \bibinfo {editor} {\bibfnamefont {F.~J.~W.}\
  \bibnamefont {Hahne}}}\ (\bibinfo  {publisher} {Springer},\ \bibinfo
  {address} {Berlin},\ \bibinfo {year} {1983})\ pp.\ \bibinfo {pages}
  {1--139}\BibitemShut {NoStop}%
\bibitem [{\citenamefont {Pelissetto}\ and\ \citenamefont
  {Vicari}(2002)}]{pelissetto2002}%
  \BibitemOpen
  \bibfield  {author} {\bibinfo {author} {\bibfnamefont {A.}~\bibnamefont
  {Pelissetto}}\ and\ \bibinfo {author} {\bibfnamefont {E.}~\bibnamefont
  {Vicari}},\ }\href@noop {} {\bibfield  {journal} {\bibinfo  {journal} {Phys.
  Rep.}\ }\textbf {\bibinfo {volume} {368}},\ \bibinfo {pages} {549} (\bibinfo
  {year} {2002})}\BibitemShut {NoStop}%
\bibitem [{\citenamefont {Sengers}\ and\ \citenamefont
  {Shanks}(2009)}]{sengers2009}%
  \BibitemOpen
  \bibfield  {author} {\bibinfo {author} {\bibfnamefont {J.~V.}\ \bibnamefont
  {Sengers}}\ and\ \bibinfo {author} {\bibfnamefont {J.~G.}\ \bibnamefont
  {Shanks}},\ }\href@noop {} {\bibfield  {journal} {\bibinfo  {journal}
  {J.~Stat. Phys.}\ }\textbf {\bibinfo {volume} {137}},\ \bibinfo {pages} {857}
  (\bibinfo {year} {2009})}\BibitemShut {NoStop}%
\bibitem [{\citenamefont {Rowlinson}\ and\ \citenamefont
  {Swinton}(1982)}]{rowlinson1982}%
  \BibitemOpen
  \bibfield  {author} {\bibinfo {author} {\bibfnamefont {J.~S.}\ \bibnamefont
  {Rowlinson}}\ and\ \bibinfo {author} {\bibfnamefont {F.~L.}\ \bibnamefont
  {Swinton}},\ }\href@noop {} {\emph {\bibinfo {title} {Liquids and Liquid
  Mixtures}}},\ \bibinfo {edition} {3rd}\ ed.\ (\bibinfo  {publisher}
  {Butterworth},\ \bibinfo {address} {London},\ \bibinfo {year}
  {1982})\BibitemShut {NoStop}%
\bibitem [{\citenamefont {Anisimov}\ \emph {et~al.}(1995)\citenamefont
  {Anisimov}, \citenamefont {Gorodetskii}, \citenamefont {Kulikov},\ and\
  \citenamefont {Sengers}}]{anisimov1995}%
  \BibitemOpen
  \bibfield  {author} {\bibinfo {author} {\bibfnamefont {M.~A.}\ \bibnamefont
  {Anisimov}}, \bibinfo {author} {\bibfnamefont {E.~E.}\ \bibnamefont
  {Gorodetskii}}, \bibinfo {author} {\bibfnamefont {V.~D.}\ \bibnamefont
  {Kulikov}}, \ and\ \bibinfo {author} {\bibfnamefont {J.~V.}\ \bibnamefont
  {Sengers}},\ }\href@noop {} {\bibfield  {journal} {\bibinfo  {journal} {Phys.
  Rev. E}\ }\textbf {\bibinfo {volume} {51}},\ \bibinfo {pages} {1199}
  (\bibinfo {year} {1995})}\BibitemShut {NoStop}%
\bibitem [{\citenamefont {Lee}\ and\ \citenamefont {Yang}(1952)}]{lee1952}%
  \BibitemOpen
  \bibfield  {author} {\bibinfo {author} {\bibfnamefont {T.~D.}\ \bibnamefont
  {Lee}}\ and\ \bibinfo {author} {\bibfnamefont {C.~N.}\ \bibnamefont {Yang}},\
  }\href@noop {} {\bibfield  {journal} {\bibinfo  {journal} {Phys. Rev.}\
  }\textbf {\bibinfo {volume} {87}},\ \bibinfo {pages} {410} (\bibinfo {year}
  {1952})}\BibitemShut {NoStop}%
\bibitem [{\citenamefont {Fisher}(1967)}]{fisher1967}%
  \BibitemOpen
  \bibfield  {author} {\bibinfo {author} {\bibfnamefont {M.~E.}\ \bibnamefont
  {Fisher}},\ }\href@noop {} {\bibfield  {journal} {\bibinfo  {journal} {Rep.
  Progr. Phys.}\ }\textbf {\bibinfo {volume} {30(II)}},\ \bibinfo {pages} {615}
  (\bibinfo {year} {1967})}\BibitemShut {NoStop}%
\bibitem [{\citenamefont {Sengers}\ and\ \citenamefont
  {Levelt~Sengers}(1978)}]{sengerscroxton}%
  \BibitemOpen
  \bibfield  {author} {\bibinfo {author} {\bibfnamefont {J.~V.}\ \bibnamefont
  {Sengers}}\ and\ \bibinfo {author} {\bibfnamefont {J.~M.~H.}\ \bibnamefont
  {Levelt~Sengers}},\ }in\ \href@noop {} {\emph {\bibinfo {booktitle} {Progress
  in Liquid Physics}}},\ \bibinfo {editor} {edited by\ \bibinfo {editor}
  {\bibfnamefont {C.~A.}\ \bibnamefont {Croxton}}}\ (\bibinfo  {publisher}
  {Wiley},\ \bibinfo {address} {New York},\ \bibinfo {year} {1978})\
  Chap.~\bibinfo {chapter} {4}, pp.\ \bibinfo {pages} {103--174}\BibitemShut
  {NoStop}%
\bibitem [{\citenamefont {Levelt~Sengers}\ \emph {et~al.}(1983)\citenamefont
  {Levelt~Sengers}, \citenamefont {Kamgar-Parsi}, \citenamefont {Balfour},\
  and\ \citenamefont {Sengers}}]{leveltsengers1983}%
  \BibitemOpen
  \bibfield  {author} {\bibinfo {author} {\bibfnamefont {J.~M.~H.}\
  \bibnamefont {Levelt~Sengers}}, \bibinfo {author} {\bibfnamefont
  {B.}~\bibnamefont {Kamgar-Parsi}}, \bibinfo {author} {\bibfnamefont {F.~W.}\
  \bibnamefont {Balfour}}, \ and\ \bibinfo {author} {\bibfnamefont {J.~V.}\
  \bibnamefont {Sengers}},\ }\href@noop {} {\bibfield  {journal} {\bibinfo
  {journal} {J.~Phys. Chem. Ref. Data}\ }\textbf {\bibinfo {volume} {12}},\
  \bibinfo {pages} {1} (\bibinfo {year} {1983})}\BibitemShut {NoStop}%
\bibitem [{\citenamefont {Behnejad}, \citenamefont {Sengers},\ and\
  \citenamefont {Anisimov}(2010)}]{behnejad2010}%
  \BibitemOpen
  \bibfield  {author} {\bibinfo {author} {\bibfnamefont {H.}~\bibnamefont
  {Behnejad}}, \bibinfo {author} {\bibfnamefont {J.~V.}\ \bibnamefont
  {Sengers}}, \ and\ \bibinfo {author} {\bibfnamefont {M.~A.}\ \bibnamefont
  {Anisimov}},\ }in\ \href@noop {} {\emph {\bibinfo {booktitle} {Applied
  Thermodynamics of Fluids}}},\ \bibinfo {editor} {edited by\ \bibinfo {editor}
  {\bibfnamefont {A.~R.~H.}\ \bibnamefont {Goodwin}}, \bibinfo {editor}
  {\bibfnamefont {J.~V.}\ \bibnamefont {Sengers}}, \ and\ \bibinfo {editor}
  {\bibfnamefont {C.~J.}\ \bibnamefont {Peters}}}\ (\bibinfo  {publisher} {RSC
  Publishing},\ \bibinfo {address} {Cambridge, UK},\ \bibinfo {year} {2010})\
  Chap.~\bibinfo {chapter} {10}, pp.\ \bibinfo {pages} {321--367}\BibitemShut
  {NoStop}%
\bibitem [{\citenamefont {Anisimov}\ \emph {et~al.}(1992)\citenamefont
  {Anisimov}, \citenamefont {Kiselev}, \citenamefont {Sengers},\ and\
  \citenamefont {Tang}}]{anisimov1992}%
  \BibitemOpen
  \bibfield  {author} {\bibinfo {author} {\bibfnamefont {M.~A.}\ \bibnamefont
  {Anisimov}}, \bibinfo {author} {\bibfnamefont {S.~B.}\ \bibnamefont
  {Kiselev}}, \bibinfo {author} {\bibfnamefont {J.~V.}\ \bibnamefont
  {Sengers}}, \ and\ \bibinfo {author} {\bibfnamefont {S.}~\bibnamefont
  {Tang}},\ }\href@noop {} {\bibfield  {journal} {\bibinfo  {journal} {Physica
  A}\ }\textbf {\bibinfo {volume} {188}},\ \bibinfo {pages} {487} (\bibinfo
  {year} {1992})}\BibitemShut {NoStop}%
\bibitem [{\citenamefont {Anisimov}, \citenamefont {Agayan},\ and\
  \citenamefont {Collings}(1998)}]{anisimov1998}%
  \BibitemOpen
  \bibfield  {author} {\bibinfo {author} {\bibfnamefont {M.~A.}\ \bibnamefont
  {Anisimov}}, \bibinfo {author} {\bibfnamefont {V.~A.}\ \bibnamefont
  {Agayan}}, \ and\ \bibinfo {author} {\bibfnamefont {P.~J.}\ \bibnamefont
  {Collings}},\ }\href@noop {} {\bibfield  {journal} {\bibinfo  {journal}
  {Phys. Rev. E}\ }\textbf {\bibinfo {volume} {57}},\ \bibinfo {pages} {582}
  (\bibinfo {year} {1998})}\BibitemShut {NoStop}%
\bibitem [{\citenamefont {Fisher}(1971)}]{fisher1971}%
  \BibitemOpen
  \bibfield  {author} {\bibinfo {author} {\bibfnamefont {M.~E.}\ \bibnamefont
  {Fisher}},\ }in\ \href@noop {} {\emph {\bibinfo {booktitle} {Proceedings of
  the International school of Physics ``Enrico Fermi''}}},\ \bibinfo {editor}
  {edited by\ \bibinfo {editor} {\bibfnamefont {M.~S.}\ \bibnamefont {Green}}}\
  (\bibinfo  {publisher} {Academic Press},\ \bibinfo {address} {New York},\
  \bibinfo {year} {1971})\ pp.\ \bibinfo {pages} {1--99}\BibitemShut {NoStop}%
\bibitem [{\citenamefont {Brézin}, \citenamefont {Wallace},\ and\ \citenamefont
  {Wilson}(1972)}]{brezin1972}%
  \BibitemOpen
  \bibfield  {author} {\bibinfo {author} {\bibfnamefont {E.}~\bibnamefont
  {Brézin}}, \bibinfo {author} {\bibfnamefont {D.~J.}\ \bibnamefont {Wallace}},
  \ and\ \bibinfo {author} {\bibfnamefont {K.~G.}\ \bibnamefont {Wilson}},\
  }\href@noop {} {\bibfield  {journal} {\bibinfo  {journal} {Phys. Rev. Lett.}\
  }\textbf {\bibinfo {volume} {29}},\ \bibinfo {pages} {591} (\bibinfo {year}
  {1972})}\BibitemShut {NoStop}%
\bibitem [{\citenamefont {Schofield}(1969)}]{schofield1969a}%
  \BibitemOpen
  \bibfield  {author} {\bibinfo {author} {\bibfnamefont {P.}~\bibnamefont
  {Schofield}},\ }\href@noop {} {\bibfield  {journal} {\bibinfo  {journal}
  {Phys. Rev. Lett.}\ }\textbf {\bibinfo {volume} {22}},\ \bibinfo {pages}
  {606} (\bibinfo {year} {1969})}\BibitemShut {NoStop}%
\bibitem [{\citenamefont {Schofield}, \citenamefont {Litster},\ and\
  \citenamefont {Ho}(1969)}]{schofield1969b}%
  \BibitemOpen
  \bibfield  {author} {\bibinfo {author} {\bibfnamefont {P.}~\bibnamefont
  {Schofield}}, \bibinfo {author} {\bibfnamefont {J.~D.}\ \bibnamefont
  {Litster}}, \ and\ \bibinfo {author} {\bibfnamefont {J.~T.}\ \bibnamefont
  {Ho}},\ }\href@noop {} {\bibfield  {journal} {\bibinfo  {journal} {Phys. Rev.
  Lett.}\ }\textbf {\bibinfo {volume} {23}},\ \bibinfo {pages} {1098} (\bibinfo
  {year} {1969})}\BibitemShut {NoStop}%
\bibitem [{\citenamefont {Hohenberg}\ and\ \citenamefont
  {Barmatz}(1972)}]{hohenberg1972}%
  \BibitemOpen
  \bibfield  {author} {\bibinfo {author} {\bibfnamefont {P.~C.}\ \bibnamefont
  {Hohenberg}}\ and\ \bibinfo {author} {\bibfnamefont {M.}~\bibnamefont
  {Barmatz}},\ }\href@noop {} {\bibfield  {journal} {\bibinfo  {journal} {Phys.
  Rev. A}\ }\textbf {\bibinfo {volume} {6}},\ \bibinfo {pages} {289} (\bibinfo
  {year} {1972})}\BibitemShut {NoStop}%
\bibitem [{\citenamefont {Moldover}\ \emph {et~al.}(1979)\citenamefont
  {Moldover}, \citenamefont {Sengers}, \citenamefont {Gammon},\ and\
  \citenamefont {Hocken}}]{moldover1979}%
  \BibitemOpen
  \bibfield  {author} {\bibinfo {author} {\bibfnamefont {M.~R.}\ \bibnamefont
  {Moldover}}, \bibinfo {author} {\bibfnamefont {J.~V.}\ \bibnamefont
  {Sengers}}, \bibinfo {author} {\bibfnamefont {R.~W.}\ \bibnamefont {Gammon}},
  \ and\ \bibinfo {author} {\bibfnamefont {R.~J.}\ \bibnamefont {Hocken}},\
  }\href@noop {} {\bibfield  {journal} {\bibinfo  {journal} {Rev. Mod. Phys.}\
  }\textbf {\bibinfo {volume} {51}},\ \bibinfo {pages} {79} (\bibinfo {year}
  {1979})}\BibitemShut {NoStop}%
\bibitem [{\citenamefont {Mohr}, \citenamefont {Taylor},\ and\ \citenamefont
  {Newell}(2011)}]{codata2010}%
  \BibitemOpen
  \bibfield  {author} {\bibinfo {author} {\bibfnamefont {P.~J.}\ \bibnamefont
  {Mohr}}, \bibinfo {author} {\bibfnamefont {B.~N.}\ \bibnamefont {Taylor}}, \
  and\ \bibinfo {author} {\bibfnamefont {D.~B.}\ \bibnamefont {Newell}},\
  }\href@noop {} {\enquote {\bibinfo {title} {The 2010 {CODATA} recommended
  values of the fundamental physical constants},}\ }\bibinfo {howpublished}
  {http://physics.nist.gov/constants} (\bibinfo {year} {2011})\BibitemShut
  {NoStop}%
\bibitem [{iap(2008)}]{iapwsfundam}%
  \BibitemOpen
  \href {http://www.iapws.org/relguide/fundam.pdf} {\emph {\bibinfo {title}
  {Guideline on the Use of Fundamental Physical Constants and Basic Constants
  of Water}}},\ \bibinfo {organization} {IAPWS} (\bibinfo {year} {2008}),\ \url
  {http://www.iapws.org/relguide/fundam.pdf}\BibitemShut {NoStop}%
\bibitem [{\citenamefont {Lemmon}\ and\ \citenamefont
  {Span}(2010)}]{lemmon2010}%
  \BibitemOpen
  \bibfield  {author} {\bibinfo {author} {\bibfnamefont {E.~W.}\ \bibnamefont
  {Lemmon}}\ and\ \bibinfo {author} {\bibfnamefont {R.}~\bibnamefont {Span}},\
  }in\ \href@noop {} {\emph {\bibinfo {booktitle} {Applied Thermodynamics of
  Fluids}}},\ \bibinfo {editor} {edited by\ \bibinfo {editor} {\bibfnamefont
  {A.~R.~H.}\ \bibnamefont {Goodwin}}, \bibinfo {editor} {\bibfnamefont
  {J.~V.}\ \bibnamefont {Sengers}}, \ and\ \bibinfo {editor} {\bibfnamefont
  {C.~J.}\ \bibnamefont {Peters}}}\ (\bibinfo  {publisher} {RSC Publishing},\
  \bibinfo {address} {Cambridge, UK},\ \bibinfo {year} {2010})\ Chap.~\bibinfo
  {chapter} {12}, pp.\ \bibinfo {pages} {394--432}\BibitemShut {NoStop}%
\bibitem [{\citenamefont {Henderson}\ and\ \citenamefont
  {Speedy}(1987{\natexlab{a}})}]{henderson1987b}%
  \BibitemOpen
  \bibfield  {author} {\bibinfo {author} {\bibfnamefont {S.~J.}\ \bibnamefont
  {Henderson}}\ and\ \bibinfo {author} {\bibfnamefont {R.~J.}\ \bibnamefont
  {Speedy}},\ }\href@noop {} {\bibfield  {journal} {\bibinfo  {journal} {J.
  Phys. Chem.}\ }\textbf {\bibinfo {volume} {91}},\ \bibinfo {pages} {3069}
  (\bibinfo {year} {1987}{\natexlab{a}})}\BibitemShut {NoStop}%
\bibitem [{\citenamefont {Caldwell}(1978)}]{caldwell1978}%
  \BibitemOpen
  \bibfield  {author} {\bibinfo {author} {\bibfnamefont {D.~R.}\ \bibnamefont
  {Caldwell}},\ }\href@noop {} {\bibfield  {journal} {\bibinfo  {journal}
  {Deep-Sea Res.}\ }\textbf {\bibinfo {volume} {25}},\ \bibinfo {pages} {175}
  (\bibinfo {year} {1978})}\BibitemShut {NoStop}%
\bibitem [{\citenamefont {Henderson}\ and\ \citenamefont
  {Speedy}(1987{\natexlab{b}})}]{henderson1987a}%
  \BibitemOpen
  \bibfield  {author} {\bibinfo {author} {\bibfnamefont {S.~J.}\ \bibnamefont
  {Henderson}}\ and\ \bibinfo {author} {\bibfnamefont {R.~J.}\ \bibnamefont
  {Speedy}},\ }\href@noop {} {\bibfield  {journal} {\bibinfo  {journal} {J.
  Phys. Chem.}\ }\textbf {\bibinfo {volume} {91}},\ \bibinfo {pages} {3062}
  (\bibinfo {year} {1987}{\natexlab{b}})}\BibitemShut {NoStop}%
\bibitem [{\citenamefont {Dougherty}(2004)}]{dougherty2004}%
  \BibitemOpen
  \bibfield  {author} {\bibinfo {author} {\bibfnamefont {R.~C.}\ \bibnamefont
  {Dougherty}},\ }\href@noop {} {\bibfield  {journal} {\bibinfo  {journal}
  {Chem. Phys.}\ }\textbf {\bibinfo {volume} {298}},\ \bibinfo {pages} {307}
  (\bibinfo {year} {2004})}\BibitemShut {NoStop}%
\bibitem [{\citenamefont {Murphy}\ and\ \citenamefont {Koop}(2005)}]{mur05}%
  \BibitemOpen
  \bibfield  {author} {\bibinfo {author} {\bibfnamefont {D.~M.}\ \bibnamefont
  {Murphy}}\ and\ \bibinfo {author} {\bibfnamefont {T.}~\bibnamefont {Koop}},\
  }\href@noop {} {\bibfield  {journal} {\bibinfo  {journal} {Q.~J.~R. Meteorol.
  Soc.}\ }\textbf {\bibinfo {volume} {131}},\ \bibinfo {pages} {1539} (\bibinfo
  {year} {2005})}\BibitemShut {NoStop}%
\bibitem [{\citenamefont {Zábranský}\ \emph {et~al.}(2010)\citenamefont
  {Zábranský}, \citenamefont {Kolská}, \citenamefont {R{\r u}ži{\v c}ka},\ and\
  \citenamefont {Domalski}}]{zabransky2010}%
  \BibitemOpen
  \bibfield  {author} {\bibinfo {author} {\bibfnamefont {M.}~\bibnamefont
  {Zábranský}}, \bibinfo {author} {\bibfnamefont {Z.}~\bibnamefont {Kolská}},
  \bibinfo {author} {\bibfnamefont {V.}~\bibnamefont {R{\r u}ži{\v c}ka},
  \bibfnamefont {Jr.}}, \ and\ \bibinfo {author} {\bibfnamefont {E.~S.}\
  \bibnamefont {Domalski}},\ }\href@noop {} {\bibfield  {journal} {\bibinfo
  {journal} {J.~Phys. Chem. Ref. Data}\ }\textbf {\bibinfo {volume} {39}},\
  \bibinfo {pages} {013103} (\bibinfo {year} {2010})}\BibitemShut {NoStop}%
\bibitem [{\citenamefont {Kell}(1967)}]{kell67}%
  \BibitemOpen
  \bibfield  {author} {\bibinfo {author} {\bibfnamefont {G.~S.}\ \bibnamefont
  {Kell}},\ }\href@noop {} {\bibfield  {journal} {\bibinfo  {journal} {J.~Chem.
  Eng. Data}\ }\textbf {\bibinfo {volume} {12}},\ \bibinfo {pages} {66}
  (\bibinfo {year} {1967})}\BibitemShut {NoStop}%
\bibitem [{\citenamefont {Chen}\ and\ \citenamefont
  {Millero}(1977)}]{chen1977}%
  \BibitemOpen
  \bibfield  {author} {\bibinfo {author} {\bibfnamefont {C.-T.}\ \bibnamefont
  {Chen}}\ and\ \bibinfo {author} {\bibfnamefont {F.~J.}\ \bibnamefont
  {Millero}},\ }\href@noop {} {\bibfield  {journal} {\bibinfo  {journal} {J.
  Acoust. Soc. Am.}\ }\textbf {\bibinfo {volume} {62}},\ \bibinfo {pages} {553}
  (\bibinfo {year} {1977})}\BibitemShut {NoStop}%
\bibitem [{\citenamefont {Marczak}(1997)}]{marczak1997}%
  \BibitemOpen
  \bibfield  {author} {\bibinfo {author} {\bibfnamefont {W.}~\bibnamefont
  {Marczak}},\ }\href@noop {} {\bibfield  {journal} {\bibinfo  {journal}
  {Acustica}\ }\textbf {\bibinfo {volume} {83}},\ \bibinfo {pages} {473}
  (\bibinfo {year} {1997})}\BibitemShut {NoStop}%
\bibitem [{\citenamefont {Kamgar-Parsi}, \citenamefont {Levelt~Sengers},\ and\
  \citenamefont {Sengers}(1983)}]{kamgarparsi1983}%
  \BibitemOpen
  \bibfield  {author} {\bibinfo {author} {\bibfnamefont {B.}~\bibnamefont
  {Kamgar-Parsi}}, \bibinfo {author} {\bibfnamefont {J.~M.~H.}\ \bibnamefont
  {Levelt~Sengers}}, \ and\ \bibinfo {author} {\bibfnamefont {J.~V.}\
  \bibnamefont {Sengers}},\ }\href@noop {} {\bibfield  {journal} {\bibinfo
  {journal} {J.~Phys. Chem. Ref. Data}\ }\textbf {\bibinfo {volume} {12}},\
  \bibinfo {pages} {513} (\bibinfo {year} {1983})}\BibitemShut {NoStop}%
\bibitem [{\citenamefont {Kostrowicka~Wyczalkowska}\ \emph
  {et~al.}(2000)\citenamefont {Kostrowicka~Wyczalkowska}, \citenamefont
  {Abdulkadirova}, \citenamefont {Anisimov},\ and\ \citenamefont
  {Sengers}}]{kostrow2000}%
  \BibitemOpen
  \bibfield  {author} {\bibinfo {author} {\bibfnamefont {A.}~\bibnamefont
  {Kostrowicka~Wyczalkowska}}, \bibinfo {author} {\bibfnamefont
  {{\relax{}Kh}.~S.}\ \bibnamefont {Abdulkadirova}}, \bibinfo {author}
  {\bibfnamefont {M.~A.}\ \bibnamefont {Anisimov}}, \ and\ \bibinfo {author}
  {\bibfnamefont {J.~V.}\ \bibnamefont {Sengers}},\ }\href@noop {} {\bibfield
  {journal} {\bibinfo  {journal} {J.~Chem. Phys.}\ }\textbf {\bibinfo {volume}
  {113}},\ \bibinfo {pages} {4985} (\bibinfo {year} {2000})}\BibitemShut
  {NoStop}%
\bibitem [{\citenamefont {R{\"o}ntgen}(1892)}]{roentgen}%
  \BibitemOpen
  \bibfield  {author} {\bibinfo {author} {\bibfnamefont {W.~C.}\ \bibnamefont
  {R{\"o}ntgen}},\ }\href@noop {} {\bibfield  {journal} {\bibinfo  {journal}
  {Ann. Phys. (Leipzig)}\ }\textbf {\bibinfo {volume} {281}},\ \bibinfo {pages}
  {91} (\bibinfo {year} {1892})}\BibitemShut {NoStop}%
\bibitem [{\citenamefont {Ponyatovskii}, \citenamefont {Sinitsyn},\ and\
  \citenamefont {Pozdnyakova}(1994)}]{ponyatovskii1994}%
  \BibitemOpen
  \bibfield  {author} {\bibinfo {author} {\bibfnamefont {E.~G.}\ \bibnamefont
  {Ponyatovskii}}, \bibinfo {author} {\bibfnamefont {V.~V.}\ \bibnamefont
  {Sinitsyn}}, \ and\ \bibinfo {author} {\bibfnamefont {T.~A.}\ \bibnamefont
  {Pozdnyakova}},\ }\href@noop {} {\bibfield  {journal} {\bibinfo  {journal}
  {JETP Lett.}\ }\textbf {\bibinfo {volume} {60}},\ \bibinfo {pages} {360}
  (\bibinfo {year} {1994})}\BibitemShut {NoStop}%
\bibitem [{\citenamefont {Moynihan}(1997)}]{moynihan1997}%
  \BibitemOpen
  \bibfield  {author} {\bibinfo {author} {\bibfnamefont {C.~T.}\ \bibnamefont
  {Moynihan}},\ }\href@noop {} {\bibfield  {journal} {\bibinfo  {journal} {Mat.
  Res. Soc. Symp. Proc.}\ }\textbf {\bibinfo {volume} {455}},\ \bibinfo {pages}
  {411} (\bibinfo {year} {1997})}\BibitemShut {NoStop}%
\bibitem [{\citenamefont {Prigogine}\ and\ \citenamefont
  {Defay}(1954)}]{prigogine1954}%
  \BibitemOpen
  \bibfield  {author} {\bibinfo {author} {\bibfnamefont {I.}~\bibnamefont
  {Prigogine}}\ and\ \bibinfo {author} {\bibfnamefont {R.}~\bibnamefont
  {Defay}},\ }\href@noop {} {\emph {\bibinfo {title} {Chemical
  Thermodynamics}}}\ (\bibinfo  {publisher} {Longmans, Green \& Co.},\ \bibinfo
  {address} {London},\ \bibinfo {year} {1954})\BibitemShut {NoStop}%
\bibitem [{\citenamefont {Hrubý}\ and\ \citenamefont
  {Holten}(2004)}]{hrubyholten2004}%
  \BibitemOpen
  \bibfield  {author} {\bibinfo {author} {\bibfnamefont {J.}~\bibnamefont
  {Hrubý}}\ and\ \bibinfo {author} {\bibfnamefont {V.}~\bibnamefont {Holten}},\
  }\enquote {\bibinfo {title} {A two-structure model of thermodynamic
  properties and surface tension of supercooled water},}\ in\ \href@noop {}
  {\emph {\bibinfo {booktitle} {Proceedings of the 14th International
  Conference on the Properties of Water and Steam}}}\ (\bibinfo  {publisher}
  {Maruzen},\ \bibinfo {year} {2004})\BibitemShut {NoStop}%
\bibitem [{\citenamefont {Holten}(2004)}]{hol04}%
  \BibitemOpen
  \bibfield  {author} {\bibinfo {author} {\bibfnamefont {V.}~\bibnamefont
  {Holten}},\ }\emph {\bibinfo {title} {From supersaturated water vapour to
  supercooled liquid water: analysis and experiments}},\ \href@noop {}
  {Master's thesis},\ \bibinfo  {school} {Eindhoven University of Technology}
  (\bibinfo {year} {2004})\BibitemShut {NoStop}%
\bibitem [{\citenamefont {Mishima}(1994)}]{mishima1994}%
  \BibitemOpen
  \bibfield  {author} {\bibinfo {author} {\bibfnamefont {O.}~\bibnamefont
  {Mishima}},\ }\href@noop {} {\bibfield  {journal} {\bibinfo  {journal}
  {J.~Chem. Phys.}\ }\textbf {\bibinfo {volume} {100}},\ \bibinfo {pages}
  {5910} (\bibinfo {year} {1994})}\BibitemShut {NoStop}%
\bibitem [{\citenamefont {Whalley}, \citenamefont {Klug},\ and\ \citenamefont
  {Handa}(1989)}]{whalley1989}%
  \BibitemOpen
  \bibfield  {author} {\bibinfo {author} {\bibfnamefont {E.}~\bibnamefont
  {Whalley}}, \bibinfo {author} {\bibfnamefont {D.~D.}\ \bibnamefont {Klug}}, \
  and\ \bibinfo {author} {\bibfnamefont {Y.~P.}\ \bibnamefont {Handa}},\
  }\href@noop {} {\bibfield  {journal} {\bibinfo  {journal} {Nature}\ }\textbf
  {\bibinfo {volume} {342}},\ \bibinfo {pages} {782} (\bibinfo {year}
  {1989})}\BibitemShut {NoStop}%
\bibitem [{\citenamefont {Mishima}\ and\ \citenamefont
  {Stanley}(1998{\natexlab{b}})}]{mishima1998review}%
  \BibitemOpen
  \bibfield  {author} {\bibinfo {author} {\bibfnamefont {O.}~\bibnamefont
  {Mishima}}\ and\ \bibinfo {author} {\bibfnamefont {H.~E.}\ \bibnamefont
  {Stanley}},\ }\href@noop {} {\bibfield  {journal} {\bibinfo  {journal}
  {Nature}\ }\textbf {\bibinfo {volume} {396}},\ \bibinfo {pages} {329}
  (\bibinfo {year} {1998}{\natexlab{b}})}\BibitemShut {NoStop}%
\bibitem [{\citenamefont {Zhang}\ \emph {et~al.}(2011)\citenamefont {Zhang},
  \citenamefont {Faraone}, \citenamefont {Kamitakahara}, \citenamefont {Liu},
  \citenamefont {Mou}, \citenamefont {Leão}, \citenamefont {Chang},\ and\
  \citenamefont {Chen}}]{zhang2011}%
  \BibitemOpen
  \bibfield  {author} {\bibinfo {author} {\bibfnamefont {Y.}~\bibnamefont
  {Zhang}}, \bibinfo {author} {\bibfnamefont {A.}~\bibnamefont {Faraone}},
  \bibinfo {author} {\bibfnamefont {W.~A.}\ \bibnamefont {Kamitakahara}},
  \bibinfo {author} {\bibfnamefont {K.-H.}\ \bibnamefont {Liu}}, \bibinfo
  {author} {\bibfnamefont {C.-Y.}\ \bibnamefont {Mou}}, \bibinfo {author}
  {\bibfnamefont {J.~B.}\ \bibnamefont {Leão}}, \bibinfo {author}
  {\bibfnamefont {S.}~\bibnamefont {Chang}}, \ and\ \bibinfo {author}
  {\bibfnamefont {S.-H.}\ \bibnamefont {Chen}},\ }\href@noop {} {\bibfield
  {journal} {\bibinfo  {journal} {Proc. Natl. Acad. Sci. U.S.A.}\ }\textbf
  {\bibinfo {volume} {108}},\ \bibinfo {pages} {12206} (\bibinfo {year}
  {2011})}\BibitemShut {NoStop}%
\bibitem [{\citenamefont {Tester}\ and\ \citenamefont {Modell}(1997)}]{tmbook}%
  \BibitemOpen
  \bibfield  {author} {\bibinfo {author} {\bibfnamefont {J.~W.}\ \bibnamefont
  {Tester}}\ and\ \bibinfo {author} {\bibfnamefont {M.}~\bibnamefont
  {Modell}},\ }\href@noop {} {\emph {\bibinfo {title} {Thermodynamics and Its
  Applications}}},\ \bibinfo {edition} {3rd}\ ed.\ (\bibinfo  {publisher}
  {Prentice Hall},\ \bibinfo {address} {Upper Saddle River},\ \bibinfo {year}
  {1997})\BibitemShut {NoStop}%
\bibitem [{\citenamefont {O'Connell}\ and\ \citenamefont
  {Haile}(2005)}]{oconnellbook}%
  \BibitemOpen
  \bibfield  {author} {\bibinfo {author} {\bibfnamefont {J.~P.}\ \bibnamefont
  {O'Connell}}\ and\ \bibinfo {author} {\bibfnamefont {J.~M.}\ \bibnamefont
  {Haile}},\ }\href@noop {} {\emph {\bibinfo {title} {Thermodynamics:
  Fundamentals for Applications}}}\ (\bibinfo  {publisher} {Cambridge
  University Press},\ \bibinfo {address} {Cambridge, MA},\ \bibinfo {year}
  {2005})\BibitemShut {NoStop}%
\bibitem [{\citenamefont {Fisher}, \citenamefont {Zinn},\ and\ \citenamefont
  {Upton}(1999)}]{fisher1999}%
  \BibitemOpen
  \bibfield  {author} {\bibinfo {author} {\bibfnamefont {M.~E.}\ \bibnamefont
  {Fisher}}, \bibinfo {author} {\bibfnamefont {S.-Y.}\ \bibnamefont {Zinn}}, \
  and\ \bibinfo {author} {\bibfnamefont {P.~J.}\ \bibnamefont {Upton}},\
  }\href@noop {} {\bibfield  {journal} {\bibinfo  {journal} {Phys. Rev. B}\
  }\textbf {\bibinfo {volume} {59}},\ \bibinfo {pages} {14533} (\bibinfo {year}
  {1999})}\BibitemShut {NoStop}%
\bibitem [{\citenamefont {Moore}\ and\ \citenamefont
  {Molinero}(2011)}]{moore2011}%
  \BibitemOpen
  \bibfield  {author} {\bibinfo {author} {\bibfnamefont {E.~B.}\ \bibnamefont
  {Moore}}\ and\ \bibinfo {author} {\bibfnamefont {V.}~\bibnamefont
  {Molinero}},\ }\href@noop {} {\bibfield  {journal} {\bibinfo  {journal}
  {Nature}\ }\textbf {\bibinfo {volume} {479}},\ \bibinfo {pages} {506}
  (\bibinfo {year} {2011})}\BibitemShut {NoStop}%
\bibitem [{\citenamefont {Liu}\ \emph {et~al.}(2007)\citenamefont {Liu},
  \citenamefont {Zhang}, \citenamefont {Chen}, \citenamefont {Mou},
  \citenamefont {Poole},\ and\ \citenamefont {Chen}}]{liu07}%
  \BibitemOpen
  \bibfield  {author} {\bibinfo {author} {\bibfnamefont {D.}~\bibnamefont
  {Liu}}, \bibinfo {author} {\bibfnamefont {Y.}~\bibnamefont {Zhang}}, \bibinfo
  {author} {\bibfnamefont {C.-C.}\ \bibnamefont {Chen}}, \bibinfo {author}
  {\bibfnamefont {C.-Y.}\ \bibnamefont {Mou}}, \bibinfo {author} {\bibfnamefont
  {P.~H.}\ \bibnamefont {Poole}}, \ and\ \bibinfo {author} {\bibfnamefont
  {S.-H.}\ \bibnamefont {Chen}},\ }\href@noop {} {\bibfield  {journal}
  {\bibinfo  {journal} {Proc. Natl. Acad. Sci. U.S.A.}\ }\textbf {\bibinfo
  {volume} {104}},\ \bibinfo {pages} {9570} (\bibinfo {year}
  {2007})}\BibitemShut {NoStop}%
\bibitem [{\citenamefont {Mallamace}\ \emph {et~al.}(2007)\citenamefont
  {Mallamace}, \citenamefont {Branca}, \citenamefont {Broccio}, \citenamefont
  {Corsaro}, \citenamefont {Mou},\ and\ \citenamefont {Chen}}]{mallamace07b}%
  \BibitemOpen
  \bibfield  {author} {\bibinfo {author} {\bibfnamefont {F.}~\bibnamefont
  {Mallamace}}, \bibinfo {author} {\bibfnamefont {C.}~\bibnamefont {Branca}},
  \bibinfo {author} {\bibfnamefont {M.}~\bibnamefont {Broccio}}, \bibinfo
  {author} {\bibfnamefont {C.}~\bibnamefont {Corsaro}}, \bibinfo {author}
  {\bibfnamefont {C.-Y.}\ \bibnamefont {Mou}}, \ and\ \bibinfo {author}
  {\bibfnamefont {S.-H.}\ \bibnamefont {Chen}},\ }\href@noop {} {\bibfield
  {journal} {\bibinfo  {journal} {Proc. Natl. Acad. Sci. U.S.A.}\ }\textbf
  {\bibinfo {volume} {104}},\ \bibinfo {pages} {18387} (\bibinfo {year}
  {2007})}\BibitemShut {NoStop}%
\bibitem [{\citenamefont {Zhang}\ \emph {et~al.}(2009)\citenamefont {Zhang},
  \citenamefont {Liu}, \citenamefont {Lagi}, \citenamefont {Liu}, \citenamefont
  {Littrell}, \citenamefont {Mou},\ and\ \citenamefont {Chen}}]{zhang2009}%
  \BibitemOpen
  \bibfield  {author} {\bibinfo {author} {\bibfnamefont {Y.}~\bibnamefont
  {Zhang}}, \bibinfo {author} {\bibfnamefont {K.-H.}\ \bibnamefont {Liu}},
  \bibinfo {author} {\bibfnamefont {M.}~\bibnamefont {Lagi}}, \bibinfo {author}
  {\bibfnamefont {D.}~\bibnamefont {Liu}}, \bibinfo {author} {\bibfnamefont
  {K.~C.}\ \bibnamefont {Littrell}}, \bibinfo {author} {\bibfnamefont {C.-Y.}\
  \bibnamefont {Mou}}, \ and\ \bibinfo {author} {\bibfnamefont {S.-H.}\
  \bibnamefont {Chen}},\ }\href@noop {} {\bibfield  {journal} {\bibinfo
  {journal} {J.~Phys. Chem.~B}\ }\textbf {\bibinfo {volume} {113}},\ \bibinfo
  {pages} {5007} (\bibinfo {year} {2009})}\BibitemShut {NoStop}%
\bibitem [{\citenamefont {Mallamace}(2009)}]{mallamace2009}%
  \BibitemOpen
  \bibfield  {author} {\bibinfo {author} {\bibfnamefont {F.}~\bibnamefont
  {Mallamace}},\ }\href@noop {} {\bibfield  {journal} {\bibinfo  {journal}
  {Proc. Natl. Acad. Sci. U.S.A.}\ }\textbf {\bibinfo {volume} {106}},\
  \bibinfo {pages} {15097} (\bibinfo {year} {2009})}\BibitemShut {NoStop}%
\bibitem [{\citenamefont {Nagoe}\ \emph {et~al.}(2010)\citenamefont {Nagoe},
  \citenamefont {Kanke}, \citenamefont {Oguni},\ and\ \citenamefont
  {Namba}}]{nagoe2010}%
  \BibitemOpen
  \bibfield  {author} {\bibinfo {author} {\bibfnamefont {A.}~\bibnamefont
  {Nagoe}}, \bibinfo {author} {\bibfnamefont {Y.}~\bibnamefont {Kanke}},
  \bibinfo {author} {\bibfnamefont {M.}~\bibnamefont {Oguni}}, \ and\ \bibinfo
  {author} {\bibfnamefont {S.}~\bibnamefont {Namba}},\ }\href@noop {}
  {\bibfield  {journal} {\bibinfo  {journal} {J.~Phys. Chem.~B}\ }\textbf
  {\bibinfo {volume} {114}},\ \bibinfo {pages} {13940} (\bibinfo {year}
  {2010})}\BibitemShut {NoStop}%
\bibitem [{\citenamefont {Gallo}, \citenamefont {Rovere},\ and\ \citenamefont
  {Chen}(2010)}]{gallo2010}%
  \BibitemOpen
  \bibfield  {author} {\bibinfo {author} {\bibfnamefont {P.}~\bibnamefont
  {Gallo}}, \bibinfo {author} {\bibfnamefont {M.}~\bibnamefont {Rovere}}, \
  and\ \bibinfo {author} {\bibfnamefont {S.-H.}\ \bibnamefont {Chen}},\
  }\href@noop {} {\bibfield  {journal} {\bibinfo  {journal} {J. Phys.: Condens.
  Matter}\ }\textbf {\bibinfo {volume} {22}},\ \bibinfo {pages} {284102}
  (\bibinfo {year} {2010})}\BibitemShut {NoStop}%
\bibitem [{\citenamefont {Xu}\ and\ \citenamefont {Molinero}(2011)}]{xu2011}%
  \BibitemOpen
  \bibfield  {author} {\bibinfo {author} {\bibfnamefont {L.}~\bibnamefont
  {Xu}}\ and\ \bibinfo {author} {\bibfnamefont {V.}~\bibnamefont {Molinero}},\
  }\href@noop {} {\bibfield  {journal} {\bibinfo  {journal} {J.~Phys. Chem.~B}\
  }\textbf {\bibinfo {volume} {115}},\ \bibinfo {pages} {14210} (\bibinfo
  {year} {2011})}\BibitemShut {NoStop}%
\bibitem [{\citenamefont {Sastry}\ \emph {et~al.}(1996)\citenamefont {Sastry},
  \citenamefont {Debenedetti}, \citenamefont {Sciortino},\ and\ \citenamefont
  {Stanley}}]{sastry1996}%
  \BibitemOpen
  \bibfield  {author} {\bibinfo {author} {\bibfnamefont {S.}~\bibnamefont
  {Sastry}}, \bibinfo {author} {\bibfnamefont {P.~G.}\ \bibnamefont
  {Debenedetti}}, \bibinfo {author} {\bibfnamefont {F.}~\bibnamefont
  {Sciortino}}, \ and\ \bibinfo {author} {\bibfnamefont {H.~E.}\ \bibnamefont
  {Stanley}},\ }\href@noop {} {\bibfield  {journal} {\bibinfo  {journal} {Phys.
  Rev. E}\ }\textbf {\bibinfo {volume} {53}},\ \bibinfo {pages} {6144}
  (\bibinfo {year} {1996})}\BibitemShut {NoStop}%
\bibitem [{\citenamefont {Rebelo}, \citenamefont {Debenedetti},\ and\
  \citenamefont {Sastry}(1998)}]{rebelo1998}%
  \BibitemOpen
  \bibfield  {author} {\bibinfo {author} {\bibfnamefont {L.~P.~N.}\
  \bibnamefont {Rebelo}}, \bibinfo {author} {\bibfnamefont {P.~G.}\
  \bibnamefont {Debenedetti}}, \ and\ \bibinfo {author} {\bibfnamefont
  {S.}~\bibnamefont {Sastry}},\ }\href@noop {} {\bibfield  {journal} {\bibinfo
  {journal} {J.~Chem. Phys.}\ }\textbf {\bibinfo {volume} {109}},\ \bibinfo
  {pages} {626} (\bibinfo {year} {1998})}\BibitemShut {NoStop}%
\bibitem [{\citenamefont {Angell}(2008)}]{angell2008}%
  \BibitemOpen
  \bibfield  {author} {\bibinfo {author} {\bibfnamefont {C.~A.}\ \bibnamefont
  {Angell}},\ }\href@noop {} {\bibfield  {journal} {\bibinfo  {journal}
  {Science}\ }\textbf {\bibinfo {volume} {319}},\ \bibinfo {pages} {582}
  (\bibinfo {year} {2008})}\BibitemShut {NoStop}%
\bibitem [{\citenamefont {Stokely}\ \emph {et~al.}(2010)\citenamefont
  {Stokely}, \citenamefont {Mazza}, \citenamefont {Stanley},\ and\
  \citenamefont {Franzese}}]{stokely2010}%
  \BibitemOpen
  \bibfield  {author} {\bibinfo {author} {\bibfnamefont {K.}~\bibnamefont
  {Stokely}}, \bibinfo {author} {\bibfnamefont {M.~G.}\ \bibnamefont {Mazza}},
  \bibinfo {author} {\bibfnamefont {H.~E.}\ \bibnamefont {Stanley}}, \ and\
  \bibinfo {author} {\bibfnamefont {G.}~\bibnamefont {Franzese}},\ }\href@noop
  {} {\bibfield  {journal} {\bibinfo  {journal} {Proc. Natl. Acad. Sci.
  U.S.A.}\ }\textbf {\bibinfo {volume} {107}},\ \bibinfo {pages} {1301}
  (\bibinfo {year} {2010})}\BibitemShut {NoStop}%
\bibitem [{\citenamefont {Henriques}\ \emph {et~al.}(2005)\citenamefont
  {Henriques}, \citenamefont {Guisoni}, \citenamefont {Barbosa}, \citenamefont
  {Thielo},\ and\ \citenamefont {Barbosa}}]{henriques2005}%
  \BibitemOpen
  \bibfield  {author} {\bibinfo {author} {\bibfnamefont {V.~B.}\ \bibnamefont
  {Henriques}}, \bibinfo {author} {\bibfnamefont {N.}~\bibnamefont {Guisoni}},
  \bibinfo {author} {\bibfnamefont {M.~A.}\ \bibnamefont {Barbosa}}, \bibinfo
  {author} {\bibfnamefont {M.}~\bibnamefont {Thielo}}, \ and\ \bibinfo {author}
  {\bibfnamefont {M.~C.}\ \bibnamefont {Barbosa}},\ }\href@noop {} {\bibfield
  {journal} {\bibinfo  {journal} {Mol. Phys.}\ }\textbf {\bibinfo {volume}
  {103}},\ \bibinfo {pages} {3001} (\bibinfo {year} {2005})}\BibitemShut
  {NoStop}%
\bibitem [{\citenamefont {Brovchenko}, \citenamefont {Geiger},\ and\
  \citenamefont {Oleinikova}(2005)}]{brovchenko2005}%
  \BibitemOpen
  \bibfield  {author} {\bibinfo {author} {\bibfnamefont {I.}~\bibnamefont
  {Brovchenko}}, \bibinfo {author} {\bibfnamefont {A.}~\bibnamefont {Geiger}},
  \ and\ \bibinfo {author} {\bibfnamefont {A.}~\bibnamefont {Oleinikova}},\
  }\href@noop {} {\bibfield  {journal} {\bibinfo  {journal} {J.~Chem. Phys.}\
  }\textbf {\bibinfo {volume} {123}},\ \bibinfo {pages} {044515} (\bibinfo
  {year} {2005})}\BibitemShut {NoStop}%
\bibitem [{\citenamefont {Brovchenko}\ and\ \citenamefont
  {Oleinikova}(2008)}]{brovchenkobook}%
  \BibitemOpen
  \bibfield  {author} {\bibinfo {author} {\bibfnamefont {I.}~\bibnamefont
  {Brovchenko}}\ and\ \bibinfo {author} {\bibfnamefont {A.}~\bibnamefont
  {Oleinikova}},\ }\href@noop {} {\emph {\bibinfo {title} {Interfacial and
  Confined Water}}}\ (\bibinfo  {publisher} {Elsevier},\ \bibinfo {address}
  {Amsterdam},\ \bibinfo {year} {2008})\BibitemShut {NoStop}%
\bibitem [{\citenamefont {Molinero}\ and\ \citenamefont
  {Moore}(2009)}]{molinero2009}%
  \BibitemOpen
  \bibfield  {author} {\bibinfo {author} {\bibfnamefont {V.}~\bibnamefont
  {Molinero}}\ and\ \bibinfo {author} {\bibfnamefont {E.~B.}\ \bibnamefont
  {Moore}},\ }\href@noop {} {\bibfield  {journal} {\bibinfo  {journal}
  {J.~Phys. Chem.~B}\ }\textbf {\bibinfo {volume} {113}},\ \bibinfo {pages}
  {4008} (\bibinfo {year} {2009})}\BibitemShut {NoStop}%
\bibitem [{\citenamefont {Liu}, \citenamefont {Panagiotopoulos},\ and\
  \citenamefont {Debenedetti}(2009)}]{liu2009}%
  \BibitemOpen
  \bibfield  {author} {\bibinfo {author} {\bibfnamefont {Y.}~\bibnamefont
  {Liu}}, \bibinfo {author} {\bibfnamefont {A.~Z.}\ \bibnamefont
  {Panagiotopoulos}}, \ and\ \bibinfo {author} {\bibfnamefont {P.~G.}\
  \bibnamefont {Debenedetti}},\ }\href@noop {} {\bibfield  {journal} {\bibinfo
  {journal} {J.~Chem. Phys.}\ }\textbf {\bibinfo {volume} {131}},\ \bibinfo
  {pages} {104508} (\bibinfo {year} {2009})}\BibitemShut {NoStop}%
\bibitem [{\citenamefont {Sciortino}, \citenamefont {Saika-Voivod},\ and\
  \citenamefont {Poole}(2011)}]{sciortino2011}%
  \BibitemOpen
  \bibfield  {author} {\bibinfo {author} {\bibfnamefont {F.}~\bibnamefont
  {Sciortino}}, \bibinfo {author} {\bibfnamefont {I.}~\bibnamefont
  {Saika-Voivod}}, \ and\ \bibinfo {author} {\bibfnamefont {P.~H.}\
  \bibnamefont {Poole}},\ }\href@noop {} {\bibfield  {journal} {\bibinfo
  {journal} {Phys. Chem. Chem. Phys.}\ }\textbf {\bibinfo {volume} {13}},\
  \bibinfo {pages} {19759} (\bibinfo {year} {2011})}\BibitemShut {NoStop}%
\bibitem [{\citenamefont {Brazovskii}(1975)}]{brazovskii1975}%
  \BibitemOpen
  \bibfield  {author} {\bibinfo {author} {\bibfnamefont {S.~A.}\ \bibnamefont
  {Brazovskii}},\ }\href@noop {} {\bibfield  {journal} {\bibinfo  {journal}
  {Sov. Phys.--JETP}\ }\textbf {\bibinfo {volume} {41}},\ \bibinfo {pages} {85}
  (\bibinfo {year} {1975})}\BibitemShut {NoStop}%
\bibitem [{\citenamefont {Brazovskii}, \citenamefont {Dzyaloshinskii},\ and\
  \citenamefont {Muratov}(1987)}]{brazovskii1987}%
  \BibitemOpen
  \bibfield  {author} {\bibinfo {author} {\bibfnamefont {S.~A.}\ \bibnamefont
  {Brazovskii}}, \bibinfo {author} {\bibfnamefont {I.~E.}\ \bibnamefont
  {Dzyaloshinskii}}, \ and\ \bibinfo {author} {\bibfnamefont {A.~R.}\
  \bibnamefont {Muratov}},\ }\href@noop {} {\bibfield  {journal} {\bibinfo
  {journal} {Sov. Phys.--JETP}\ }\textbf {\bibinfo {volume} {66}},\ \bibinfo
  {pages} {625} (\bibinfo {year} {1987})}\BibitemShut {NoStop}%
\bibitem [{\citenamefont {Kats}, \citenamefont {Lebedev},\ and\ \citenamefont
  {Muratov}(1993)}]{kats1993}%
  \BibitemOpen
  \bibfield  {author} {\bibinfo {author} {\bibfnamefont {E.~I.}\ \bibnamefont
  {Kats}}, \bibinfo {author} {\bibfnamefont {V.~V.}\ \bibnamefont {Lebedev}}, \
  and\ \bibinfo {author} {\bibfnamefont {A.~R.}\ \bibnamefont {Muratov}},\
  }\href@noop {} {\bibfield  {journal} {\bibinfo  {journal} {Phys. Rep.}\
  }\textbf {\bibinfo {volume} {228}},\ \bibinfo {pages} {1} (\bibinfo {year}
  {1993})}\BibitemShut {NoStop}%
\bibitem [{\citenamefont {Anisimov}(1991)}]{anisimovbook}%
  \BibitemOpen
  \bibfield  {author} {\bibinfo {author} {\bibfnamefont {M.~A.}\ \bibnamefont
  {Anisimov}},\ }\href@noop {} {\emph {\bibinfo {title} {Critical Phenomena in
  Liquids and Liquid Crystals}}}\ (\bibinfo  {publisher} {Gordon and Breach},\
  \bibinfo {address} {Philadelphia},\ \bibinfo {year} {1991})\BibitemShut
  {NoStop}%
\bibitem [{\citenamefont {Gohin}\ \emph {et~al.}(1983)\citenamefont {Gohin},
  \citenamefont {Destrade}, \citenamefont {Gasparoux},\ and\ \citenamefont
  {Prost}}]{gohin1983}%
  \BibitemOpen
  \bibfield  {author} {\bibinfo {author} {\bibfnamefont {A.}~\bibnamefont
  {Gohin}}, \bibinfo {author} {\bibfnamefont {C.}~\bibnamefont {Destrade}},
  \bibinfo {author} {\bibfnamefont {H.}~\bibnamefont {Gasparoux}}, \ and\
  \bibinfo {author} {\bibfnamefont {J.}~\bibnamefont {Prost}},\ }\href@noop {}
  {\bibfield  {journal} {\bibinfo  {journal} {J. Physique}\ }\textbf {\bibinfo
  {volume} {44}},\ \bibinfo {pages} {427} (\bibinfo {year} {1983})}\BibitemShut
  {NoStop}%
\bibitem [{\citenamefont {Anisimov}\ \emph
  {et~al.}(1987{\natexlab{a}})\citenamefont {Anisimov}, \citenamefont {Konev},
  \citenamefont {Labko}, \citenamefont {Nikolaenko}, \citenamefont
  {Olefirenko},\ and\ \citenamefont {Yudin}}]{anisimov1987a}%
  \BibitemOpen
  \bibfield  {author} {\bibinfo {author} {\bibfnamefont {M.~A.}\ \bibnamefont
  {Anisimov}}, \bibinfo {author} {\bibfnamefont {S.~A.}\ \bibnamefont {Konev}},
  \bibinfo {author} {\bibfnamefont {V.~I.}\ \bibnamefont {Labko}}, \bibinfo
  {author} {\bibfnamefont {G.~L.}\ \bibnamefont {Nikolaenko}}, \bibinfo
  {author} {\bibfnamefont {G.~I.}\ \bibnamefont {Olefirenko}}, \ and\ \bibinfo
  {author} {\bibfnamefont {I.~K.}\ \bibnamefont {Yudin}},\ }\href@noop {}
  {\bibfield  {journal} {\bibinfo  {journal} {Mol. Cryst. Liq. Cryst.}\
  }\textbf {\bibinfo {volume} {146}},\ \bibinfo {pages} {421} (\bibinfo {year}
  {1987}{\natexlab{a}})}\BibitemShut {NoStop}%
\bibitem [{\citenamefont {Anisimov}\ \emph
  {et~al.}(1987{\natexlab{b}})\citenamefont {Anisimov}, \citenamefont {Labko},
  \citenamefont {Nikolaenko},\ and\ \citenamefont {Yudin}}]{anisimov1987b}%
  \BibitemOpen
  \bibfield  {author} {\bibinfo {author} {\bibfnamefont {M.~A.}\ \bibnamefont
  {Anisimov}}, \bibinfo {author} {\bibfnamefont {V.~I.}\ \bibnamefont {Labko}},
  \bibinfo {author} {\bibfnamefont {G.~L.}\ \bibnamefont {Nikolaenko}}, \ and\
  \bibinfo {author} {\bibfnamefont {I.~K.}\ \bibnamefont {Yudin}},\ }\href@noop
  {} {\bibfield  {journal} {\bibinfo  {journal} {JETP Lett.}\ }\textbf
  {\bibinfo {volume} {45}},\ \bibinfo {pages} {111} (\bibinfo {year}
  {1987}{\natexlab{b}})}\BibitemShut {NoStop}%
\bibitem [{\citenamefont {Saul}\ and\ \citenamefont
  {Wagner}(1989)}]{saulwagner89}%
  \BibitemOpen
  \bibfield  {author} {\bibinfo {author} {\bibfnamefont {A.}~\bibnamefont
  {Saul}}\ and\ \bibinfo {author} {\bibfnamefont {W.}~\bibnamefont {Wagner}},\
  }\href@noop {} {\bibfield  {journal} {\bibinfo  {journal} {J.~Phys. Chem.
  Ref. Data}\ }\textbf {\bibinfo {volume} {18}},\ \bibinfo {pages} {1537}
  (\bibinfo {year} {1989})}\BibitemShut {NoStop}%
\bibitem [{Note1()}]{Note1}%
  \BibitemOpen
  \bibinfo {note} {There is a sign error in the correlation of Asada \protect
  \textit {et al.}\cite {asada2002}: the coefficient $a_6$ in their Table~1
  should have a positive sign.}\BibitemShut {Stop}%
\bibitem [{\citenamefont {Duan}, \citenamefont {Thompson},\ and\ \citenamefont
  {Ward}(2008)}]{duan2008}%
  \BibitemOpen
  \bibfield  {author} {\bibinfo {author} {\bibfnamefont {F.}~\bibnamefont
  {Duan}}, \bibinfo {author} {\bibfnamefont {I.}~\bibnamefont {Thompson}}, \
  and\ \bibinfo {author} {\bibfnamefont {C.~A.}\ \bibnamefont {Ward}},\
  }\href@noop {} {\bibfield  {journal} {\bibinfo  {journal} {J.~Phys. Chem.~B}\
  }\textbf {\bibinfo {volume} {112}},\ \bibinfo {pages} {8605} (\bibinfo {year}
  {2008})}\BibitemShut {NoStop}%
\bibitem [{\citenamefont {Leyendekkers}\ and\ \citenamefont
  {Hunter}(1985)}]{ley85a}%
  \BibitemOpen
  \bibfield  {author} {\bibinfo {author} {\bibfnamefont {J.~V.}\ \bibnamefont
  {Leyendekkers}}\ and\ \bibinfo {author} {\bibfnamefont {R.~J.}\ \bibnamefont
  {Hunter}},\ }\href@noop {} {\bibfield  {journal} {\bibinfo  {journal}
  {J.~Chem. Phys.}\ }\textbf {\bibinfo {volume} {82}},\ \bibinfo {pages} {1440}
  (\bibinfo {year} {1985})}\BibitemShut {NoStop}%
\end{thebibliography}%

\end{document}